\documentclass[11pt,a4paper]{article}
\usepackage{amsmath,amssymb,mathrsfs}
\usepackage{subfigure}
\usepackage{url}

\usepackage{graphicx}
\usepackage{hyperref}

\usepackage{cite}

\allowdisplaybreaks[4]

\title{Primordial gravitational waves, precisely}
\begin{document}

\renewcommand{\theequation}{\thesection.\arabic{equation}}

%%%%%%%%%%%%%%%%%%%%%%%%%%%%%%%%%%%%%%%%%%%%%%%%%%
%%%%%%%%%%%%%%%%%%%%%%%%%%%%%%%%%%%%%%%%%%%%%%%%%%

\hfill{} IPMU18-0037

\hfill{} MPP-2018-19

\vspace{1.0 truecm}

\begin{center}

{\textbf{\LARGE
Primordial gravitational waves, precisely:\\
\vspace{3 truemm}
The role of thermodynamics in the Standard Model
}}

\bigskip

\vspace{0.5 truecm}

{\bf Ken'ichi Saikawa$^1$ and Satoshi Shirai$^2$} \\[5mm]

\begin{tabular}{lc}
&\!\! {$^1$ \em Max-Planck-Institut f\"ur Physik (Werner-Heisenberg-Institut),}\\
&{\em F\"ohringer Ring 6, D-80805 M\"unchen, Germany}\\[.4em]
&\!\! {$^2$ \em Kavli Institute for the Physics and Mathematics of the Universe (Kavli IPMU),}\\
&{\em UTIAS, WPI, The University of Tokyo, Kashiwa, Chiba 277-8568, Japan}\\[.4em]
\end{tabular}

\vspace{1.0 truecm}

{\bf Abstract}
\end{center}

\begin{quote}
In this paper, we revisit the estimation of the spectrum of primordial gravitational waves originated from inflation, 
particularly focusing on the effect of thermodynamics in the Standard Model of particle physics.
By collecting recent results of perturbative and non-perturbative analysis of thermodynamic quantities in the Standard Model, we obtain 
the effective degrees of freedom including the corrections due to non-trivial interaction properties of particles in the Standard Model for a wide temperature interval.
The impact of such corrections on the spectrum of primordial gravitational waves as well as the damping effect due to free-streaming particles is investigated by numerically solving the evolution equation of tensor perturbations in the expanding universe.
It is shown that the reevaluation of the effects of free-streaming photons and neutrinos gives rise to some additional damping features overlooked in previous studies.
We also observe that the continuous nature of the QCD crossover
results in a smooth spectrum
for modes that reenter the horizon at around the epoch of 
the QCD phase transition.
Furthermore, we explicitly show that the values of the effective degrees of freedom remain smaller than the commonly used value 106.75 even at temperature much higher than the critical
temperature of the electroweak crossover, and that the amplitude of primordial gravitational waves at a frequency range relevant to direct detection experiments
becomes $\mathcal{O}(1)\,\%$ larger than previous estimates that do not include such corrections.
This effect can be relevant to future high-sensitivity gravitational wave experiments such as ultimate DECIGO. 
Our results on the temperature evolution of the effective degrees of freedom are 
made available as tabulated data and fitting functions, which can also be used in the analysis of other cosmological relics.
\end{quote}

\thispagestyle{empty}

{
\vfill\flushleft
\noindent\rule{6 truecm}{0.4pt}\\
{\small  E-mail addresses: \tt saikawa@mpp.mpg.de, satoshi.shirai@ipmu.jp}
}

\newpage

\tableofcontents

\renewcommand{\thepage}{\arabic{page}}
\renewcommand{\thefootnote}{\arabic{footnote}}
\setcounter{footnote}{0}
%%%%%%%%%%%%%%%%%%%%%%%%%%%%%%%%%%%%%%%%%%%%%%%%%%

%%%%%%%%%%%%%%%%%%%%%%%%%%%%%%%%%%%%%%%%%%%%%%%%%%
\section{Introduction}
\label{sec:introduction}
\setcounter{equation}{0}
%%%%%%%%%%%%%%%%%%%%%%%%%%%%%%%%%%%%%%%%%%%%%%%%%%

The recent detections of gravitational waves (GWs) in the 
Laser Interferometer Gravitational-Wave Observatory (LIGO) and Virgo~\cite{Abbott:2016blz,Abbott:2016nmj,Abbott:2017vtc,Abbott:2017oio,TheLIGOScientific:2017qsa,Abbott:2017gyy}
opened up new possibilities to investigate various astrophysical phenomena which cannot be probed 
by the conventional methods based on electromagnetic waves.
Although the ongoing detectors are only sensitive to strong transient events such as merging black holes,
future GW experiments are expected to detect much weaker signatures of GWs produced in the early universe [see, e.g. Refs.~\cite{Maggiore:1999vm,Guzzetti:2016mkm,Caprini:2018mtu} for reviews].
Several space-borne interferometers such as Laser Interferometer Space Antenna (LISA)~\cite{Audley:2017drz} 
and Deci-hertz Interferometer Gravitational Wave Observatory (DECIGO)~\cite{Seto:2001qf,Kawamura:2006up}
are planned to be launched in the future with the aim of detecting the primordial GW background.
Given these perspectives, it will become more important to improve the precision of theoretical calculations of the
primordial GWs in order to distinguish various models describing the history of the early universe.

The existence of the primordial GW background is one of the most crucial predictions of the inflationary scenario of the early universe~\cite{Grishchuk:1974ny,Starobinsky:1979ty}.
In order to estimate the spectrum of the inflationary GWs, basically we need to know two factors:
One is the power spectrum of primordial tensor perturbations generated during inflation,
and the other is the expansion rate of the universe from the end of inflation until today.
The former defines the initial
magnitude of the GW signature, and it is directly associated with the detailed properties of inflationary models~\cite{Turner:1993vb,Turner:1996ck,Smith:2005mm}.
On the other hand, the latter describes how the density of the primordial GWs has been diluted in subsequent stages
of the cosmic expansion.
In other words, the primordial GWs ``know" how the universe has evolved after inflation.
This fact implies that there is a possibility to obtain detailed information about the early history of the universe 
as well as the inflationary models by observing the spectrum of the primordial GWs.

It is known that in the standard slow-roll inflationary models
the spectrum of the inflationary GW background is almost flat for a broad range of frequencies,
since the amplitude of GWs gets a common dilution factor if the corresponding mode reenters the horizon during the radiation dominated era.
The deviation from the flat spectrum can be associated with either the tilt of the primordial power spectrum or the fact that 
the hot plasma produced after inflation does not exactly behave like an ideal gas of massless particles,  
which can be caused by non-trivial interaction properties of elementary particles.
In order to evaluate the latter effect correctly, we need to know the concrete theory of particle physics.
The Standard Model (SM), which formulates all fundamental interactions of quarks and leptons in terms of 
electromagnetic, weak, and strong forces, can be regarded as a ``benchmark" model to describe the properties of the primordial plasma.
The non-trivial interactions of the SM plasma can affect the spectrum of the primordial GWs, which has to be
quantitatively taken into account.

The intervening effects which alter the nature of the
spectrum of GWs during their propagation are represented as a transfer function, and we have to follow several steps in order to calculate it correctly.
First, Schwarz~\cite{Schwarz:1997gv} explicitly pointed out that the deviation of the behavior of fluid components of the universe from that of a simple ideal gas of relativistic particles 
is imprinted in the spectrum of GWs, and that the observation of the spectrum of GWs can be used to probe the equation of state of the early universe [see also Ref.~\cite{Seto:2003kc}]. 
Second, Weinberg~\cite{Weinberg:2003ur} 
considered the effects of neutrinos, which decouple from the thermal bath and start to free-stream when the universe becomes cooled below $\sim\mathcal{O}(1)\,\mathrm{MeV}$. It was 
shown that such neutrinos damp the amplitude of GWs by 35.5\,\% in the frequency range between $\sim 10^{-16}\,\mathrm{Hz}$ and $\sim 10^{-10}\,\mathrm{Hz}$.\footnote{It appears that the damping effect due to the free-streaming particles was already pointed out in earlier works [see, e.g. Refs.~\cite{Vishniac:1982,Rebhan:1994zw}]. 
Weinberg's paper was the first one to point out that this effect might indeed be relevant to observations.}
This damping effect due to the free-streaming particles was further investigated by 
several authors~\cite{Bashinsky:2005tv,Dicus:2005rh,Mangilli:2008bw,Stefanek:2012hj,Dent:2013asa,Baym:2017xvh}.
A more precise computational method was developed by Watanabe and Komatsu~\cite{Watanabe:2006qe},
and it was pointed out 
that the change of the effective degrees of freedom in the SM elementary particles modifies the spectrum of GWs. 
In addition to such an effect, Boyle and Steinhardt~\cite{Boyle:2005se} explored general possibilities that the 
equation of state might be modified by some non-trivial features of interacting particles. 
Finally, Kuroyanagi, Chiba, and Sugiyama~\cite{Kuroyanagi:2008ye} performed full numerical calculations of 
the inflationary GW spectrum over broad frequency ranges by taking into account the scalar field dynamics during inflation, the process during the reheating after inflation, the damping 
due to the free-streaming neutrinos, and the change of the effective degrees of freedom.

It should be noted that the current estimate of the transfer function is still incomplete in spite of several developments described above. 
The key ingredients considered here are the equation of state of the primordial plasma,
\begin{equation}
w(T) \equiv \frac{p(T)}{\rho(T)}, \label{w_definition}
\end{equation}
and the effective degrees of freedom,
\begin{equation}
g_{*\rho}(T) \equiv \frac{\rho(T)}{\left[\frac{\pi^2 T^4}{30}\right]},\quad g_{*s}(T) \equiv \frac{s(T)}{\left[\frac{2\pi^2 T^3}{45}\right]}, \label{gstar_definition}
\end{equation}
where $p(T)$, $\rho(T)$, and $s(T)$ are the pressure, energy density, and entropy density of the primordial plasma at temperature $T$.
In most previous studies of primordial GWs (and also in those of other cosmological relics), 
the values of $g_{*\rho}(T)$ and $g_{*s}(T)$ were estimated based on the ideal gas approximation,
and the equation of state was assumed to be $1/3$ in the radiation dominated universe.
However, these approximations do not always hold once we take account of the effect of interactions of elementary particles.
It is reasonable to expect that $g_{*\rho}$ and $g_{*s}$ as well as $w$ can deviate from commonly used values
due to the non-trivial properties of particle interactions even in the SM of particle physics,
and in principle such corrections can be imprinted on the spectrum of primordial GWs.
The main question addressed in this paper is to what extent the shape of the spectrum of primordial GWs
can be modified if we systematically include the corrections on the thermodynamic quantities of the primordial plasma
arising from particle interactions in the SM.

In order to estimate the thermodynamic properties of interacting particles precisely, we must use state-of-the-art methods of quantum field theory.
In the literature, there has been a lot of work on the estimation of thermodynamic quantities in the SM based on the finite temperature field theory,
and we aim to collect all the results to reconstruct the relevant quantities such as the equation of state parameter
and the effective degrees of freedom at arbitrary temperature.
We show that the improvement in estimation of thermodynamic properties of high temperature plasma in the SM
has several impacts on the calculation of the spectrum of primordial GWs, and some of them are 
relevant to future high-sensitivity GW experiments.
As a by-product, we also obtain some detailed information on the temperature evolution of $g_{*\rho}(T)$ and $g_{*s}(T)$ and their uncertainty 
for a wide temperature interval, which is made available as tabulated data and fitting functions.

The organization of this paper is as follows. In Sec.~\ref{sec:IGW}, we describe some basic properties of inflationary GWs
and discuss several damping effects that can take place during their evolution after inflation.
In Sec.~\ref{sec:EoS_in_the_SM}, we estimate the effective degrees of freedom $g_{*\rho}(T)$ and $g_{*s}(T)$ 
and the equation of state parameter $w(T)$ at arbitrary temperature by collecting
recent developments including the analysis of the neutrino decoupling, 
quantum chromodynamics (QCD) phase transition, and electroweak phase transition.
After obtaining the updated values of the effective degrees of freedom in the SM,
in Sec.~\ref{sec:spectrum_of_GW} we calculate the spectrum of GWs based on them
and highlight difference from the results obtained in previous studies.
Sec.~\ref{sec:conclusion} is devoted to conclusions and discussion.
Some technical details are presented in appendixes.
Appendix~\ref{app:supplementary_material} deals with our tabulated data of $g_{*\rho}(T)$ and $g_{*s}(T)$ obtained in Sec.~\ref{sec:EoS_in_the_SM}
and that of the transfer function of GWs obtained in Sec.~\ref{sec:spectrum_of_GW}.
Appendix~\ref{app:thermodynamic} is devoted to a short review of thermodynamic quantities used in the main text.
In appendix~\ref{app:fitting_functions}, we give fitting functions for $g_{*\rho}(T)$ and $g_{*s}(T)$ 
that remain consistent with the actual data obtained in Sec.~\ref{sec:EoS_in_the_SM} within a range of uncertainty.
In appendix~\ref{app:sensitivity}, we describe some details about sensitivity of ultimate DECIGO mentioned in Sec.~\ref{sec:spectrum_of_GW}.

%%%%%%%%%%%%%%%%%%%%%%%%%%%%%%%%%%%%%%%%%%%%%%%%%%
\section{Inflationary gravitational waves}
\label{sec:IGW}
\setcounter{equation}{0}
%%%%%%%%%%%%%%%%%%%%%%%%%%%%%%%%%%%%%%%%%%%%%%%%%%

In this section, we describe how to estimate the spectrum of the inflationary GW background and enumerate several post-inflationary events that can affect its shape.
First, we consider the evolution of GWs in the expanding universe and derive the relation among the energy density of GWs,
the primordial tensor power spectrum and the transfer function in Sec.~\ref{sec:IGW_energy_density}.
After that, in Sec.~\ref{sec:IGW_damping} we discuss various damping effects of the primordial tensor perturbations encoded in the transfer function.

%%%%%%%%%%%%%%%%%%%%%%%%%%%%%%%%%%%%%%%%%%%%%%%%%%
\subsection{\label{sec:IGW_energy_density} Energy density of gravitational waves}
%%%%%%%%%%%%%%%%%%%%%%%%%%%%%%%%%%%%%%%%%%%%%%%%%%

We work in a spatially flat Friedmann-Robertson-Walker (FRW) background with a metric,
\begin{equation}
ds^2 = -dt^2 + a^2(t)(\delta_{ij}+h_{ij})dx^idx^j,
\end{equation}
where GWs are represented by spatial metric perturbations ($|h_{ij}|\ll 1$) that satisfy the transverse-traceless conditions, $h^i_i=\partial^ih_{ij}=0$.
The evolution of GWs is described by the linearized Einstein equation,
\begin{equation}
\ddot{h}_{ij} + 3H\dot{h}_{ij} - \frac{\nabla^2}{a^2}h_{ij} = 16\pi G\Pi^{\rm TT}_{ij}, \label{GW_waveequation_xspace}
\end{equation}
where a dot represents a derivative with respect to cosmic time $t$, and $G$ is the Newton's gravitational constant.
$\Pi^{\rm TT}_{ij}$ are the transverse-traceless part of the anisotropic stress $\Pi_{ij}$,
which are defined in terms of the spatial components of the stress-energy tensor $T_{ij}$, those of the metric tensor $g_{ij}$,
and the background homogeneous pressure $p$, 
\begin{equation}
a^2\Pi_{ij} = T_{ij}-pg_{ij}.
\end{equation}
It is convenient to decompose $h_{ij}$ into their Fourier modes,
\begin{equation}
h_{ij}(t,{\bf x}) = \sum_{\lambda}\int\frac{d^3k}{(2\pi)^3}h^{\lambda}(t,{\bf k})\epsilon^{\lambda}_{ij}({\bf k})e^{i{\bf k\cdot x}},
\end{equation}
where $\lambda=+,\times$ specifies two independent polarization states,
and $\epsilon^{\lambda}_{ij}({\bf k})$ are the spin-2 polarization tensors satisfying the normalization conditions $\sum_{ij}\epsilon^{\lambda}_{ij}(\epsilon^{\lambda'}_{ij})^*=2\delta^{\lambda\lambda'}$.
In terms of the Fourier modes, Eq.~\eqref{GW_waveequation_xspace} can be rewritten as
\begin{equation}
\ddot{h}^{\lambda}_{\bf k} + 3H\dot{h}^{\lambda}_{\bf k} + \frac{k^2}{a^2}h^{\lambda}_{\bf k} = 16\pi G\Pi^{\lambda}_{\bf k}, \label{GW_waveequation_kspace}
\end{equation}
where $h^{\lambda}_{\bf k}(t) \equiv  h^{\lambda}(t,{\bf k})$, and $\Pi^{\lambda}_{\bf k}$ are corresponding Fourier components of the source term $\Pi^{\rm TT}_{ij}$.

The GW background is originated from the quantum fluctuations during inflation, which are assumed to be classicalized after
the corresponding modes cross outside the horizon ($k<aH$).
For the modes far outside the horizon ($k\ll aH$), we can ignore the source term and the third term of the left-hand side of Eq.~\eqref{GW_waveequation_kspace},
which implies that $h^{\lambda}_{\bf k}\propto \text{const}$. At some point after inflation, these modes reenter the horizon ($k>aH$) and start to evolve
according to Eq.~\eqref{GW_waveequation_kspace}.
Based on these facts, we write the solution of Eq.~\eqref{GW_waveequation_kspace} as
\begin{equation}
h^{\lambda}_{\bf k} = h^{\lambda}_{{\bf k},\mathrm{prim}}\chi(t,k), \label{h_prim_T}
\end{equation}
where $h^{\lambda}_{{\bf k},\mathrm{prim}}$ represents the amplitude of the primordial tensor perturbations,
and $\chi(t,k)$ is the transfer function.
The magnitude of the transfer function is normalized such that $\chi(t,k) \to 1$ for $k\ll aH$.
If we ignore the source therm ($\Pi^{\lambda}_{\bf k}=0$), the evolution of the modes deep inside the horizon ($k\gg aH$)
is well approximated by the WKB solution, $\chi(\tau,k)\propto a^{-1}e^{\pm ik\tau}$, where $\tau$ is conformal time defined by $d\tau = dt/a$.

The energy density of the relic GWs is given by~\cite{Maggiore:1900zz}
\begin{equation}
\rho_{\rm gw}(t) = \frac{1}{32\pi G}\langle\dot{h}_{ij}(t,{\bf x})\dot{h}_{ij}(t,{\bf x})\rangle_{\rm av} = \frac{1}{32\pi G}\sum_{\lambda}\int\frac{d^3 k}{(2\pi)^3}2 |\dot{h}^{\lambda}_{\bf k}|^2,
\end{equation}
where $\langle\dots\rangle_{\rm av}$ denotes an spatial average. The spectrum of GWs is described in terms of the fraction of their energy density
per logarithmic frequency interval,
\begin{equation}
\Omega_{\rm gw}(t,k) = \frac{1}{\rho_{\rm crit}(t)}\frac{d\rho_{\rm gw}(t,k)}{d\ln k}, \label{Omega_gw_def}
\end{equation}
where $\rho_{\rm crit} = 3H^2/8\pi G$ is the critical energy density of the universe. Substituting Eq.~\eqref{h_prim_T}, we obtain
\begin{equation}
\Omega_{\rm gw}(\tau,k) = \frac{1}{12 a^2(\tau) H^2(\tau)}\mathcal{P}_T(k)\left[\chi'(\tau,k)\right]^2, \label{Omega_gw_PT_T}
\end{equation}
where a prime represents a derivative with respect to conformal time $\tau$, i.e. $' = \frac{d}{d\tau}=a\frac{d}{dt}$.
Here we introduced the primordial tensor power spectrum $\mathcal{P}_T(k)$, which is determined by the Hubble parameter 
at the time when the corresponding mode crosses the horizon during inflation ($k=aH$),
\begin{equation}
\mathcal{P}_T(k) \equiv \frac{k^3}{\pi^2}\sum_{\lambda}|h^{\lambda}_{{\bf k},\mathrm{prim}}|^2 = \left.\frac{2H^2}{\pi^2 M_{\rm Pl}^2}\right|_{k\, =\, aH},
\end{equation}
where $M_{\rm Pl}\simeq 2.435\times 10^{18}\mathrm{GeV}$ is the reduced Planck mass.

Let us estimate the typical amplitude of GWs at the present time $\tau=\tau_0$ for the modes that reenter the horizon during the radiation dominated era.
From the fact that the transfer function can be approximated by the WKB solution ($\chi\propto a^{-1}e^{\pm ik\tau}$) after the modes reenter the horizon,
and that it should approach to $\chi\to 1$ at early times, we can make the following replacement
\begin{equation}
\left[\chi'(\tau,k)\right]^2 \approx \frac{k^2}{2}\left(\frac{a(\tau_{\rm hc})}{a(\tau)}\right)^2 = \frac{a^4(\tau_{\rm hc})H^2(\tau_{\rm hc})}{2a^2(\tau)}, \label{transfer_function_approximate}
\end{equation}
where $\tau_{\rm hc}$ represents the conformal time of the horizon crossing, and we used $k=a(\tau_{\rm hc})H(\tau_{\rm hc})$ in the second equality.
The factor $1/2$ arises from the time average of the rapidly oscillating function.
If the horizon crossing happens during the radiation dominated era, we can also use the following relation,
\begin{equation}
\frac{H^2_{\rm hc}}{H_0^2} = \frac{\frac{\pi^2}{30}g_{*\rho,{\rm hc}}T_{\rm hc}^4}{\rho_{\rm crit,0}} 
= \Omega_{\gamma} \left(\frac{g_{*\rho,{\rm hc}}}{2}\right)\left(\frac{g_{*s,{\rm hc}}}{g_{*s,\mathrm{fin}}}\right)^{-\frac{4}{3}}\left(\frac{a_0}{a_{\rm hc}}\right)^4, \label{H_hc_to_H_0}
\end{equation}
where the subscript ``$0$" and ``${\rm hc}$" represent the quantity at the time $\tau_0$ and $\tau_{\rm hc}$, respectively,
and $g_{*\rho}$ and $g_{*s}$ are the effective degrees of freedom for the energy density and for the entropy density, respectively.
$\Omega_{\gamma} = \rho_{\gamma,0}/\rho_{\rm crit,0}$ is the fraction of the energy density of photons $\rho_{\gamma,0}=(\pi^2/15)T_{\gamma,0}^4$ at the present time.
Using the measured value $T_{\gamma,0} = 2.72548(57)\,\mathrm{K}$~\cite{Fixsen:2009ug} for current cosmic microwave background (CMB) temperature, 
we obtain $\Omega_{\gamma}h^2 = 2.4728(21)\times 10^{-5}$, 
where $h$ is the renormalized Hubble parameter, $H_0= 100\,h\,\mathrm{km}\,\mathrm{sec}^{-1}\mathrm{Mpc}^{-1}$.
In the last equality in Eq.~\eqref{H_hc_to_H_0}, we used the conservation of entropy, $g_{*s}a^3T^3=\text{const}$.
Note that we have introduced the notation $g_{*s,\mathrm{fin}}$ rather than $g_{*s,0}$ in Eq.~\eqref{H_hc_to_H_0},
where $g_{*s,\mathrm{fin}}$ is the value of $g_{*s}$ evaluated after the neutrino decoupling.
We use this notation in order to remind that the value of $g_{*s}$ is not directly evaluated at the present time, 
but indirectly estimated at the epoch of the neutrino decoupling. Actually, this procedure gives rise to some ambiguities as we discuss in Sec.~\ref{sec:photons_and_leptons}. 

Substituting Eqs.~\eqref{transfer_function_approximate} and~\eqref{H_hc_to_H_0} into Eq.~\eqref{Omega_gw_PT_T}, we obtain
\begin{equation}
\Omega_{\rm gw}(\tau_0,k) \approx \frac{1}{24}\Omega_{\gamma} \left(\frac{g_{*\rho,{\rm hc}}}{2}\right)\left(\frac{g_{*s,{\rm hc}}}{g_{*s,\mathrm{fin}}}\right)^{-\frac{4}{3}} \mathcal{P}_T(k). \label{Omega_gw_gs_PT}
\end{equation}
We see that the spectrum shown in Eq.~\eqref{Omega_gw_gs_PT} is almost flat except for the (weak) scale dependences in $\mathcal{P}_T(k)$, $g_{*\rho,{\rm hc}}$ and $g_{*s,{\rm hc}}$.
It should be noted that the amplitude of GWs for a given mode is related to the Hubble parameter $H_{\rm inf}$ or the energy scale of inflation $V^{1/4}_{\rm inf}$
when the corresponding mode exits the horizon,
\begin{align}
\Omega_{\rm gw}(\tau_0,k)h^2 &\approx \frac{1}{3}\Omega_{\gamma}h^2 \left(\frac{g_{*\rho,{\rm hc}}}{2}\right)\left(\frac{g_{*s,{\rm hc}}}{g_{*s,\mathrm{fin}}}\right)^{-\frac{4}{3}} \left(\frac{H_{\rm inf}}{2\pi M_{\rm Pl}}\right)^2 \nonumber\\
&\approx 1.29\times 10^{-17}\left(\frac{g_{*s,\mathrm{fin}}}{3.931}\right)^{\frac{4}{3}}\left(\frac{g_{*\rho,{\rm hc}}}{106.75}\right)\left(\frac{g_{*s,{\rm hc}}}{106.75}\right)^{-\frac{4}{3}}\bigg(\frac{V^{1/4}_{\rm inf}}{10^{16}\,\mathrm{GeV}}\bigg)^4.
\end{align}
Furthermore, the frequency $f$ of GWs can be related to the temperature $T_{\rm hc}$ at which the corresponding mode reenters the horizon,
\begin{equation}
f = \frac{k}{2\pi a_0} = \frac{H_{\rm hc}}{2\pi}\frac{a_{\rm hc}}{a_0} \approx 2.65\,\mathrm{Hz}\left(\frac{g_{*s,\mathrm{fin}}}{3.931}\right)^{\frac{1}{3}}\left(\frac{g_{*\rho,{\rm hc}}}{106.75}\right)^{\frac{1}{2}}\left(\frac{g_{*s,{\rm hc}}}{106.75}\right)^{-\frac{1}{3}}\left(\frac{T_{\rm hc}}{10^8\,\mathrm{GeV}}\right).
\end{equation}

We emphasize that Eq.~\eqref{Omega_gw_gs_PT} is derived based on the assumption that the transfer function is exactly 
given by the WKB solution after the horizon crossing.
Strictly speaking, this is not the case if the equation of state parameter $w$ of the background fluid
deviates from the value for pure radiation $1/3$ around the time of the horizon crossing.
Accordingly, the amplitude of GWs can be smaller (or larger) than the right-hand side of Eq.~\eqref{Omega_gw_gs_PT} under certain conditions.
We further elaborate on this point in Sec.~\ref{sec:spectrum_of_GW}.

%%%%%%%%%%%%%%%%%%%%%%%%%%%%%%%%%%%%%%%%%%%%%%%%%%
\subsection{\label{sec:IGW_damping} Damping effects}
%%%%%%%%%%%%%%%%%%%%%%%%%%%%%%%%%%%%%%%%%%%%%%%%%%

Various phenomena occurring after inflation can affect the transfer function, which results in a non-trivial shape of the spectrum of GWs at the present time.
One important effect is the collisionless damping caused by free-streaming neutrinos.\footnote{In this paper, we only consider the damping effects due to the SM neutrinos and photons.
If there exist additional free-streaming particles, they can cause similar damping effects. Such possibilities are explicitly analyzed in Refs.~\cite{Jinno:2012xb,Jinno:2013xqa}.
We also note that the damping effect due to cold dark matter is negligibly small since the fraction of its kinetic energy is small or
its contribution to the energy density is small compared to the total energy density of the universe~\cite{Flauger:2017ged}.}
Neutrinos decouple from the thermal bath when the temperature drops below $\sim 2\,\mathrm{MeV}$,
and the subsequent evolution of their distribution function is described by the collisionless Boltzmann equation.
The existence of primordial GWs causes a tensor type perturbation in the neutrino distribution function,
which contributes to the transverse-traceless part of the anisotropic stress and results in the energy flow from neutrinos to GWs, or vice versa.

At linear order in perturbation theory, the contribution to the anisotropic stress is simply given by~\cite{Weinberg:2003ur,Watanabe:2006qe}
\begin{equation}
\Pi_{\bf k}^{\lambda} = -4\rho_{\nu}(\tau)\int^{\tau}_{\tau_{\nu {\rm dec}}} d\tau' \frac{j_2[k(\tau-\tau')]}{k^2(\tau-\tau')^2}\frac{dh^{\lambda}_{\bf k}(\tau')}{d\tau'}, \label{neutrino_anisotropic_stress}
\end{equation}
where $\rho_{\nu}(\tau)$ is the homogeneous energy density of neutrinos, $\tau_{\nu {\rm dec}}$ is the conformal time of the neutrino decoupling,
and $j_n(z)$ is the spherical Bessel function of the first kind.
The evolution equation of GWs [Eq.~\eqref{GW_waveequation_kspace}] becomes an integro-differential equation, which can be solved numerically.
It was shown that the effect of the neutrino free-streaming leads to the damping of the amplitude of primordial GWs by 35.5\,\% in the frequency range
$10^{-16}\,\mathrm{Hz}\lesssim f \lesssim 10^{-10}\,\mathrm{Hz}$.
Note that this damping effect does not work for the modes that reenter the horizon before the neutrino decoupling ($f\gtrsim 10^{-10}\,\mathrm{Hz}$),
since these modes are rapidly oscillating already at the time of the neutrino decoupling and there is no net energy conversion between neutrinos and GWs.
This effect also becomes less important for the modes that reenter the horizon after the time of mater-radiation equality,
since the energy density of neutrinos becomes small compared to the total energy density of the universe during the matter dominated era.

In principle, the damping of primordial GWs can also be caused by photons, which decouple from the thermal bath at $T \sim 3000\,\mathrm{K}$.
We can describe this contribution analogous to Eq.~\eqref{neutrino_anisotropic_stress}:
\begin{equation}
\Pi_{\bf k}^{\lambda} = -4\rho_{\gamma}(\tau)\int^{\tau}_{\tau_{\rm ls}} d\tau' \frac{j_2[k(\tau-\tau')]}{k^2(\tau-\tau')^2}\frac{dh^{\lambda}_{\bf k}(\tau')}{d\tau'}, \label{photon_anisotropic_stress}
\end{equation}
where $\tau_{\rm ls}$ is the conformal time of the photon last scattering.
Although this effect must exist in the standard cosmology, it was overlooked in most previous studies on the spectrum of primordial GWs.
This is because the photon last scattering occurs after the time of matter-radiation equality and the damping effect soon becomes irrelevant
as the energy density of photons becomes small compared to the total energy density of the universe.
Nevertheless, we take account of this contribution in the analysis in this paper for the sake of completeness.
We actually see that the inclusion of the contribution of photons to the anisotropic stress leads to an additional damping of the amplitude of GWs
by $\lesssim 14\,\%$ at $f \sim 10^{-17}\,\mathrm{Hz}$ as shown in Sec.~\ref{sec:spectrum_of_GW}.

Another important effect is the change of the equation of state of the universe~\cite{Schwarz:1997gv,Seto:2003kc}.
To see this effect, let us consider the evolution of the energy density of GWs relative to that of the total energy density of the universe.
From Eq.~\eqref{Omega_gw_def}, we have
\begin{equation}
\frac{d\ln\Omega_{\rm gw}(\tau,k)}{d\ln a} = \frac{d\ln\tilde{\rho}_{\rm gw}(\tau,k)}{d\ln a} - \frac{d\ln\rho_{\rm crit}(\tau)}{d\ln a}, \label{dlnOmega_gw_dlna}
\end{equation}
where $\tilde{\rho}_{\rm gw} \equiv d\rho_{\rm gw}/d\ln k$.
The evolution of the total energy density of the universe is determined by the equation of state [see Eq.~\eqref{energy_conservation}],
\begin{equation}
\frac{d\ln\rho_{\rm crit}(\tau)}{d\ln a} = -3\left(1+w(a)\right). \label{dlnrho_crit_dlna}
\end{equation}
On the other hand, if there is no energy flow due to the anisotropic stress, the energy density of GWs after the horizon crossing is simply diluted as
\begin{equation}
\frac{d\ln\tilde{\rho}_{\rm gw}(\tau,k)}{d\ln a} \approx -4. \label{dlnrho_gw_dlna}
\end{equation}
Note that the above equation does not exactly hold around the time of the horizon crossing
and that there would be some modifications according to the behavior of the equation of state parameter $w$ at that time.
Although we stick to the approximation~\eqref{dlnrho_gw_dlna} in this section, 
we quantify the corresponding modifications via numerical methods in Sec.~\ref{sec:spectrum_of_GW}.
Combining Eqs.~\eqref{dlnOmega_gw_dlna}-\eqref{dlnrho_gw_dlna}, we obtain
\begin{equation}
\Omega_{\rm gw}(\tau,k) \approx \exp\left[\int^a_{a_{\rm hc}}\left(3w(a)-1\right)d\ln a\right] \Omega_{\rm gw}(\tau_{\rm hc},k). \label{Omega_gw_damp_w}
\end{equation}
From Eqs.~\eqref{Omega_gw_PT_T} and~\eqref{transfer_function_approximate}, we see that the scale dependence of $\Omega_{\rm gw}$ at the time of the horizon crossing is determined by the
primordial tensor power spectrum: $\Omega_{\rm gw}(k,\tau_{\rm hc})\sim \mathcal{P}_T(k)$.
Therefore, some feature of the present spectrum of GWs that is not caused by the primordial power spectrum can be characterized by the change of the equation of state parameter $w$.

Any non-trivial evolution of the equation of state after inflation can lead to the modification of the spectrum of GWs.
For example, the inflaton field may undergo an oscillation around the minimum of its potential before it decays into radiations to reheat the universe after inflation.
Such a dynamics of the inflaton field at the reheating stage causes some modification of
the equation of state and results in a damping feature in the spectrum of GWs~\cite{Seto:2003kc,Nakayama:2008ip,Nakayama:2008wy,Kuroyanagi:2011fy,Jinno:2014qka,Kuroyanagi:2014qza}.

The equation of state of the universe can vary even in the radiation dominated era, which leads to a deviation from the flat spectrum of GWs.
We have already seen this fact in Eq.~\eqref{Omega_gw_gs_PT}, which explicitly shows that there exists non-trivial frequency dependence 
according to the values of $g_{*\rho,\mathrm{hc}}$ and $g_{*s,\mathrm{hc}}$.
Indeed, it is straightforward to show that the frequency dependence caused by $g_{*\rho,\mathrm{hc}}$ and $g_{*s,\mathrm{hc}}$ 
is equivalent to what shown in Eq.~\eqref{Omega_gw_damp_w}.
For the modes $k_1$ and $k_2$ that reenter the horizon at $\tau_{\rm hc,1}$ and $\tau_{\rm hc,2}$, respectively,
we obtain (see Appendix~\ref{app:thermodynamic} for a more explicit proof)
\begin{align}
\frac{\Omega_{\rm gw}(\tau_0,k_2)}{\Omega_{\rm gw}(\tau_0,k_1)} &\approx \exp\left[\int^{a_{\rm hc,1}}_{a_{\rm hc,2}}\left(3w(a)-1\right)d\ln a\right]\left[\frac{\Omega_{\rm gw}(\tau_{\rm hc,2},k_2)}{\Omega_{\rm gw}(\tau_{\rm hc,1},k_1)}\right] \nonumber\\
&= \left(\frac{g_{*\rho,\mathrm{hc,2}}}{g_{*\rho,\mathrm{hc,1}}}\right)\left(\frac{g_{*s,\mathrm{hc,2}}}{g_{*s,\mathrm{hc,1}}}\right)^{-\frac{4}{3}}\left[\frac{\Omega_{\rm gw}(\tau_{\rm hc,2},k_2)}{\Omega_{\rm gw}(\tau_{\rm hc,1},k_1)}\right]. \label{Omega_gw_damp_gs}
\end{align}
If the primordial tensor power spectrum is completely flat, we have $\Omega_{\rm gw}(\tau_{\rm hc,1},k_1) = \Omega_{\rm gw}(\tau_{\rm hc,2},k_2)$,
and the variation of the amplitude of GWs at the interval between $k_1$ and $k_2$ is mostly determined by
the integration of the equation of state parameter, or equivalently, the change of the effective degrees of freedom.

The change of the effective degrees of freedom can be understood by noting that the contribution of a particle with mass $m$ to 
radiations is exponentially suppressed when the temperature $T$ becomes smaller than $m$.
For this reason, we expect that the values of $g_{*\rho,\mathrm{hc}}$ and $g_{*s,\mathrm{hc}}$ change at $T_{\rm hc}\sim m$,
and that there exits a damping of the amplitude of GWs at the corresponding frequency.
If the values of $g_{*\rho}$ and $g_{*s}$ are almost the same at high temperatures, which is the case in the SM,
the damping factor can be approximated as $(g_{*\rho,\mathrm{hc}})(g_{*s,\mathrm{hc}})^{-4/3}\approx (g_{*\rho,\mathrm{hc}})^{-1/3}$.

In addition to the effect mentioned above, the change of the equation of state and effective degrees of freedom can also be caused 
due to the interaction properties of elementary particles composing radiations.
If the interaction among particles is weak enough such that we can treat them as ideal gases
and we ignore the change of the effective degrees of freedom caused by the finite particle masses, the equation of state is exactly given by $w = 1/3$.
However, this description breaks down if we take account of the interaction among elementary particles.
In Sec.~\ref{sec:EoS_in_the_SM}, we explicitly show that the inclusion of particle interactions causes shifts in the values of $g_{*\rho}$ and $g_{*s}$
as well as that of the equation of state parameter $w$ even if the number of particle species of the system remains unchanged.
This effect leads to a further non-trivial change in the spectrum of GWs, and we discuss it more quantitatively in Sec.~\ref{sec:spectrum_of_GW}.

Once we take the interaction properties of elementary particles seriously, we may also speculate that the perfect fluid description of
background radiations does not hold exactly. 
When we deal with such imperfect fluids, we have to include correction terms in the energy-momentum tensor characterized by three quantities,
heat conduction, shear viscosity, and bulk viscosity.
Although the heat conduction and bulk viscosity do not affect the evolution of GWs as far as we consider the linear order in perturbation theory, 
the existence of shear viscosity (parameterized by the shear viscosity coefficient $\eta$) modifies the propagation equation of GWs.
In the flat FRW universe, we have~\cite{Weinberg:1972kfs}
\begin{equation}
\ddot{h}_{ij} + \left(3H + 16\pi G\eta\right)\dot{h}_{ij} - \frac{\nabla^2}{a^2}h_{ij} = 0.
\end{equation}
This fact implies that the amplitude of GWs decays in a viscous medium~\cite{Hawking:1966qi} at a rate $\Gamma_g = 16\pi G\eta$.
However, this damping is inefficient in the expanding universe for the following reasons.
On dimensional grounds, the shear viscosity should behave as $\eta \sim \rho/\Gamma$ with $\Gamma$ being the interaction rate of constituent particles.
Using this estimate in the damping rate and assuming the radiation dominated background ($H^2 \sim G\rho$), 
we obtain $\Gamma_g/H \sim G\rho/\Gamma H \sim H/\Gamma$.
Therefore, the damping rate due to the shear viscosity is negligible compared with the expansion rate as long as 
interactions of constituent particle proceed much faster than the cosmic expansion.\footnote{It was argued that macroscopic hydrodynamic
fluctuations can produce GWs, whose production rate is proportional to the shear viscosity coefficient $\eta$~\cite{Ghiglieri:2015nfa}.
Although such a GW background may exist even in the SM, here we do not consider this effect, since
the spectrum shows a peak at a GHz frequency range and it becomes negligibly small at lower frequencies considered in this paper.}

In contrast to the shear viscosity, the bulk viscosity does not modify the propagation of GWs, 
but it modifies the equation for the background expansion [i.e. the evolution of the scale factor $a(t)$].
In the FRW universe, the effect of the bulk viscosity (parameterized by the bulk viscosity coefficient $\zeta$)
is described by making the following replacement for the pressure~\cite{Weinberg:1972kfs},
\begin{equation}
p \to p - 3H\zeta.
\end{equation}
Hence, the whole effect results in the shift of the equation of state parameter~\cite{Boyle:2005se},
\begin{equation}
w - \frac{1}{3} = -\frac{3H\zeta}{\rho} = - \frac{8\pi G\zeta}{H}.
\end{equation}
However, by using a rough estimate $\zeta \sim \rho/\Gamma$ we again see that the above correction is negligible as long as $\Gamma \gg H$.
Furthermore, it is generically expected that the bulk viscosity coefficient $\zeta$ is much smaller than 
the shear viscosity coefficient $\eta \sim \rho/\Gamma$ for high temperature radiations~\cite{Arnold:2006fz},
and the net effect can be further suppressed.
For these reasons, we can safely neglect the correction due to the bulk viscosity.

Based on the above discussions, we treat the primordial plasma as a perfect fluid, but we still allow a deviation from the ideal gas description.
In the next section, we explore the latter possibility in more detail and quantify its effect by taking account of the particle interactions in the SM.

%%%%%%%%%%%%%%%%%%%%%%%%%%%%%%%%%%%%%%%%%%%%%%%%%%
\section{Equation of state in the Standard Model}
\label{sec:EoS_in_the_SM}
\setcounter{equation}{0}
%%%%%%%%%%%%%%%%%%%%%%%%%%%%%%%%%%%%%%%%%%%%%%%%%%

In order to estimate the equation of state of the early universe filled by the gases of interacting SM particles, we need to follow several steps according to the thermal history predicted by the SM.
In particular, there are a couple of drastic changes according to two kinds of phase transitions: the QCD phase transition and electroweak phase transition.
We note that the nature of these phase transitions might not be so drastic as expected:
The result of lattice QCD simulation indicates that the nature of the QCD phase transition is not a sharp transition but a smooth crossover~\cite{Aoki:2006we},
and the observed value of the Higgs boson mass $m_h \approx 125\,\mathrm{GeV}$~\cite{Aad:2015zhl} implies that the nature of the electroweak symmetry breaking is also
a smooth crossover rather than a strong first order phase transition~\cite{Kajantie:1996mn}.
Hereinafter, we basically rely on these facts and follow an approach based on the equilibrium thermodynamics, 
referring to two kinds of phase transitions as the QCD crossover and electroweak crossover.

In the literature, there has been a lot of progress on the calculation of thermodynamic properties of the primordial plasma in the SM.
At temperatures far above the QCD crossover, we can adopt the perturbative method to estimate the pressure of the quark-gluon plasma.
This was investigated in Refs.~\cite{Shuryak:1977ut,Chin:1978gj,Kapusta:1979fh,Toimela:1982hv,Arnold:1994ps,Arnold:1994eb,Zhai:1995ac,Braaten:1995jr,Kajantie:2002wa},
and the pressure of QCD with massless quarks has been estimated up to $\mathcal{O}(g_s^6\ln(1/g_s))$, where $g_s$ is the gauge coupling of the strong interactions.
Furthermore, at temperatures above the electroweak crossover,
it is also possible to estimate the pressure of the SM plasma via the perturbative method, and the detailed calculations were performed in Refs.~\cite{Gynther:2005dj,Gynther:2005av,Laine:2015kra}.
These perturbative approaches break down when the temperature of the universe becomes comparable to the critical temperatures of 
the QCD crossover and electroweak crossover, and 
we need to adopt some non-perturbative methods to investigate the thermodynamics of interacting plasma.
Recently, there has been some progress in estimation of the equation of state in such non-perturbative phases.
These include the investigation of the equation of state of $2+1$~\cite{Borsanyi:2010cj,Borsanyi:2013bia,Bazavov:2014pvz} and $2+1+1$~\cite{Borsanyi:2016ksw} flavor lattice QCD in the continuum limit, 
and that of the electroweak crossover with the physical Higgs mass of $125\,\mathrm{GeV}$ on the lattice~\cite{DOnofrio:2014rug,DOnofrio:2015gop}.

In this section, we collect the results obtained in the literature and combine them 
to estimate the effective degrees of freedom for the energy density $g_{*\rho}(T)$ and entropy density $g_{*s}(T)$ 
of the SM at arbitrary temperatures.
Note that the energy density $\rho(T)$ and entropy density $s(T)$ can be derived from the pressure 
$p(T)$ by using thermodynamic equations [see Eqs.~\eqref{s_rho_p_relation} and~\eqref{rho_p_relation}].
Therefore, practically we need to know $p(T)$ and its derivative with respect to $T$ by taking account of the interaction among particles in the SM.
To this end, it becomes convenient to introduce the following quantity:
\begin{equation}
\Delta(T) \equiv \frac{\rho(T)-3p(T)}{T^4} = T\frac{d}{dT}\left\{\frac{p(T)}{T^4}\right\},
\end{equation}
where the second equality follows from the thermodynamic equation [Eq.~\eqref{rho_p_relation}].
The quantity $\Delta(T)$ is sometimes called the trace anomaly, since it is related to the trace of the stress-energy tensor $T^{\mu}_{\mu} = \rho - 3p$,
which vanishes at classical level for scale-invariant systems.
Once we obtain $p(T)$ and $\Delta(T)$, the energy density and entropy density can be estimated as
\begin{equation}
\rho(T) = T^4\left[\Delta(T) + \frac{3p(T)}{T^4}\right],\quad s(T) = T^3\left[\Delta(T) + \frac{4p(T)}{T^4}\right], \label{rho_and_s_from_p_and_delta}
\end{equation}
and they are translated into $g_{*\rho}(T)$ and $g_{*s}(T)$ via Eq.~\eqref{gstar_definition}.

As mentioned above, we could obtain all relevant quantities from a single function $p(T)$
if we knew the form of $p(T)$ exactly as a differentiable function at all temperatures.
However, in some cases it becomes difficult to obtain the value of $p(T)$ unambiguously, 
as we discuss in the following subsections.
For this reason, we take $\Delta(T)$ as a fundamental quantity and estimate $p(T)$ by integrating $\Delta(T)$ with appropriate boundary conditions.

In the literature, there are pioneering works to apply the equation of state in the SM to the calculation of 
the Weakly Interacting Massive Particle (WIMP) dark matter abundance~\cite{Hindmarsh:2005ix,Drees:2015exa}
and that of the axion dark matter abundance~\cite{Wantz:2009it,Borsanyi:2016ksw}.
We basically follow a similar approach, but take different steps to estimate the effective degrees of freedom and their uncertainty at high temperatures
up to those far above the electroweak crossover,
as our main focus in this paper is to apply it to the calculation of the spectrum of primordial GWs.\footnote{See Refs.~\cite{Schwarz:1997gv,Schettler:2010dp} for some earlier attempts
to apply the equation of state for strongly interacting particles to the calculation
of the spectrum of primordial GWs.}

%%%%%%%%%%%%%%%%%%%%%%%%%%%%%%%%%%%%%%%%%%%%%%%%%%
\subsection{Photons and leptons}
\label{sec:photons_and_leptons}
%%%%%%%%%%%%%%%%%%%%%%%%%%%%%%%%%%%%%%%%%%%%%%%%%%

Let us start by considering the contribution of photons and leptons.
First of all, we need to take account of the effects of neutrino decoupling and $e^+e^-$ annihilation,
which are relevant at temperatures below a few $\mathrm{MeV}$.
Neutrinos decouple from the thermal bath when their weak interaction rate becomes smaller than the expansion rate of the universe at around $T \sim 2\,\mathrm{MeV}$,
and after that $e^+e^-$ annihilation takes place when the temperature of the universe becomes comparable to the electron mass $m_e \simeq 0.511\,\mathrm{MeV}$.

The energy density of radiations for $T\ll1\,\mathrm{MeV}$ is conventionally parameterized by
the effective neutrino degrees of freedom $N_{\rm eff}$, 
\begin{equation}
\rho = \left[1+\frac{7}{8}\left(\frac{4}{11}\right)^{\frac{4}{3}}N_{\rm eff}\right]\rho_{\gamma}, \label{Neff_definition}
\end{equation}
where $\rho_{\gamma} = (\pi^2/15)T^4$ is the energy density of photons.
If we assume that the neutrino decoupling has happened instantaneously
and that they do not couple with the electromagnetic plasma when $e^+e^-$ are annihilated, it is straightforward to obtain $N_{\rm eff}=3$.
However, neutrinos are still slightly interacting at that epoch, and the value of $N_{\rm eff}$ deviates from $3$.
Detailed calculations~\cite{Mangano:2001iu,Mangano:2005cc,deSalas:2016ztq} showed that the residual interactions
lead to a distortion in the neutrino spectra, and that neutrinos (photons) have higher (lower) energy density
with respect to those obtained in the instantaneous decoupling approximation.
To evaluate the value of $N_{\rm eff}$ in the non-instantaneous decoupling scenario, 
it is convenient to recast Eq.~\eqref{Neff_definition} in the following form~\cite{Mangano:2001iu}, 
\begin{align}
N_{\rm eff} & = \frac{8}{7}\left(\frac{11}{4}\right)^{\frac{4}{3}}\left(\frac{\rho - \rho_{\gamma}}{\rho_{\gamma}}\right) \nonumber\\
& = \left(\frac{z_0}{z_{\rm fin}}\right)^4\left[3+\frac{\delta\rho_{\nu_e}}{\rho_{\nu}^0}+\frac{\delta\rho_{\nu_{\mu}}}{\rho_{\nu}^0}+\frac{\delta\rho_{\nu_{\tau}}}{\rho_{\nu}^0}\right],
\end{align}
where $\rho_{\nu}^0$ denotes the energy density of a single specie of neutrinos with
an equilibrium distribution, $\delta\rho_{\nu_i}$ represent deviations from $\rho_{\nu}^0$
for three neutrino species $i=e,\mu,\tau$,
$z \equiv Ta$ is the dimensionless photon temperature with a scale factor $a$ being normalized such that $a \to 1/T$ at large temperatures,
and $z_0$ and $z_{\rm fin}$ represent asymptotic values for the case of instantaneous decoupling and that of full numerical calculation, respectively.
Recent results of numerical calculations including the effect of neutrino oscillations~\cite{deSalas:2016ztq}
showed that $\delta\rho_{\nu_i}/\rho_{\nu}^0$ increase by $0.5\textendash0.7\%$ and that the photon temperature becomes
$z_{\rm fin} = 1.39779$, which is smaller than $z_0 = (11/4)^{1/3} \simeq 1.40102$ obtained in the instantaneous decoupling approximation.
These values result in the following estimate for the effective neutrino degrees of freedom:
\begin{equation}
N_{\rm eff} \simeq 3.045. \label{N_eff_value}
\end{equation}
This corresponds to the asymptotic value of the effective degrees of freedom for the energy density $g_{*\rho,\mathrm{fin}} \simeq 3.383$.

In principle, it should also be possible to estimate an asymptotic value of the effective degrees of freedom for the entropy density, $g_{*s,\mathrm{fin}}$.
To the best of our knowledge, however, there is no clear consensus as to the value of $g_{*s,\mathrm{fin}}$ compared to the value of $g_{*\rho,\mathrm{fin}}$ (or $N_{\rm eff}$)
in the literature. 
Here we adopt the following naive argument based on the conservation of entropy.
In the numerical calculations in Refs.~\cite{Mangano:2001iu,Mangano:2005cc,deSalas:2016ztq}, the initial time was set to be 
$a_{\rm in}^{-1} = 10\,\mathrm{MeV}$, at which the dimensionless photon temperature was given by $z_{\rm in} = 1.00003$~\cite{Dolgov:1998sf}.
At that time, neutrinos are regarded to be in equilibrium with the electromagnetic plasma, and hence we can at least evaluate
the entropy density at the initial time straightforwardly.
Then, with the definition of the dimensionless variable $z = Ta$, the conservation of entropy of the system implies
\begin{equation}
g_{*s,\mathrm{fin}} = \left(\frac{z_{\rm in}}{z_{\rm fin}}\right)^3\left(g_{*s,\mathrm{EM}}(T_{\rm in}) + \frac{7}{8}\cdot 6\right), \label{gssfin_estimate}
\end{equation}
where $g_{*s,\mathrm{EM}}(T_{\rm in})$ parameterizes the contributions of the electromagnetic plasma to the entropy density
at the temperature $T_{\rm in} = z_{\rm in} \cdot 10\,\mathrm{MeV}$.
This quantity can be estimated as
\begin{equation}
g_{*s,\mathrm{EM}}(T_{\rm in}) = 2 + \frac{s_e^0(T_{\rm in})+s_{\rm QED}^0(T_{\rm in})}{\left[\frac{2\pi^2T_{\rm in}^3}{45}\right]}, \label{gssEM_Tin}
\end{equation}
where
\begin{equation}
s_i^0(T) = T^3\left[\Delta_{F,i}^0(T) + \frac{4p^0_{F,i}(T)}{T^4}\right] \label{s_fermion_free}
\end{equation}
with $i=e$ is the contribution of electrons/positrons to the entropy density,
and $p^0_{F,i}(T)$ and $\Delta_{F,i}^0(T)$ represent the contributions of fermion species $i$
with mass $m_i$ and internal degrees of freedom $g_i$
to the pressure and trace anomaly, respectively, in the ideal gas approximation:
\begin{align}
p_{F,i}^0(T) &= \frac{g_i}{6\pi^2}m_i^4 I_{F1}\left(\frac{m_i}{T}\right),\label{p_fermion_free} \\
\Delta_{F,i}^0(T) &= \frac{g_i}{6\pi^2}\frac{m_i^4}{T^4}\left[\frac{m_i}{T}I_{F2}\left(\frac{m_i}{T}\right)-4I_{F1}\left(\frac{m_i}{T}\right)\right]. \label{delta_fermion_free}
\end{align}
The functions $I_{F1}(y)$ and $I_{F2}(y)$ are given by
\begin{equation}
I_{F1}(y) \equiv \int^{\infty}_1 \frac{(x^2-1)^{\frac{3}{2}}}{e^{yx}+1}dx,\qquad 
I_{F2}(y) \equiv \int^{\infty}_1 \frac{e^{yx}(x^2-1)^{\frac{3}{2}}}{(e^{yx}+1)^2}xdx. \label{definition_IF}
\end{equation}
Note that $I_{F1}(y)$ and $I_{F2}(y)$ exhibit asymptotic behaviors,
\begin{equation}
I_{F1}(y) \xrightarrow{y\to 0} \frac{7\pi^4}{120y^4},\quad I_{F2}(y) \xrightarrow{y\to 0} \frac{7\pi^4}{30y^5}.
\end{equation}
In Eq.~\eqref{gssEM_Tin}, we also introduced corrections from quantum electrodynamics (QED),
\begin{equation}
s_{\rm QED}(T) = T^3\left[\Delta_{\rm QED}(T) + \frac{4p_{\rm QED}(T)}{T^4}\right].
\end{equation}
Here we adopt leading order QED corrections suggested in Refs.~\cite{Heckler:1994tv,Fornengo:1997wa,Mangano:2001iu}:
\begin{align}
p_{\rm QED}(T) &= -\int^{\infty}_0\frac{dk}{2\pi^2}\left[\frac{k^2}{E_k}\frac{\delta m_e^2(T)}{e^{E_k/T}+1}+\frac{k}{2}\frac{\delta m_{\gamma}^2(T)}{e^{k/T}-1}\right],\\
\Delta_{\rm QED}(T) &= T\frac{d}{dT}\left\{\frac{p_{\rm QED}(T)}{T^4}\right\}, \label{delta_QED}
\end{align}
where $E_k \equiv \sqrt{k^2+m_e^2}$,
\begin{align}
\delta m_e^2(T) &= \frac{2\pi\alpha T^2}{3} + \frac{4\alpha}{\pi}\int^{\infty}_0dk\frac{k^2}{E_k}\frac{1}{e^{E_k/T}+1},\label{finite_T_correction_m_e} \\
\delta m_{\gamma}^2(T) &= \frac{8\alpha}{\pi}\int^{\infty}_0dk \frac{k^2}{E_k}\frac{1}{e^{E_k/T}+1}
\end{align}
are finite temperature contributions to the electron mass and photon mass, respectively,
and $\alpha = e^2/(4\pi)$ is the fine structure constant.
In Eq.~\eqref{finite_T_correction_m_e}, we have neglected the momentum-dependent term as suggested in Ref.~\cite{Mangano:2001iu}.
Substituting the values $z_{\rm in} = 1.00003$ and $z_{\rm fin} = 1.39779$ to Eq.~\eqref{gssfin_estimate}, we obtain
$g_{\rm *s,\mathrm{fin}}\simeq 3.931$.

Although the results of the numerical calculations mentioned above appear to be reasonable,
we warn that there is a subtlety in estimating accuracy of them.
Actually, we find that the value of $g_{*s,\mathrm{fin}}$ can increase by $0.1\%$ 
when we add the contribution of muons with a mass $m_{\mu} \simeq 105.7\,\mathrm{MeV}$ to $g_{*s,\mathrm{EM}}(T_{\rm in})$.
Since the effect of muons was neglected in Refs.~\cite{Dolgov:1998sf,Mangano:2001iu,Mangano:2005cc,deSalas:2016ztq}, 
it is possible that all numerical results are subjected to similar uncertainty.
Regarding this fact, here we adopt the above estimate with $0.1\%$ uncertainty:
\begin{equation}
g_{*s,\mathrm{fin}} = 3.931 \pm 0.004. \label{gss_fin_value}
\end{equation}
For the same reason, we use the following value for the effective degrees of freedom for the energy density and their uncertainty:
\begin{equation}
g_{*\rho,\mathrm{fin}} = 3.383 \pm 0.003.
\end{equation}

If we assume that neutrinos have decoupled completely from the thermal bath at the epoch of $e^+e^-$ annihilation
and that the electromagnetic plasma behaves as an ideal gas,
the temperature dependence of $\rho(T)$ during this process can be described by the following formula [see, e.g. Ref.~\cite{Weinberg:2008zzc}]:
\begin{equation}
g_{*\rho}^0(T) = 2 + 6 \cdot \frac{7}{8}\cdot\left(\frac{4}{11}\right)^{\frac{4}{3}}\mathcal{S}^{\frac{4}{3}}\left(\frac{m_e}{T}\right) + \frac{\rho^0_e(T)}{\left[\frac{\pi^2T^4}{30}\right]}, \label{gsr_instantaneous_decoupling}
\end{equation}
where
\begin{equation}
\mathcal{S}(y) \equiv 1+ \frac{15y^5}{2\pi^4}I_{F2}(y),
\end{equation}
and
\begin{equation}
\rho_i^0(T) = T^4\left[\Delta_{F,i}^0(T) + \frac{3p_{F,i}^0(T)}{T^4}\right] \label{rho_fermion_free}
\end{equation}
with $i=e$ is the contribution of electrons/positrons to the energy density in the ideal gas approximation.
The function $\mathcal{S}(m_e/T)$ describes the temperature dependence of the ratio between the effective neutrino temperature $T_{\nu}$ and photon temperature,
\begin{equation}
\frac{T_{\nu}}{T} = \left(\frac{4}{11}\right)^{\frac{1}{3}}\mathcal{S}^{\frac{1}{3}}\left(\frac{m_e}{T}\right),
\end{equation}
and this relation follows from entropy conservation of the electromagnetic plasma and the scaling behavior of the effective neutrino temperature $T_{\nu} \propto 1/a$.
Similarly, the temperature dependence of $s(T)$ in the instantaneous decoupling scenario is given by
\begin{equation}
g_{*s}^0(T) = 2 + 6 \cdot \frac{7}{8}\cdot\frac{4}{11}\mathcal{S}\left(\frac{m_e}{T}\right) +  \frac{s^0_e(T)}{\left[\frac{2\pi^2T^3}{45}\right]}. \label{gss_instantaneous_decoupling}
\end{equation}
Note that the function $\mathcal{S}(y)$ takes the following asymptotic values:
\begin{equation}
\mathcal{S}(\infty) = 1,\quad \mathcal{S}(0) = \frac{11}{4}.
\end{equation}

Equations~\eqref{gsr_instantaneous_decoupling} and~\eqref{gss_instantaneous_decoupling}
should be modified if we take account of the relic interactions between neutrinos and electromagnetic plasma.
In principle, it is possible to follow the evolution of $\rho(T)$ and $s(T)$ including all necessary physics by solving
coupled Boltzmann equations numerically, which is out of the scope of this paper.
Instead, here we modify Eqs.~\eqref{gsr_instantaneous_decoupling} and~\eqref{gss_instantaneous_decoupling}
such that $g_{*\rho}(T)$ and $g_{*s}(T)$ exhibit appropriate asymptotic behavior:
\begin{align}
g_{*\rho}(T) &= 2a_{r1} + \frac{21}{4}\left(\frac{4}{11}\right)^{\frac{4}{3}}a_{r2}\mathcal{S}^{\frac{4}{3}}\left(\frac{m_e}{T}\right) + \frac{\rho^0_e(T)+\rho_{\rm QED}(T)+\rho_{\mu}^0(T)+\rho_{\rm hadrons}(T)}{\left[\frac{\pi^2T^4}{30}\right]}, \label{gsr_nu_decoupling_corrected} \\
g_{*s}(T) &= 2a_{s1} + \frac{21}{11}a_{s2}\mathcal{S}\left(\frac{m_e}{T}\right) + \frac{s^0_e(T)+s_{\rm QED}(T)+s_{\mu}^0(T)+s_{\rm hadrons}(T)}{\left[\frac{2\pi^2T^3}{45}\right]}, \label{gss_nu_decoupling_corrected}
\end{align}
where the coefficients $a_{r1}$, $a_{r2}$, $a_{s1}$, and $a_{s2}$ parameterize deviations from the instantaneous decoupling results, and
\begin{equation}
\rho_{\rm QED}(T) = T^4\left[\Delta_{\rm QED}(T) + \frac{3p_{\rm QED}(T)}{T^4}\right]
\end{equation} 
is the QED corrections to the energy density.
The values of $a_{r1}$ and $a_{r2}$ can be fixed by requiring that the first two terms in the right-hand side of Eq.~\eqref{gsr_nu_decoupling_corrected}
approach asymptotically to $g_{*\rho,\mathrm{fin}}$ for $T\ll m_e$ and to $2+\frac{7}{8}\cdot 6$ for $T \gg m_e$.
Adopting these conditions, we obtain
\begin{align}
a_{r1} &= 1 + \frac{g_{*\rho,\mathrm{fin}}-2\left(1+\frac{21}{8}\left(\frac{4}{11}\right)^{\frac{4}{3}}\right)}{2\left(1-\left(\frac{4}{11}\right)^{\frac{4}{3}}\right)} = 1.014,\label{a_r1}\\
a_{r2} &= 1 - \frac{4}{21}\frac{g_{*\rho,\mathrm{fin}}-2\left(1+\frac{21}{8}\left(\frac{4}{11}\right)^{\frac{4}{3}}\right)}{1-\left(\frac{4}{11}\right)^{\frac{4}{3}}} = 0.9947, \label{a_r2}
\end{align}
where we have substituted $g_{*\rho,\mathrm{fin}}=3.383$ in the last equality.
Similarly, requiring that the first two terms in the right-hand side of Eq.~\eqref{gss_nu_decoupling_corrected}
approach asymptotically to $g_{*s,\mathrm{fin}}$ for $T\ll m_e$ and to $2+\frac{7}{8}\cdot 6$ for $T \gg m_e$, we obtain
\begin{align}
a_{s1} &= 1 + \frac{11}{14}\left(g_{*s,\mathrm{fin}}-\frac{43}{11}\right) = 1.017, \label{a_s1}\\
a_{s2} &= 1 - \frac{44}{147}\left(g_{*s,\mathrm{fin}}-\frac{43}{11}\right)  = 0.9935, \label{a_s2}
\end{align}
where we have substituted $g_{*s,\mathrm{fin}}=3.931$ in the last equality.
In Eqs.~\eqref{gsr_nu_decoupling_corrected} and~\eqref{gss_nu_decoupling_corrected},
we also added the contributions of muons [given by Eqs.~\eqref{rho_fermion_free} and~\eqref{s_fermion_free} with $i=\mu$] and those of hadrons
($\rho_{\rm hadrons}$ and $s_{\rm hadrons}$) in order to accomplish a smooth connection to the calculations at higher temperatures.
We discuss more details about the contributions of hadrons in the next subsection.

\begin{figure}[htbp]
\centering
$\begin{array}{cc}
\subfigure{
\includegraphics[width=75mm]{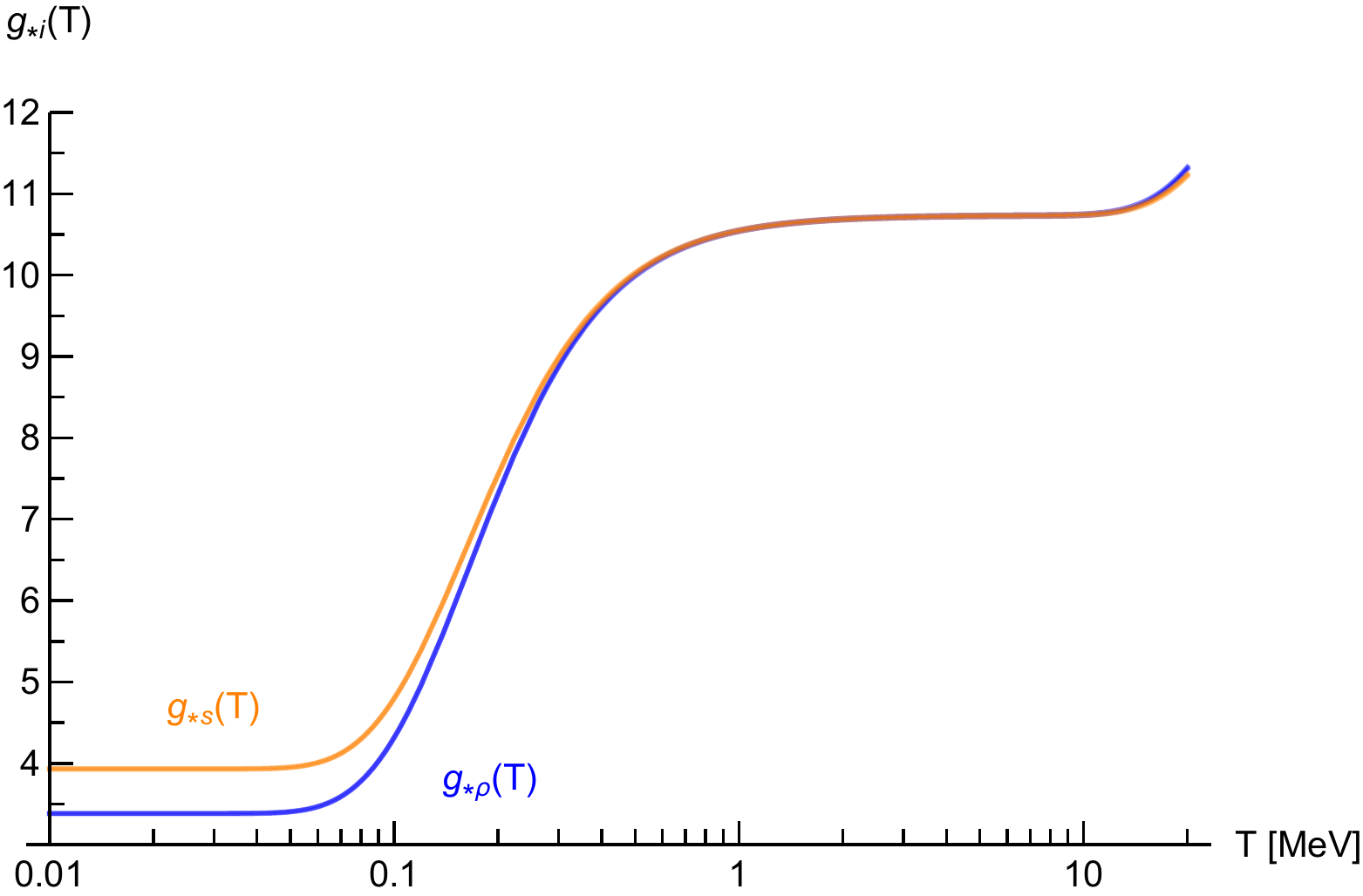}}
\hspace{5mm}
\subfigure{
\includegraphics[width=80mm]{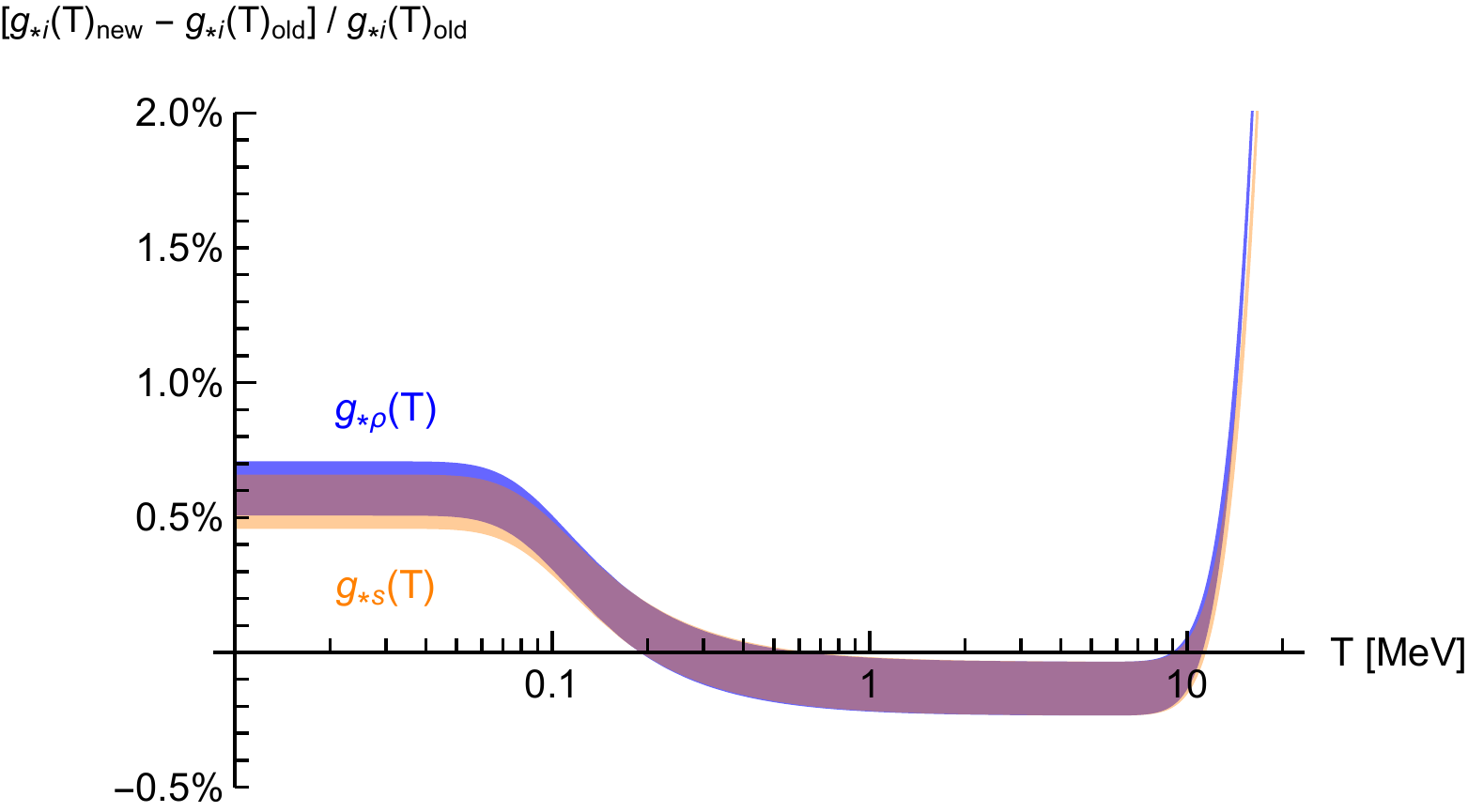}}
\vspace{10mm}
\end{array}$
\caption{Temperature dependence of $g_{*i}(T)$ ($i=\rho, s$) at $T<20\,\mathrm{MeV}$.
In the left panel, the values of $g_{*\rho}(T)$ (blue line) and $g_{*s}(T)$ (orange line) evaluated by using the modified functions 
[Eqs.~\eqref{gsr_nu_decoupling_corrected} and~\eqref{gss_nu_decoupling_corrected}] are plotted.
In the right panel, we show the relative difference of $g_{*i}(T)$ between
the results of the instantaneous decoupling scenario without the contributions of muons, hadrons, 
and QED corrections [Eqs.~\eqref{gsr_instantaneous_decoupling} and~\eqref{gss_instantaneous_decoupling},
dubbed as ``$g_{*i}(T)_{\rm old}$"]
and the modified functions [dubbed as ``$g_{*i}(T)_{\rm new}$"].
The blue and orange bands indicate $\pm 0.1\,\%$ uncertainty of $g_{*\rho}(T)_{\rm new}$
and $g_{*s}(T)_{\rm new}$, respectively (see text).}
\label{fig:gsnudec}
\end{figure}

In Fig.~\ref{fig:gsnudec}, we compare the simplistic results~\eqref{gsr_instantaneous_decoupling} and~\eqref{gss_instantaneous_decoupling} 
with modified functions~\eqref{gsr_nu_decoupling_corrected} and~\eqref{gss_nu_decoupling_corrected}.
Although the corrections look small at low temperatures, they become relevant for $T\gtrsim 10\,\mathrm{MeV}$.
Actually, we see that the values of $g_{*\rho}$ and $g_{*s}$ increase by $\mathcal{O}(1)\%$ at $T=20\,\mathrm{MeV}$
due to the contributions of muons and light hadrons.

For temperatures below $10\,\mathrm{MeV}$, we evaluate the effective degrees of freedom by using Eqs.~\eqref{gsr_nu_decoupling_corrected} and~\eqref{gss_nu_decoupling_corrected}
with coefficients specified by Eqs.~\eqref{a_r1},~\eqref{a_r2},~\eqref{a_s1}, and~\eqref{a_s2},
and take their uncertainty by simply multiplying the functions~\eqref{gsr_nu_decoupling_corrected} and~\eqref{gss_nu_decoupling_corrected} by a factor 1.001 or 0.999.
Here we neglect the effect of the neutrino masses, though they can lead to further modifications at $T \lesssim 0.1\,\mathrm{eV}$ if some of them are close to the current 
cosmological bound on the sum $\sum_j m_j < 0.17\,\mathrm{eV}$ reported by 
the Planck Collaboration~\cite{Ade:2015xua,Aghanim:2016yuo} (including data on the Baryon Acoustic Oscillations).
For $T>10\,\mathrm{MeV}$, on the other hand, neutrinos are kept in equilibrium with the thermal bath via weak interactions,
and hence in such higher temperatures we evaluate the energy density and entropy density in terms of 
thermodynamic quantities $p(T)$ and $\Delta(T)$ [i.e. via Eq.~\eqref{rho_and_s_from_p_and_delta}]
instead of using Eqs.~\eqref{gsr_nu_decoupling_corrected} and~\eqref{gss_nu_decoupling_corrected}.
In particular, we estimate the pressure $p(T)$ by integrating the trace anomaly $\Delta (T)$ as we mentioned in the beginning of this section.
Note that the contribution of taus with a mass $m_{\tau} \simeq 1.777\,\mathrm{GeV}$ becomes relevant for $T \gtrsim \mathcal{O}(0.1)\,\mathrm{GeV}$.
We include this contribution and calculate $p(T)$ and $\Delta(T)$ at higher temperatures in Sec.~\ref{sec:full_EoS} 
after we discuss the effects of the QCD crossover and electroweak crossover in the next subsections.

%%%%%%%%%%%%%%%%%%%%%%%%%%%%%%%%%%%%%%%%%%%%%%%%%%
\subsection{Strongly interacting particles}
\label{sec:strongly_interacting_particles}
%%%%%%%%%%%%%%%%%%%%%%%%%%%%%%%%%%%%%%%%%%%%%%%%%%

The most drastic change happens at the epoch of the QCD crossover.
Here we can consider three different phases: the hadronic phase [$T\lesssim \mathcal{O}(100)\,\mathrm{MeV}$],
the non-perturbative phase [$T\sim \mathcal{O}(100)\,\mathrm{MeV}$], and the perturbative phase
[$T\gg \mathcal{O}(100)\,\mathrm{MeV}$].
In this subsection, we discuss how to estimate the contribution of strongly interacting particles in each phase.
Interpolation between different phases should be considered carefully, and we quantify the uncertainty
arising from the interpolation procedure in Sec.~\ref{sec:full_EoS}.

The thermodynamic properties of hadronic plasma
at temperatures below the critical temperature of the QCD crossover
can be described by the hadron resonance gas model,
in which the system is approximated by free hadrons and resonances~\cite{Hagedorn:1965st,Dashen:1969ep,Venugopalan:1992hy}.
The hadron resonance gas model has shown a good agreement with lattice QCD results~\cite{Borsanyi:2013bia,Bellwied:2015lba} and 
successfully reproduced the experimental results observed in heavy ion collisions
[see, e.g. Refs.~\cite{BraunMunzinger:2003zd,Braun-Munzinger:2015hba} and references therein].
In this model, thermodynamic quantities are expressed as sums over single free particle contributions with mass $m_i$
and internal degrees of freedom $g_i$. The pressure and trace anomaly read
\begin{align}
p_{\rm hadrons}(T) &= \sum_i p_i^0(T),\label{p_HRG}\\
\Delta_{\rm hadrons}(T) &= \sum_i \Delta_i^0(T), \label{delta_HRG}
\end{align}
where
\begin{align}
p_i^0(T) = 
\left\{
\begin{array}{ll}
p_{B,i}^0(T) = {\displaystyle\frac{g_i}{6\pi^2}m_i^4 I_{B1}\left(\frac{m_i}{T}\right)} & \text{for bosons}, \\ [2.0ex]
p_{F,i}^0(T) = {\displaystyle\frac{g_i}{6\pi^2}m_i^4 I_{F1}\left(\frac{m_i}{T}\right)} & \text{for fermions}, \\
\end{array}
\right. \label{p_tree_level}
\end{align}
\begin{align}
\Delta_i^0(T) = \left\{
\begin{array}{ll}
\Delta_{B,i}^0(T) = {\displaystyle \frac{g_i}{6\pi^2}\frac{m_i^4}{T^4}\left[\frac{m_i}{T}I_{B2}\left(\frac{m_i}{T}\right)-4I_{B1}\left(\frac{m_i}{T}\right)\right]} & \text{for bosons}, \\ [2.0ex]
\Delta_{F,i}^0(T) = {\displaystyle \frac{g_i}{6\pi^2}\frac{m_i^4}{T^4}\left[\frac{m_i}{T}I_{F2}\left(\frac{m_i}{T}\right)-4I_{F1}\left(\frac{m_i}{T}\right)\right]} & \text{for fermions}, \\
\end{array}
\right. \label{delta_tree_level}
\end{align}
$I_{F1}(y)$ and $I_{F2}(y)$ are defined in Eq.~\eqref{definition_IF}, and $I_{B1}(y)$ and $I_{B2}(y)$ are given by
\begin{equation}
I_{B1}(y) \equiv \int^{\infty}_1 \frac{(x^2-1)^{\frac{3}{2}}}{e^{yx}-1}dx,\qquad 
I_{B2}(y) \equiv \int^{\infty}_1 \frac{e^{yx}(x^2-1)^{\frac{3}{2}}}{(e^{yx}-1)^2}xdx. \label{definition_IB}
\end{equation}
These functions exhibit the following asymptotic behaviors,
\begin{equation}
I_{B1}(y) \xrightarrow{y\to 0} \frac{\pi^4}{15y^4},\quad I_{B2}(y) \xrightarrow{y\to 0} \frac{4\pi^4}{15y^5}.
\end{equation}
In the sums in Eqs.~\eqref{p_HRG} and~\eqref{delta_HRG}, we include all known baryons and mesons
whose masses are less than $2.5\,\mathrm{GeV}$ as enumerated in the Particle Data Group's review~\cite{Patrignani:2016xqp}.
The energy density and entropy density can be estimated as $\rho_{\rm hadrons}(T) = T^4[\Delta_{\rm hadrons}(T)+3p_{\rm hadrons}(T)/T^4]$
and $s_{\rm hadrons}(T) = T^3[\Delta_{\rm hadrons}(T)+4p_{\rm hadrons}(T)/T^4]$, respectively,
and they were used in Eqs.~\eqref{gsr_nu_decoupling_corrected} and~\eqref{gss_nu_decoupling_corrected}.

\begin{figure}[htbp]
\centering
$\begin{array}{c}
\subfigure{
\includegraphics[width=160mm]{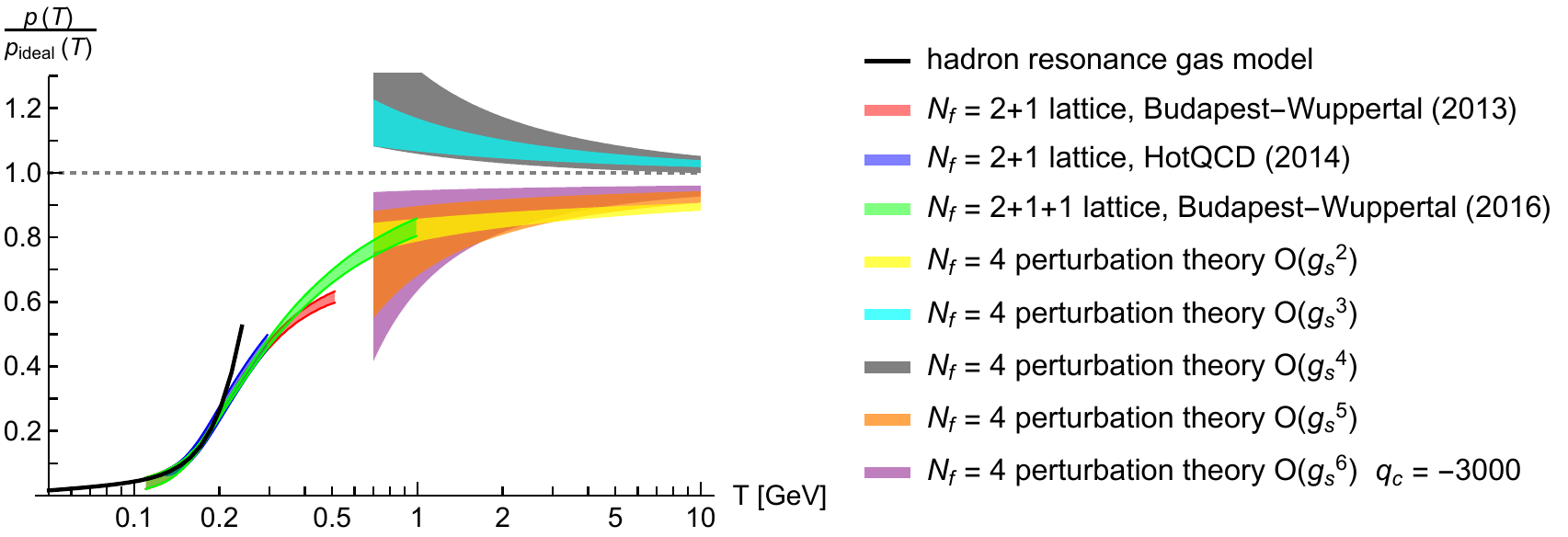}}
\vspace{10mm}
\\
\subfigure{
\includegraphics[width=160mm]{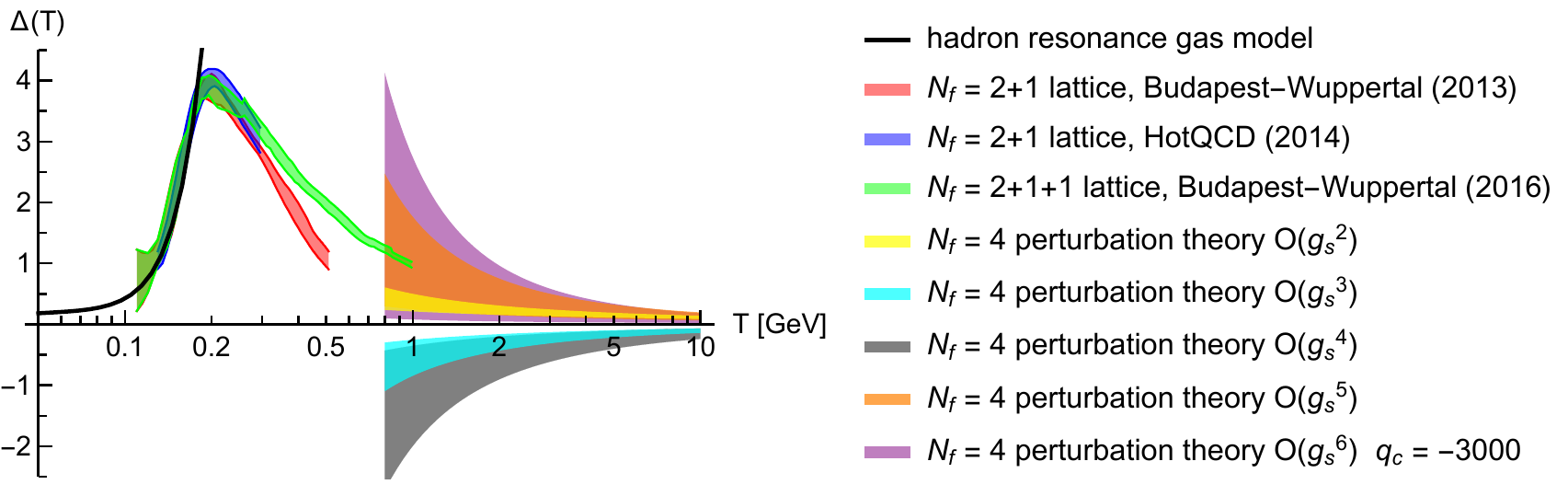}}
\end{array}$
  \caption{
  Contributions of strongly interacting particles to the pressure (top panel) and the trace anomaly (bottom panel).
  In the top panel, $p(T)$ is divided by the value in the ideal gas approximation of $N_f = 4$ QCD (the Stefan-Boltzmann result),
  $p_{\rm ideal}(T) = 58\cdot\frac{\pi^2}{90}T^4$.
  The black solid line represents the estimate based on the hadron resonance gas model [Eq.~\eqref{p_HRG} or Eq.~\eqref{delta_HRG}].
  The results of lattice QCD simulations (including error bars) are shown in red [Budapest-Wuppertal (2013), Ref.~\cite{Borsanyi:2013bia}], 
  blue [HotQCD (2014), Ref.~\cite{Bazavov:2014pvz}], and green [Budapest-Wuppertal (2016), Ref.~\cite{Borsanyi:2016ksw}]. 
  The results of perturbative calculations [Eqs.~\eqref{p_QCD_perturbative} and~\eqref{delta_QCD_perturbative}]
  are also shown in yellow [$\mathcal{O}(g_s^2)$], cyan [$\mathcal{O}(g_s^3)$], gray [$\mathcal{O}(g_s^4)$],
  orange [$\mathcal{O}(g_s^5)$], and purple [$\mathcal{O}(g_s^6)$].
  Here, a label $\mathcal{O}(g_s^n)$ corresponds to the result obtained by 
  including terms up to $n$-th order in the coupling expansion in Eq.~\eqref{p_QCD_perturbative}.
  The bands for the perturbative results indicate the variation of $p(T)$ or $\Delta(T)$ as the $\overline{\text{MS}}$
  renormalization scale is varied in the range $\mu = (1\dots 4)\pi T$.
  For the $\mathcal{O}(g_s^6)$ result, we used the value $q_c(N_f=4) = -3000$ for the unknown constant (see text).}
  \label{fig:QCDpressure}
\end{figure}

The description based on the hadron resonance gas model breaks down at around the critical temperature of the QCD crossover.
At that stage, interactions among constituent particles become very strong, and we must use the non-perturbative method, i.e. lattice QCD simulations, 
to estimate the thermodynamic properties of quark-gluon plasma.
The equation of state in $2+1$ flavor lattice QCD with physical quark masses was calculated by 
the ``Budapest-Wuppertal" collaboration~\cite{Borsanyi:2010cj,Borsanyi:2013bia},
and their result was confirmed by the ``HotQCD" collaboration in Ref.~\cite{Bazavov:2014pvz}.
Recently, the contribution of the charm quark was investigated in $2+1+1$ flavor lattice QCD~\cite{Borsanyi:2016ksw}, which enables us
to estimate the equation of state for higher temperatures up to $1\,\mathrm{GeV}$.
In this paper, we adopt the result of Ref.~\cite{Borsanyi:2016ksw} to calculate the effective degrees of freedom in the non-perturbative phase.\footnote{
For $2+1$ flavor lattice QCD results plotted in Fig.~\ref{fig:QCDpressure}, we used the tabulated data attached to Ref.~\cite{Borsanyi:2013bia}
[dubbed as ``Budapest-Wuppertal (2013)"] and table I in Ref.~\cite{Bazavov:2014pvz} [dubbed as ``HotQCD (2014)"].
For $2+1+1$ flavor lattice QCD results, we extract the data from Fig.~S9 (or Fig.~S7 in the arXiv version) of Ref.~\cite{Borsanyi:2016ksw}
[dubbed as ``Budapest-Wuppertal (2016)"].}

For temperatures much higher than the critical temperature of the QCD crossover,
we can use the perturbative method of finite temperature QCD.
The pressure for QCD with massless quarks at high temperature $p_{\rm QCD}(T)$ was computed
by using effective field theory approach~\cite{Ginsparg:1980ef,Appelquist:1981vg,Braaten:1995jr},
which provides a systematic framework for combining perturbative and non-perturbative results in a consistent way.
The coefficients of the expansion in the strong gauge coupling $g_s$ have been computed at orders
$g_s^2$~\cite{Shuryak:1977ut,Chin:1978gj}, $g_s^3$~\cite{Kapusta:1979fh}, $g_s^4\ln(1/g_s)$~\cite{Toimela:1982hv},
$g_s^4$~\cite{Arnold:1994ps,Arnold:1994eb}, $g_s^5$~\cite{Zhai:1995ac,Braaten:1995jr}, and $g_s^6\ln(1/g_s)$~\cite{Kajantie:2002wa}. 
Here we quote the results obtained in Ref.~\cite{Kajantie:2002wa}:
\begin{equation}
p_{\rm QCD}(T) = \frac{8\pi^2}{45}T^4 \left[p_0 + p_2\frac{\alpha_s}{\pi}+p_3\left(\frac{\alpha_s}{\pi}\right)^{\frac{3}{2}}
+ p_4\left(\frac{\alpha_s}{\pi}\right)^2 + p_5\left(\frac{\alpha_s}{\pi}\right)^{\frac{5}{2}} + p_6\left(\frac{\alpha_s}{\pi}\right)^3 \right], \label{p_QCD_perturbative}
\end{equation}
where $\alpha_s = g_s^2/4\pi$,
\begin{align}
p_0 &= 1 + \frac{21}{32}N_f, \\
p_2 &= -\frac{15}{4}\left(1+ \frac{5}{12}N_f\right), \label{p_QCD_order_g2} \\
p_3 &= 30\left(1+\frac{1}{6}N_f\right)^{\frac{3}{2}}, \\
p_4 &= 237.2 + 15.96 N_f - 0.4150 N_f^2 + \frac{135}{2}\left(1+\frac{1}{6}N_f\right)\ln\left[\frac{\alpha_s}{\pi}\left(1+\frac{1}{6}N_f\right)\right] \nonumber\\
&\quad -\frac{165}{8}\left(1+\frac{5}{12}N_f\right)\left(1-\frac{2}{33}N_f\right)\ln\left(\frac{\mu}{2\pi T}\right), \\
p_5 &= \left(1+\frac{1}{6}N_f\right)^{\frac{1}{2}}\left[-799.1-21.96N_f -1.926N_f^2 + \frac{495}{2}\left(1+\frac{1}{6}N_f\right)\left(1-\frac{2}{33}N_f\right)\ln\left(\frac{\mu}{2\pi T}\right)\right], \\
p_6 &= \left[-659.2 - 65.89N_f - 7.653 N_f^2 + \frac{1485}{2}\left(1+\frac{1}{6}N_f\right)\left(1-\frac{2}{33}N_f\right)\ln\left(\frac{\mu}{2\pi T}\right)\right]\nonumber\\
&\quad \times \ln\left[\frac{\alpha_s}{\pi}\left(1+\frac{1}{6}N_f\right)\right] -475.6\ln\left(\frac{\alpha_s}{\pi}\right) - \frac{1815}{16}\left(1+\frac{5}{12}N_f\right)\left(1-\frac{2}{33}N_f\right)^2\ln^2\left(\frac{\mu}{2\pi T}\right) \nonumber\\
&\quad + \left(2932.9 + 42.83N_f - 16.48N_f^2 + 0.2767N_f^3\right)\ln\left(\frac{\mu}{2\pi T}\right) + q_c(N_f), \label{p_QCD_order_g6}
\end{align}
$N_f$ is the number of massless quarks, and $\mu$ is the renormalization scale in the modified minimal subtraction ($\overline{\text{MS}}$) scheme.
For the evaluation of the renormalization group running of gauge coupling $g_s$, we adopt four-loop beta function coefficients obtained in Ref.~\cite{vanRitbergen:1997va}.
The contribution to the trace anomaly can be evaluated as
\begin{equation}
\Delta_{\rm QCD}(T) = T\frac{d}{dT}\left\{\frac{p_{\rm QCD}(T)}{T^4}\right\}. \label{delta_QCD_perturbative}
\end{equation}

The coefficient $q_c(N_f)$ in Eq.~\eqref{p_QCD_order_g6}, which might depend on $N_f$, is unknown, 
and it is related to the serious infrared problem of finite temperature field theory (known as ``Linde's problem")~\cite{Linde:1980ts,Gross:1980br}.
A power counting argument implies that the contributions of $(l+1)$-loop gluon diagrams to the free energy (or pressure) 
contain infrared divergences of the form $\sim g_s^6 T^4 \ln(T/m)$ for $l=3$ and $\sim g_s^6T^4(g_s^2T/m)^{l-3}$ for $l>3$,
where $m$ is an infrared cutoff scale. If the system were described by QED, the cutoff scale could be identified as the screening mass
$m \sim g_s T$, which would regularize all the infrared divergences and make the perturbative expansion in terms of $g_s$ totally controllable.
However, this is not the case in QCD due to the self-coupling of the spacelike gluons, which give rise to the magnetic mass $m\sim g_s^2T$
for the cutoff scale. Consequently, all contributions with $l>3$ become of $\mathcal{O}(g_s^6)$, independently of the order of perturbation theory.
In other words, we cannot determine the contribution of $\mathcal{O}(g_s^6)$ in perturbative approach, and such a contribution is parameterized 
by the coefficient $q_c(N_f)$ in Eq.~\eqref{p_QCD_order_g6}.

In Fig.~\ref{fig:QCDpressure}, we show the contributions of strongly interacting particles to
the pressure and trace anomaly.
We see that the results based on the hadron resonance gas model [Eqs.~\eqref{p_HRG} and~\eqref{delta_HRG}]
agree with those obtained from lattice QCD simulations at $T \gtrsim 0.1\,\mathrm{GeV}$.
Both $p(T)$ and $\Delta(T)$ blow up at higher temperatures in the hadron resonance gas model,
indicating that the system should be described as interacting quark-gluon plasma rather than
a collection of free hadrons and resonances and that we have to adopt lattice data.
In Fig.~\ref{fig:QCDpressure}, we plot three different sets of lattice data,
in which the analysis was performed with physical quark masses and the results were shown in the continuum limit.
These include the results of $2+1$ flavor lattice QCD in Refs.~\cite{Borsanyi:2013bia,Bazavov:2014pvz}
and those of $2+1+1$ flavor lattice QCD in Ref.~\cite{Borsanyi:2016ksw}.
The results of $2+1$ flavor lattice QCD are consistent with those of $2+1+1$ flavor lattice QCD at low temperatures,
while they disagree at higher temperatures as the contribution of charm quarks
(mass $m_c \simeq 1.3\,\mathrm{GeV}$) becomes important.
This $2+1+1$ flavor lattice QCD results should be connected with the perturbative results [Eqs.~\eqref{p_QCD_perturbative} and~\eqref{delta_QCD_perturbative}] 
at some higher temperature.\footnote{It should be noted that the perturbative results for $\Delta_{\rm QCD}(T)$ shown in Fig.~\ref{fig:QCDpressure}
and used in the analysis in Sec.~\ref{sec:full_EoS} do not represent a genuine perturbative expansion of the trace anomaly but 
a direct differentiation of pressure $p_{\rm QCD}(T)$ obtained in the perturbative expansion.
Namely, ``$\mathcal{O}(g_s^n)$" result of $\Delta_{\rm QCD}(T)$ shown in the lower panel of Fig.~\ref{fig:QCDpressure}
is obtained by performing the temperature derivative in Eq.~\eqref{delta_QCD_perturbative}, where
$p_{\rm QCD}(T)$ includes terms up to $\mathcal{O}(g_s^n)$ and the renormalization group running of $g_s$ is also treated as a function of $T$.
Strictly speaking, the latter should be regarded as higher order effects, but we include them in $\Delta_{\rm QCD}(T)$ since
their effects are not negligible in the temperature range considered here.}
However, the convergence of the perturbative expansion turns out to be very poor at $T \sim \mathcal{O}(1\textendash 10)\,\mathrm{GeV}$.
In fact, the prediction of the perturbation theory changes significantly at each order of weak coupling expansion, as shown in Fig.~\ref{fig:QCDpressure}.
Furthermore, we must adjust the value of the unknown constant $q_c(N_f)$ at order $g_s^6$.
Although the choice $q_c(N_f=4) = -3000$~\cite{Borsanyi:2016ksw} appears to coincide with the $2+1+1$ flavor lattice QCD result as shown in Fig.~\ref{fig:QCDpressure},
large uncertainty of the perturbative results around $T\sim\mathcal{O}(1)\,\mathrm{GeV}$ leaves the possibility of other choices,
which we investigate more carefully when we match the perturbative results to those of the $2+1+1$ flavor lattice QCD simulations in Sec.~\ref{sec:full_EoS}.

Before going to the matching procedure between the perturbative and non-perturbative phases in QCD,
we must consider the effect of quark mass thresholds,
which should be included in order to describe the contributions of heavier quarks correctly.
The perturbative results shown in Eqs.~\eqref{p_QCD_perturbative}-\eqref{p_QCD_order_g6}
are valid only for massless quarks, and we cannot use them when the temperature becomes comparable to
the masses of heavier quarks, since these heavier states start to contribute to the pressure.
The corrections to the QCD pressure with massive quarks were estimated up to $\mathcal{O}(g_s^2)$ in Ref.~\cite{Laine:2006cp},
and it was shown that higher order corrections to the quark mass dependence are insignificant compared with
those to the pressure as a whole. 
Based on this fact, in Ref.~\cite{Borsanyi:2016ksw} it was argued that
the quark mass threshold can be sufficiently described by using a tree-level correction factor.
Here we follow the approach of Ref.~\cite{Borsanyi:2016ksw} and describe the effect of quark masses as follows.
First, we write the pressure of 4-flavor QCD including the effect of the charm quark mass as
\begin{equation}
p_{\rm QCD}^{(u+d+s+c)}(T) = \frac{p_{\rm QCD,SB}(T,3)+p_{F,c}^0(T)}{p_{\rm QCD,SB}(T,4)} p_{\rm QCD}(T)|_{N_f \,=\, 4}, \label{p_QCD(u+d+s+c)}
\end{equation}
where 
\begin{equation}
p_{\rm QCD,SB}(T,N_f) = \frac{\pi^2}{90}T^4 \left(16+\frac{21}{2}N_f\right)
\end{equation}
is the pressure in the Stefan-Boltzmann limit of the $N_f$-flavor theory,
$p_{F,c}^0(T)$ is the contribution of free charm quarks [see Eq.~\eqref{p_fermion_free}],
and $p_{\rm QCD}(T)|_{N_f \,=\, 4}$ corresponds to Eq.~\eqref{p_QCD_perturbative} evaluated at $N_f = 4$.
Next, we describe the effect of the bottom quark mass $m_b \approx 4.2\,\mathrm{GeV}$ as~\cite{Borsanyi:2016ksw}
\begin{equation}
p_{\rm QCD}^{(u+d+s+c+b)}(T) = \frac{p_{\rm QCD,SB}(T,4)+p_{F,b}^0(T)}{p_{\rm QCD,SB}(T,4)}p_{\rm QCD}^{(u+d+s+c)}(T),\label{p_QCD(u+d+s+c+b)}
\end{equation}
where $p_{F,b}^0(T)$ is the contribution of free bottom quarks evaluated by using Eq.~\eqref{p_fermion_free}.
Finally, the contribution to the trace anomaly can be estimated as
\begin{equation}
\Delta_{\rm QCD}^{(u+d+s+c+b)}(T) = T\frac{d}{dT}\left\{\frac{p_{\rm QCD}^{(u+d+s+c+b)}(T)}{T^4}\right\}. \label{delta_QCD(u+d+s+c+b)}
\end{equation}
Equations~\eqref{p_QCD(u+d+s+c+b)} and~\eqref{delta_QCD(u+d+s+c+b)} allow us to model the variation of the pressure
and trace anomaly due to the charm and bottom quark masses, and they can be used to extrapolate the results of lattice QCD simulations
to higher temperatures.

In addition to the contributions of charm and bottom quarks discussed above, we also have to include
the contribution of top quarks (mass $m_t\gtrsim 170\,\mathrm{GeV} $).
Here we do not add a top quark mass threshold effect just by extrapolating the 4-flavor result as in Eq.~\eqref{p_QCD(u+d+s+c+b)},
since at temperature much higher than $m_t$ the thermodynamic quantities of QCD can be exactly described 
by using the result with massless quarks [Eqs.~\eqref{p_QCD_perturbative} and~\eqref{delta_QCD_perturbative}].
For this reason, we model the trace anomaly such that it approaches Eq.~\eqref{delta_QCD(u+d+s+c+b)} for $T\ll m_t$ and
Eq.~\eqref{delta_QCD_perturbative} with $N_f=6$ for $T\gg m_t$,
\begin{equation}
\Delta_{\rm QCD}^{(u+d+s+c+b+t)}(T) = c_t(T)\left.\Delta_{\rm QCD}(T)\right|_{N_f\,=\,6} + \left(1-c_t(T)\right)\Delta_{\rm QCD}^{(u+d+s+c+b)}(T), \label{delta_QCD(u+d+s+c+b+t)}
\end{equation}
where 
\begin{equation}
c_t(T) = \frac{p_{F,t}^0(T)|_{m_t \,=\, m_t(v_T)}}{\frac{7\pi^2}{60}T^4}, \label{c_topmass}
\end{equation}
and $p_{F,t}^0(T)$ is the contribution of free top quarks evaluated by using Eq.~\eqref{p_fermion_free}.\footnote{If we consider 
a trace anomaly given by a derivative of the following combination,
\begin{equation}
p_{\rm QCD}^{(u+d+s+c+b+t)}(T) = c_t(T)\left.p_{\rm QCD}(T)\right|_{N_f\,=\,6} + \left(1-c_t(T)\right)p_{\rm QCD}^{(u+d+s+c+b)}(T),
\end{equation}
we have
\begin{align}
\Delta_{\rm QCD}^{(u+d+s+c+b+t)}(T) &= c_t(T)\left.\Delta_{\rm QCD}(T)\right|_{N_f\,=\,6} + \left(1-c_t(T)\right)\Delta_{\rm QCD}^{(u+d+s+c+b)}(T) \nonumber\\
&\quad + T\left(\frac{dc_t(T)}{dT}\right)\left(\frac{\left.p_{\rm QCD}(T)\right|_{N_f\,=\,6}-p_{\rm QCD}^{(u+d+s+c+b)}(T)}{T^4}\right). \label{delta_QCD(u+d+s+c+b+t)_another}
\end{align}
For a tree-level approximation, $\left.p_{\rm QCD}(T)\right|_{N_f\,=\,6}-p_{\rm QCD}^{(u+d+s+c+b)}(T) \approx \frac{7\pi^2}{60}T^4$, we see that
the last term in the right-hand side of the above equation represents the contribution of free top quarks, $\Delta_{F,t}^0(T) = T\frac{d}{dT}\{p_{F,t}^0(T)T^{-4}\}$.
Since such a contribution is automatically included once we evaluate the Higgs potential through the electroweak crossover [see Eq.~\eqref{l_t_phi_T}],
we dropped the term proportional to $dc_t(T)/dT$ in Eq.~\eqref{delta_QCD(u+d+s+c+b+t)_another} in order to avoid double counting.}
The correction factor $c_t(T)$ approaches a limit $c_t(T) \to 0$ for $T\ll m_t$ and $c_t(T) \to 1$ for $T \gg m_t$ and smoothly interpolates between 
$\Delta_{\rm QCD}^{(u+d+s+c+b)}(T)$ and $\left.\Delta_{\rm QCD}(T)\right|_{N_f\,=\,6}$.
We also note that the top quark mass $m_t(v_T)$ in Eq.~\eqref{c_topmass} is evaluated by taking account of 
the temperature variation of a Higgs expectation value $v_T$ [see Eq.~\eqref{phi_dependent_masses}].
The evolution of the expectation value $v_T$ can be traced together with the evaluation of the Higgs potential, which we discuss in the next subsection.

%%%%%%%%%%%%%%%%%%%%%%%%%%%%%%%%%%%%%%%%%%%%%%%%%%
\subsection{Electroweak crossover}
\label{sec:electroweak_crossover}
%%%%%%%%%%%%%%%%%%%%%%%%%%%%%%%%%%%%%%%%%%%%%%%%%%

Across the electroweak crossover, the universe evolved from high temperature symmetric phase
to low temperature broken phase, in which the Higgs field has a non-vanishing expectation value.
In this subsection, we discuss the additional contributions to thermodynamic quantities arising through the electroweak crossover.

Following Ref.~\cite{Laine:2006cp}, we estimate the thermodynamic quantities in the low temperature broken phase 
by using the finite temperature one-loop effective potential of the Higgs field [hereinafter, referred to as ``one-loop Coleman-Weinberg (CW) method"].
Namely, we evaluate the free energy density $f(\varphi,T)$ in the presence of a Higgs expectation value $\varphi$ at temperature $T$ as
\begin{align}
f(\varphi,T) & = -\frac{1}{2}\nu^2\varphi^2 + \frac{1}{4}\lambda\varphi^4 + l_h(\varphi,T) + l_W(\varphi,T) + l_Z(\varphi,T) + l_t(\varphi,T), \label{f_phi_4d_1loop}
\end{align}
where $\nu^2$ and $\lambda$ are the quadratic Higgs coupling and quartic Higgs coupling, respectively,
\begin{align}
l_h(\varphi,T) &= -\frac{m_h^4(\varphi)}{64\pi^2}\left(\ln\frac{\mu^2}{m_h^2(\varphi)}+\frac{3}{2}\right) - \frac{1}{6\pi^2}m_h^4(\varphi)I_{B1}\left(\frac{m_h(\varphi)}{T}\right), \label{l_h_phi_T}\\
l_W(\varphi,T) &= -\frac{3m_W^4(\varphi)}{32\pi^2}\left(\ln\frac{\mu^2}{m_W^2(\varphi)}+\frac{5}{6}\right) - \frac{1}{\pi^2}m_W^4(\varphi)I_{B1}\left(\frac{m_W(\varphi)}{T}\right), \label{l_W_phi_T}\\
l_Z(\varphi,T) &= -\frac{3m_Z^4(\varphi)}{64\pi^2}\left(\ln\frac{\mu^2}{m_Z^2(\varphi)}+\frac{5}{6}\right) - \frac{1}{2\pi^2}m_Z^4(\varphi)I_{B1}\left(\frac{m_Z(\varphi)}{T}\right), \label{l_Z_phi_T}\\
l_t(\varphi,T) &= +\frac{3m_t^4(\varphi)}{16\pi^2}\left(\ln\frac{\mu^2}{m_t^2(\varphi)}+\frac{3}{2}\right) - \frac{2}{\pi^2}m_t^4(\varphi)I_{F1}\left(\frac{m_t(\varphi)}{T}\right) \label{l_t_phi_T}
\end{align}
are one-loop contributions of Higgs bosons (subscript $h$), $W$ bosons ($W$), $Z$ bosons ($Z$), and top quarks ($t$),
and $I_{F1}(y)$ and $I_{B1}(y)$ are integrals defined in Eq.~\eqref{definition_IF} and Eq.~\eqref{definition_IB}, respectively.
The $\varphi$-dependent masses in Eqs.~\eqref{l_h_phi_T}-\eqref{l_t_phi_T} are given by
\begin{equation}
m_h^2(\varphi) = -\nu^2 + 3\lambda\varphi^2, \quad m_W^2(\varphi) = \frac{1}{4}g^2\varphi^2,\quad m_Z^2(\varphi) = \frac{1}{4}(g^2+g'^2)\varphi^2, \quad m_t^2(\varphi) = \frac{1}{2}y_t^2\varphi^2,
\label{phi_dependent_masses}
\end{equation}
where $g$ is the SU(2)$_L$ gauge coupling, $g'$ is the U(1)$_Y$ gauge coupling, and $y_t$ is the top Yukawa coupling.
Let $v_T$ and $v_0$ be values of $\varphi$ that minimize $f(\varphi,T)$ and $f(\varphi,0)$, respectively.
The pressure in the low temperature broken phase can be estimated as
\begin{equation}
p_{\rm electroweak,low}(T) = f(v_0,0) - f(v_T,T). \label{p_EW_low}
\end{equation}
The contribution to the trace anomaly is given by
\begin{equation}
\Delta_{\rm electroweak,low}(T) = T\frac{d}{dT}\left\{\frac{p_{\rm electroweak,low}(T)}{T^4}\right\} = -\frac{1}{T^3}\frac{df(v_T,T)}{dT} - \frac{4}{T^4}\left[f(v_0,0)-f(v_T,T)\right]. \label{delta_EW_low}
\end{equation}
We evaluate $\Delta_{\rm electroweak,low}(T)$ by minimizing $f(\varphi,T)$ numerically for a given value of $T$.
For coupling parameters $\nu(\mu)$, $\lambda(\mu)$, $g^2(\mu)$, $g'^2(\mu)$, $y_t^2(\mu)$ defined in the $\overline{\text{MS}}$ scheme, 
we adopt three-loop order running evaluated in Ref.~\cite{Buttazzo:2013uya}.
The renormalization scale $\mu$ is simply taken to be a fixed value $\mu = m_Z$, where $m_Z \simeq 91.19\,\mathrm{GeV}$ is the $Z$ boson mass,
and this scale is varied by a factor $0.5\dots2$ in order to see the dependence on the choice of the renormalization scale, which can be regarded as 
theoretical uncertainty.

The pressure and trace anomaly in the high temperature crossover/symmetric phase were also estimated in Refs.~\cite{Gynther:2005dj,Gynther:2005av,Laine:2015kra}
based on the dimensional reduction to the effective 3d theory~\cite{Kajantie:1995dw}, which is essentially 
the same with the effective field theory approach used to estimate the pressure of QCD at high temperature~\cite{Ginsparg:1980ef,Appelquist:1981vg,Braaten:1995jr}.
The trace anomaly in this regime can be split into three terms~\cite{Laine:2015kra}:
\begin{equation}
\Delta_{\rm electroweak,high}(T) = \Delta_1(T) + \Delta_2(T) + \Delta_3(T), \label{delta_EW_high}
\end{equation}
where\footnote{
Equation~\eqref{delta_1} is slightly different from the expression of $\Delta_1(T)$ given in Ref.~\cite{Laine:2015kra}.
Here, we dropped the terms proportional to $g_s^4$ and $g_s^5$ in $\Delta_1(T)$, 
since they are included in the pure QCD contribution $\Delta_{\rm QCD}(T)$, which we treat separately.}
\begin{align}
\Delta_1(T) &= \frac{1}{(4\pi)^2}\left[\frac{266+163n_G-40n_G^2}{288}g^4 - \frac{144+375n_G+1000n_G^2}{7776}g'^4-\frac{g^2g'^2}{32}\right. \nonumber\\
&\quad -y_t^2\left(\frac{7y_t^2}{32}-\frac{5g_s^2}{6}-\frac{15g^2}{64}-\frac{85g'^2}{576}\right) -\lambda\left(\lambda + \frac{y_t^2}{2} - \frac{g'^2+3g^2}{8}\right) \nonumber\\
&\quad \left.+ \frac{\nu^2}{T^2}\left(y_t^2 + 2\lambda - \frac{g'^2+3g^2}{4}\right) - \frac{2\nu^4}{T^4} \right] \nonumber\\
&\quad -\frac{4}{(4\pi)^3}\left[3g^5\left(\frac{5}{6}+\frac{n_G}{3}\right)^{\frac{3}{2}}\left(\frac{43}{24}-\frac{n_G}{3}\right) 
- g'^5\left(\frac{1}{6}+\frac{5n_G}{9}\right)^{\frac{3}{2}}\left(\frac{1}{24}+\frac{5n_G}{9}\right)\right], \label{delta_1}
\end{align}
\begin{align}
\Delta_2(T) & = \Delta_{2E}(T) + \Delta_{2M}(T) + \Delta_{2G}(T), \label{delta_2} 
\end{align}
\begin{align}
\Delta_{2E}(T) & = -\frac{\nu^2}{3T^2}\left\{1-\frac{3}{2(4\pi)^2}\left[(g'^2+3g^2)\left(3\ln\left(\frac{\mu}{4\pi T}\right) + \gamma_E + \frac{5}{3} + \frac{2\zeta'(-1)}{\zeta(-1)}\right) \right.\right. \nonumber\\
&\quad \left.\left.+ 4y_t^2\ln\left(\frac{\mu e^{\gamma_E}}{8\pi T}\right) + 8\lambda\ln\left(\frac{\mu e^{\gamma_E}}{4\pi T}\right)\right]\right\}, 
\end{align}
\begin{align}
\Delta_{2M}(T) & = \frac{2\nu^2}{(4\pi)^3T^2}\left[\frac{g'^2m_{E1}+3g^2m_{E2}}{T} + \frac{g'^4T}{16m_{E1}} + \frac{3g'^2g^2T}{4(m_{E1}+m_{E2})} \right. \nonumber\\
&\quad \left.+ \frac{g^4T}{2m_{E2}}\left(\frac{35}{24}-\frac{3}{2}\ln\left(\frac{\mu}{2m_{E2}}\right)\right)\right],
\end{align}
\begin{align}
\Delta_{2G}(T) & = -\frac{2\nu^2}{T^4}\left\{1+\frac{3}{2(4\pi)^2}\left[(g'^2+3g^2-8\lambda)\ln\left(\frac{\mu e^{\gamma_E}}{4\pi T}\right)-4y_t^2\ln\left(\frac{\mu e^{\gamma_E}}{\pi T}\right)\right]\right\}\langle\phi^{\dagger}\phi\rangle_{\rm 3d} \nonumber\\
&\quad + \frac{\nu^2}{3T^2}\frac{3}{2(4\pi)^2} (g'^2+3g^2) 4\ln\left(\frac{g_3^2}{\mu}\right)
+ \frac{2\nu^2}{(4\pi)^3T^2}\frac{g^4T}{2m_{E2}}\left(\ln\left(\frac{\mu}{g_3^2}\right)+\frac{1}{2}\ln\left(\frac{\mu}{2m_{E2}}\right)\right),
\end{align}
\begin{align}
\Delta_3(T) & = \frac{4p_{0R}}{T^4},
\end{align}
$n_G = 3$ is the number of generations, $\gamma_E \simeq 0.577216$ is the Euler-Mascheroni constant, 
$\zeta(s)$ denotes the Riemann zeta function, $g_3^2 = g^2T$ is the coupling parameter in the 3d theory,
\begin{equation}
m_{E1} = \sqrt{\left(\frac{1}{6}+\frac{5n_G}{9}\right)g'^2 T^2}, \quad m_{E2} = \sqrt{\left(\frac{5}{6}+\frac{n_G}{3}\right)g^2 T^2}
\end{equation}
are the Debye mass parameters, and $p_{0R}$
is a renormalized expression of the zero-temperature pressure [associated with $f(v_0,0)$ in Eq.~\eqref{p_EW_low}].
The three terms in the right-hand side of Eq.~\eqref{delta_EW_high} have different origins and hence represent different effects.
According to the terminology of Ref.~\cite{Laine:2015kra},
$\Delta_1(T)$ represents breaking of scale invariance by quantum corrections, $\Delta_2(T)$ represents the temperature evolution of the Higgs condensate,
and $\Delta_3(T)$ represents the vacuum term. Mathematically, this distinction arises due to the fact that 
$T$ can appear in $p(T)/T^4$ as $\mu/T$, $\nu^2/T^2$, or $p_{0R}/T^4$ because of dimensional reasons.
$\Delta_1(T)$ and $\Delta_2(T)$ correspond to the derivative of $p(T)/T^4$ with respect to $\mu/T$ and $\nu^2/T^2$, respectively.
In Eq.~\eqref{delta_2}, we also split $\Delta_2(T)$ into three terms for practical reasons discussed below.
These three terms correspond to three different momentum scales appearing in the effective 3d theory~\cite{Braaten:1995jr,Kajantie:1995dw,Gynther:2005dj,Gynther:2005av,Laine:2015kra}.
$\Delta_{2E}(T)$ contains contributions from the ``superheavy" scale $k\sim T$, $\Delta_{2M}(T)$ those from the ``heavy" scale $k\sim gT$,\footnote{Hereafter we use an abbreviation $g^2$ 
to represent the expansion parameter of the perturbative analysis in the symmetric phase,
i.e. $g^2 \in \{g^2,g'^2,g_s^2,y_t^2,\lambda,\nu^2/T^2\}$.}
and $\Delta_{2G}(T)$ those from ``light" scale $k\sim g^2T$.
The vacuum contribution $\Delta_3(T)$ is estimated by solving the one-loop renormalization group equation~\cite{Laine:2015kra},
\begin{equation}
\mu\frac{dp_{0R}}{d\mu} = \frac{\nu^4}{8\pi^2},
\end{equation}
from the scale $\mu = m_Z$ to the thermal scale $\mu \sim \pi T$.

$\Delta_{2G}(T)$ contains the 3d Higgs condensate $\langle\phi^{\dagger}\phi\rangle_{\rm 3d}$, which represents non-perturbative dynamics around the electroweak crossover
and should be calculated on the lattice~\cite{DOnofrio:2014rug,DOnofrio:2015gop}.
Here we adopt the latest result obtained in Ref.~\cite{DOnofrio:2015gop} for the temperature evolution of the Higgs condensate in the crossover phase.
The evolution of $\langle\phi^{\dagger}\phi\rangle_{\rm 3d}$ can also be estimated in perturbation theory at temperature higher than the crossover temperature~\cite{Laine:2015kra}.
We simply connect the lattice data with the perturbative result shown in Eq.~(A.1) of Ref.~\cite{Laine:2015kra} at $T = 170\,\mathrm{GeV}$
to see the temperature evolution of the Higgs condensate in the high temperature crossover/symmetric phase. 

In order to compare the result obtained in the low temperature broken phase with that obtained in the high temperature crossover/symmetric phase,
we need to take an additional step.
It is known that the one-loop effective potential [Eq.~\eqref{f_phi_4d_1loop}] becomes inaccurate at high temperature because of infrared singularities,
and that it must be corrected by performing the resummation of the infrared contributions from higher order diagrams~\cite{Dolan:1973qd,Takahashi:1985vx,Fendley:1987ef,Carrington:1991hz,Arnold:1992rz}.
The effective potential in such a high temperature regime was constructed systematically in Ref.~\cite{Farakos:1994kx} by using the effective 3d theory approach~\cite{Ginsparg:1980ef,Appelquist:1981vg,Braaten:1995jr}.
With the help of the effective potential $V_{\rm 3d}(\varphi_{\rm 3d})$ in the 3d theory,
we can write the contribution to the free energy density as $f_{\rm 3d}(\varphi_{\rm 3d},T) = TV_{\rm 3d}(\varphi_{\rm 3d})$.
As in Eqs.~\eqref{p_EW_low} and~\eqref{delta_EW_low}, the pressure and trace anomaly might be estimated by finding a Higgs expectation value in the 3d theory $\varphi_{\rm 3d} =  v_{\mathrm{3d},T}$
that minimizes $V_{\rm 3d}(\varphi_{\rm 3d})$~\cite{Farakos:1994xh,Kajantie:1995kf},
where we use Eq.~(54) of Ref.~\cite{Farakos:1994kx} to evaluate $V_{\rm 3d}(\varphi_{\rm 3d})$.\footnote{
We use the results obtained in Ref.~\cite{Kajantie:1995dw} for parameters in the 3d effective theory, as they contain the contribution of top Yukawa coupling
that was omitted in Ref.~\cite{Farakos:1994kx}. We also added the extra term in Eq.~(B.10) of Ref.~\cite{Kajantie:1995kf} to reproduce an ``improved" form discussed in that paper.}
However, this procedure encounters the following obstacle.
To renormalize the pressure such that it vanishes at $T=0$, we need to subtract the zero-temperature free energy density as in Eq.~\eqref{p_EW_low},
while it is impossible to define the 3d effective potential $V_{\rm 3d}(\varphi_{\rm 3d})$ at $T=0$ since it relies on the high temperature expansion.
Note that it is inappropriate just to use a 4d result $f(v_0,0)$ as a subtraction term.
$f(v_0,0)$ is evaluated by setting the renormalization scale as $\mu\sim m_Z$, while we should choose the thermal scale $\mu \sim \pi T$
to evaluate the effective potential $V_{\rm 3d}(\varphi_{\rm 3d})$ at high temperature.
Since the result for pressure should be independent of the choice of the renormalization scale, most of the $\mu$-dependence must cancel
when we subtract the zero-temperature term.
Such a cancellation does not happen if we use $f(v_0,0)$ as a subtraction term, as it is evaluated at a different scale.

In order to remedy the difficulty mentioned above in a consistent way, we may have to develop some new prescription, which is beyond the scope of this paper.
Instead, here we estimate the trace anomaly at intermediate temperature in the following way.
Based on the fact that the 3d effective potential $V_{\rm 3d}(\varphi_{\rm 3d})$ was derived by integrating out the superheavy ($\sim T$) and heavy ($\sim gT$) modes,
we expect that the free energy density $f_{\rm 3d}(v_{\rm 3d},T)$ only contains the contributions of light ($\sim g^2T$) modes.
Therefore, we replace $\Delta_{2G}(T) $ in Eq.~\eqref{delta_EW_high} with a temperature derivative of $f_{\rm 3d}(v_{\mathrm{3d},T},T)/T^4$ and write the trace anomaly as
\begin{equation}
\Delta_{\rm electroweak,low,3d}(T) = -\frac{1}{T^3}\frac{df_{\rm 3d}(v_{\mathrm{3d},T},T)}{dT} + \frac{4f_{\rm 3d}(v_{\mathrm{3d},T},T)}{T^4} + \Delta_{2E}(T) + \Delta_{2M}(T) + \Delta_3(T). \label{delta_EW_low_3d}
\end{equation}
Here we have dropped $\Delta_1(T)$ since its contribution is small compared with other terms in the relevant temperature range.

\begin{figure}[htbp]
\centering
$\begin{array}{c}
\subfigure{
\includegraphics[width=140mm]{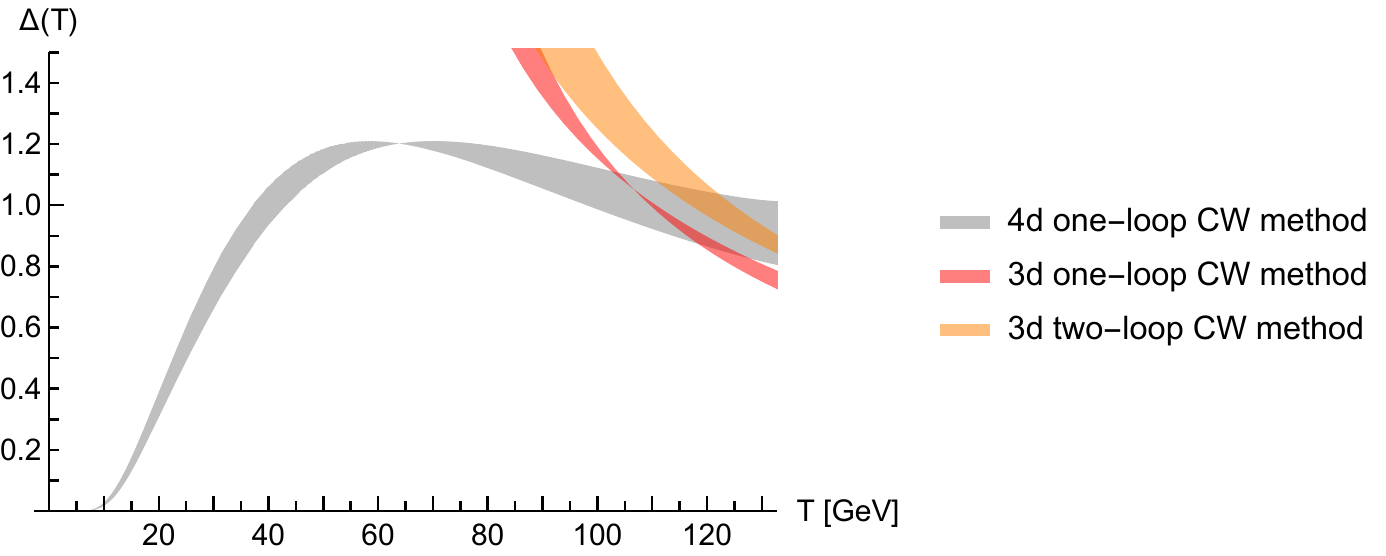}}
\vspace{10mm}
\\
\subfigure{
\includegraphics[width=140mm]{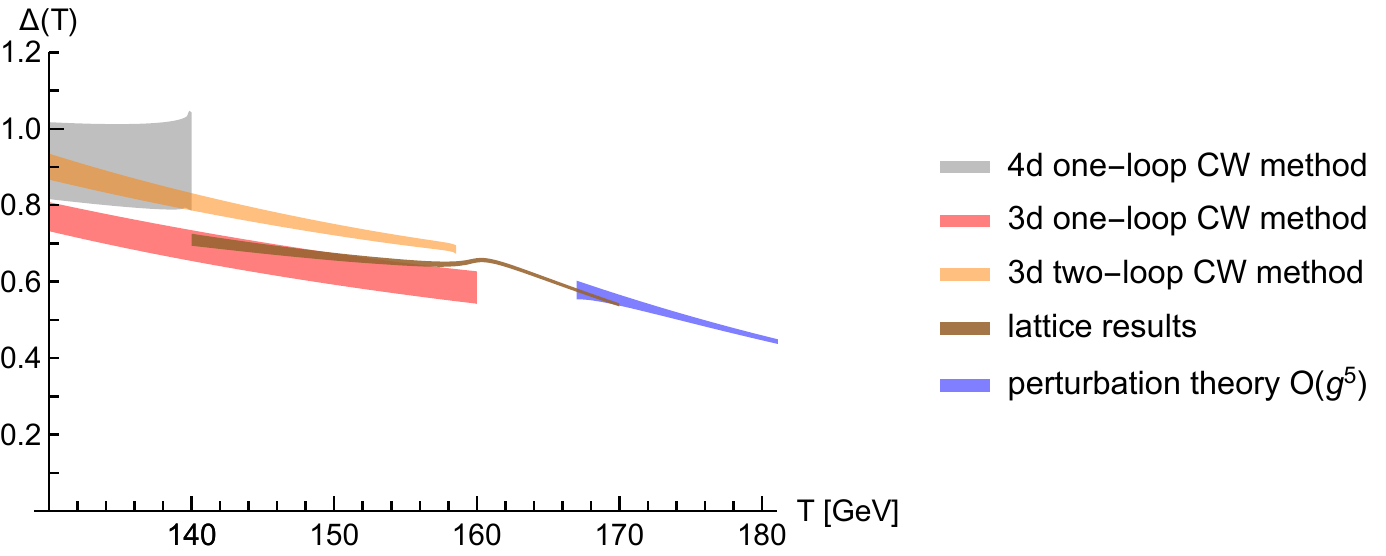}}
\end{array}$
  \caption{Contributions of ``weakly" interacting particles (i.e. those of the SM particles except for the pure QCD contribution $\Delta_{\rm QCD}$) to the trace anomaly.
  The gray and orange bands represent the result based on the one-loop CW method [Eq.~\eqref{delta_EW_low}] and
  that based on the two-loop potential in the effective 3d theory [Eq.~\eqref{delta_EW_low_3d}], respectively.
  We also show the one-loop version of Eq.~\eqref{delta_EW_low_3d} in a red band for the sake of comparison.
  The brown and blue bands are drawn by using Eq.~\eqref{delta_EW_high} with the lattice data from Ref.~\cite{DOnofrio:2015gop}
  and the perturbative results from Ref.~\cite{Laine:2015kra}.}
  \label{fig:EWtraceanomaly}
\end{figure}

In Fig.~\ref{fig:EWtraceanomaly}, we compare the results for the trace anomaly evaluated by using Eqs.~\eqref{delta_EW_low},~\eqref{delta_EW_high}, and~\eqref{delta_EW_low_3d}.
In evaluating Eqs.~\eqref{delta_EW_high} and~\eqref{delta_EW_low_3d}, we take the renormalization scale as $\mu = a\pi T$ and vary 
the factor $a$ from $0.5$ to $2$ to see theoretical uncertainty.
We see that the perturbative results~\cite{Gynther:2005dj,Gynther:2005av,Laine:2015kra} agree with the lattice data~\cite{DOnofrio:2015gop} at high temperature.
At low temperature, we also observe that there exists a region in which two results obtained based on Eqs.~\eqref{delta_EW_low} and~\eqref{delta_EW_low_3d} overlap each other.
However, the result of 3d two-loop CW method~\eqref{delta_EW_low_3d} appears to be slightly larger than the lattice results.
The discrepancy can be caused by the fact that the 3d effective potential $V_{\rm 3d}(\varphi_{\rm 3d})$ is derived based on the perturbative analysis in the broken phase,
which might not include all the relevant infrared contributions.

Here we simply take the discrepancy of $\Delta(T)$ among three different methods [Eqs.~\eqref{delta_EW_low},~\eqref{delta_EW_high}, and~\eqref{delta_EW_low_3d}] as 
theoretical uncertainty, and estimate the contributions of the electroweak sector to the trace anomaly in the following way.
We artificially switch $\Delta_{\rm electroweak,low}(T)$ [Eq.~\eqref{delta_EW_low}] to $\Delta_{\rm electroweak,low,3d}(T)$ [Eq.~\eqref{delta_EW_low_3d}] at a temperature $T_a$,
and $\Delta_{\rm electroweak,low,3d}(T)$ [Eq.~\eqref{delta_EW_low_3d}] to $\Delta_{\rm electroweak,high}(T)$ [Eq.~\eqref{delta_EW_high}] at another temperature $T_b$.
We choose two different sets for these switching temperatures, $(T_a,T_b) = (125\,\mathrm{GeV},155\,\mathrm{GeV})$ and $(135\,\mathrm{GeV},145\,\mathrm{GeV})$,
and estimate the pressure by integrating $\Delta(T)$ for each set of $(T_a,T_b)$ in order to check how such a difference affects the final result.
Fortunately, the corresponding uncertainty turns out to be small compared to that arising from the QCD sector discussed in the next subsection.

%%%%%%%%%%%%%%%%%%%%%%%%%%%%%%%%%%%%%%%%%%%%%%%%%%
\subsection{Full equation of state and its uncertainty}
\label{sec:full_EoS}
%%%%%%%%%%%%%%%%%%%%%%%%%%%%%%%%%%%%%%%%%%%%%%%%%%

Making use of all the ingredients discussed in the previous subsections, now we proceed to construct
the full equation of state in the SM at arbitrary temperatures. 
The main concern here is how to combine the results obtained at different temperatures.
We must take account of several ambiguities in the estimation of thermodynamic quantities in the SM,
such as the poor convergence of the perturbative analysis of the pressure of QCD at $T \sim \mathcal{O}(1\textendash 10)\,\mathrm{GeV}$
and the small discrepancy between the trace anomaly evaluated in the broken phase and that evaluated in the symmetric phase across the electroweak crossover.
Because of the existence of such ambiguities, it is impossible to evaluate the pressure directly at arbitrary temperatures.
Therefore, instead of relying on the direct approach, here we evaluate the pressure indirectly 
by integrating the trace anomaly from some temperature $T_*$ at which the value of $p(T_*)/T_*^4$ can be determined less ambiguously:
\begin{equation}
\frac{p(T)}{T^4} = \frac{p(T_*)}{T_*^4}+\int^T_{T_*}\frac{dT}{T}\Delta(T). \label{pressure_integration}
\end{equation}
In particular, we consider the following two possibilities for the choice of the fiducial temperature $T_*$:
\begin{enumerate}
\item \emph{Integration from low temperature}. Assuming all particles (including neutrinos) are in thermal equilibrium at $T_*=10\,\mathrm{MeV}$,
we expect that the value of $p(T_*)/T_*^4$ can be estimated by using the analytical results shown in Sec.~\ref{sec:photons_and_leptons}.
Then we integrate $\Delta(T)$ to estimate $p(T)/T^4$ at higher temperatures.
\item \emph{Integration from high temperature}. We also expect that the thermodynamic quantities can be accurately estimated based on the perturbation theory
at temperature much higher than that of the electroweak crossover. As an alternative approach, we estimate the value of $p(T_*)/T_*^4$ at $T_* = 10^{17}\,\mathrm{GeV}$
and integrate $\Delta(T)$ to estimate $p(T)/T^4$ at lower temperatures.
\end{enumerate}
In both cases, we expect that the value of $p(T)/T^4$ becomes more uncertain as we proceed with the integration, 
since it picks up the uncertainty of $\Delta(T)$ at each temperature.
We interpolate these two results at some intermediate temperature where the errors accumulated from low and high temperature show some overlap.

On the low temperature side, we estimate the value of $p(T_*)/T_*^4$ at $T_*=10\,\mathrm{MeV}$ by using 
Eqs.~\eqref{gsr_nu_decoupling_corrected} and~\eqref{gss_nu_decoupling_corrected}, which read
\begin{equation}
\left.\frac{p(T_*)}{T_*^4}\right|_{T_*\,=\,10\,\mathrm{MeV}} = \left.\frac{2\pi^2}{45}\left(g_{*s}(T_*) - \frac{3}{4}g_{*\rho}(T_*)\right)\right|_{T_*\,=\,10\,\mathrm{MeV}} = 1.177 \pm 0.001. \label{p_over_T4_low}
\end{equation}
With this boundary condition, we perform the integration in Eq.~\eqref{pressure_integration}. The integrand is given by
\begin{equation}
\Delta(T) = \Delta_{\rm leptons}(T) + \Delta_{\rm strong}(T) + \Delta_{\rm electroweak}(T),
\end{equation}
where 
\begin{equation}
\Delta_{\rm leptons}(T) = \left\{
\begin{array}{ll}
{\displaystyle \Delta_{F,e}^0(T) + \Delta_{F,\mu}^0(T) + \Delta_{F,\tau}^0(T) + \Delta_{\rm QED}(T)} & {\displaystyle \text{for} \quad T \le 120\,\mathrm{MeV},} \\ [0.5ex]
{\displaystyle \Delta_{F,e}^0(T) + \Delta_{F,\mu}^0(T) + \Delta_{F,\tau}^0(T)} & {\displaystyle \text{for} \quad T > 120\,\mathrm{MeV}}
\end{array}
\right.
\label{delta_leptons}
\end{equation}
represents the contribution of charged leptons [free electrons, muons, taus, and QED corrections~\eqref{delta_QED}], 
$\Delta_{\rm strong}(T)$ that of strongly interacting particles (explained below), and 
\begin{align}
\Delta_{\rm electroweak}(T) &= \left\{
\begin{array}{ll}
{\displaystyle \Delta_{\rm electroweak,low}(T)} & {\displaystyle \text{for}\quad T \le T_a}, \\ [0.5ex]
{\displaystyle \Delta_{\rm electroweak,low,3d}(T)} & {\displaystyle\text{for}\quad T_a < T \le T_b}, \\ [0.5ex]
{\displaystyle \Delta_{\rm electroweak,high}(T)} & {\displaystyle\text{for}\quad T > T_b}
\end{array}
\right. \label{delta_electroweak}
\end{align}
that of the electroweak crossover [see Eqs.~\eqref{delta_EW_low}, \eqref{delta_EW_low_3d}, and~\eqref{delta_EW_high}].
In Eq.~\eqref{delta_leptons}, we simply drop the QED corrections  for $T>120\,\mathrm{MeV}$ since they are negligible compared with
large uncertainty from the pressure of QCD and should be replaced with electroweak interactions at higher temperatures.
In Eq.~\eqref{delta_electroweak}, we artificially switch the formulae at temperatures $T_a$ and $T_b$ as mentioned in Sec.~\ref{sec:electroweak_crossover}
and choose two different sets, $(T_a,T_b) = (125\,\mathrm{GeV},155\,\mathrm{GeV})$ and $(135\,\mathrm{GeV},145\,\mathrm{GeV})$, to see the corresponding uncertainty.
In addition, we vary the renormalization scale $\mu$ in the range $(0.5\dots 2)\pi T$ in evaluating $\Delta_{\rm electroweak,low}(T)$ and $\Delta_{\rm electroweak,high}(T)$ 
and $(0.5\dots 2)m_Z$ in evaluating $\Delta_{\rm electroweak,low}(T)$, and add the resulting variation of $p(T)/T^4$ and $\Delta(T)$ to the error of the final results.

We believe that the contribution of strongly interacting particles $\Delta_{\rm strong}(T)$ is the dominant source of uncertainty, and it should be treated carefully.
In Ref.~\cite{Borsanyi:2016ksw}, the result of lattice QCD simulations was connected to the perturbative result [Eq.~\eqref{p_QCD(u+d+s+c+b)}] at $T = 500\,\mathrm{MeV}$ by 
fixing $\mu = 2\pi T$ and adjusting $q_c(N_f = 4)$ such that the uncertainty of the perturbative result due to the variation of $q_c(N_f = 4)$
is matched with the error in lattice data at that temperature. However, this matching procedure is somewhat artificial, leaving room for some additional systematic uncertainty.
Indeed, the uncertainty of the perturbative result turns out to be huge at the relevant temperature as we have seen 
in Fig.~\ref{fig:QCDpressure} by varying the renormalization scale $\mu$, 
and hence the perturbative result in the temperature range $T = \mathcal{O}(0.1\textendash 10)\,\mathrm{GeV}$ should be regarded as 
just a fitting function to extrapolate the lattice result rather than a realistic estimate.
In other words, the result of the pressure in the temperature range $T = \mathcal{O}(0.1\textendash 10)\,\mathrm{GeV}$ can be significantly affected 
by the method to extrapolate the lattice result.

Regarding the issue mentioned above, we model the contribution of strongly interacting particles $\Delta_{\rm strong}(T)$ as follows.
First, we use the following function to describe different phases across the QCD crossover,
\begin{equation}
\Delta_{\rm strong}(T) = \left\{
\begin{array}{ll}
{\displaystyle \Delta_{\rm hadrons}(T)} & {\displaystyle\text{for} \quad T \le 120\,\mathrm{MeV},} \\ [0.5ex]
{\displaystyle \Delta_{\rm lattice}^{(2+1+1)}(T)} & {\displaystyle\text{for} \quad 120\,\mathrm{MeV} < T \le T_s,} \\ [0.5ex]
{\displaystyle \Delta_{\rm QCD}^{(u+d+s+c+b+t)}(T)} & {\displaystyle\text{for} \quad  T > T_s,} \\
\end{array}
\right. \label{delta_strong}
\end{equation}
where $\Delta_{\rm hadrons}(T)$ is given by Eq.~\eqref{delta_HRG}, $\Delta_{\rm lattice}^{(2+1+1)}(T)$ denotes the lattice data from Ref.~\cite{Borsanyi:2016ksw},
and $\Delta_{\rm QCD}^{(u+d+s+c+b+t)}(T)$ is given by Eq.~\eqref{delta_QCD(u+d+s+c+b+t)}.
Second, we estimate the uncertainty of $\Delta_{\rm strong}(T)$ at $T\le 120\,\mathrm{MeV}$
by multiplying the error of $\Delta_{\rm lattice}^{(2+1+1)}$ at $T = 120\,\mathrm{MeV}$
by $\Delta_{\rm hadrons}(T)/\Delta_{\rm hadrons}(T=120\,\mathrm{MeV})$.
Third, for the matching between the lattice result and perturbative result, we consider two possibilities for the switching temperature, $T_s = 500\,\mathrm{MeV}$ and $T_s = 1\,\mathrm{GeV}$.
Then, we perform two different methods to connect $\Delta_{\rm lattice}^{(2+1+1)}(T)$ with $\Delta_{\rm QCD}^{(u+d+s+c+b+t)}(T)$ at $T = T_s$.
On the one hand we fix $\mu = 2\pi T$ and vary the value of $q_c(N_f=4)$ to propagate the error in lattice data at $T=T_s$,
while on the other hand we fix $q_c(N_f=4) = - 3000$ and vary $\mu$.
In both cases, we fix $q_c(N_f=6) = - 3000$ for $\left.\Delta_{\rm QCD}\right|_{N_f\,=\,6}$ included in $\Delta_{\rm QCD}^{(u+d+s+c+b+t)}(T)$.
Finally, we perform the integration~\eqref{pressure_integration} in each case (two cases for $T_s$ times two cases for the matching method)
and identify the uncertainty of $p(T)/T^4$ as a difference between the largest value and the smallest value in the results of four different trials.

On the high temperature side, we estimate the value of $p(T_*)/T_*^4$ at $T_*=10^{17}\,\mathrm{GeV}$ 
based on the perturbative result for the pressure of the SM. Here we use the $\mathcal{O}(g_s^6)$ result
for the contribution of QCD [Eq.~\eqref{p_QCD_perturbative}] and include other contributions up to the terms of $\mathcal{O}(g^4)$
extracted from Ref.~\cite{Gynther:2005dj}. The uncertainty of the result is identified by varying the value of $q_c(N_f = 6)$ in the range 
$q_c(N_f = 6) \in (-5000\dots+5000)$ and the renormalization scale $\mu$ in the range $\mu \in (0.5\dots 2)\pi T_*$.
As a result, we obtain
\begin{equation}
\left.\frac{p(T_*)}{T_*^4}\right|_{T_*\,=\,10^{17}\,\mathrm{GeV}} = 11.542 \pm 0.002. \label{p_over_T4_high}
\end{equation}
Using the above value as a boundary condition, we perform the integration~\eqref{pressure_integration} down to lower temperatures.
Instead of taking steps mentioned below Eq.~\eqref{delta_strong}, here we simply use Eq.~\eqref{delta_QCD(u+d+s+c+b+t)}
for the contribution of QCD, take $q_c(N_f = 4) = q_c(N_f = 6)$ for simplicity, and identify the uncertainty of the result
by varying the value of $q_c(N_f = 6)$ in the range $q_c(N_f = 6) \in (-5000\dots+5000)$ and $\mu$ in the range $\mu \in (0.5\dots 2)\pi T$
[or in the range $\mu \in (0.5\dots 2)m_Z$ in evaluating $\Delta_{\rm electroweak,low}(T)$].

\begin{figure}[htbp]
\centering
$\begin{array}{c}
\subfigure{
\includegraphics[width=160mm]{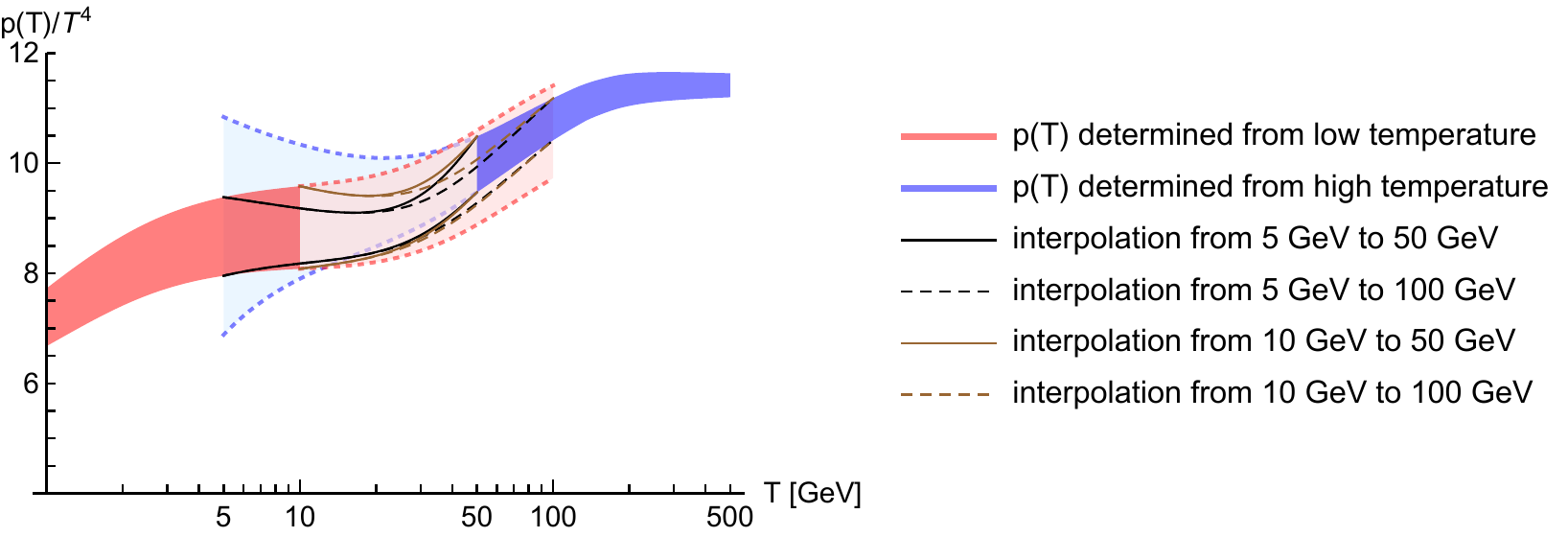}}
\vspace{10mm}
\\
\subfigure{
\includegraphics[width=160mm]{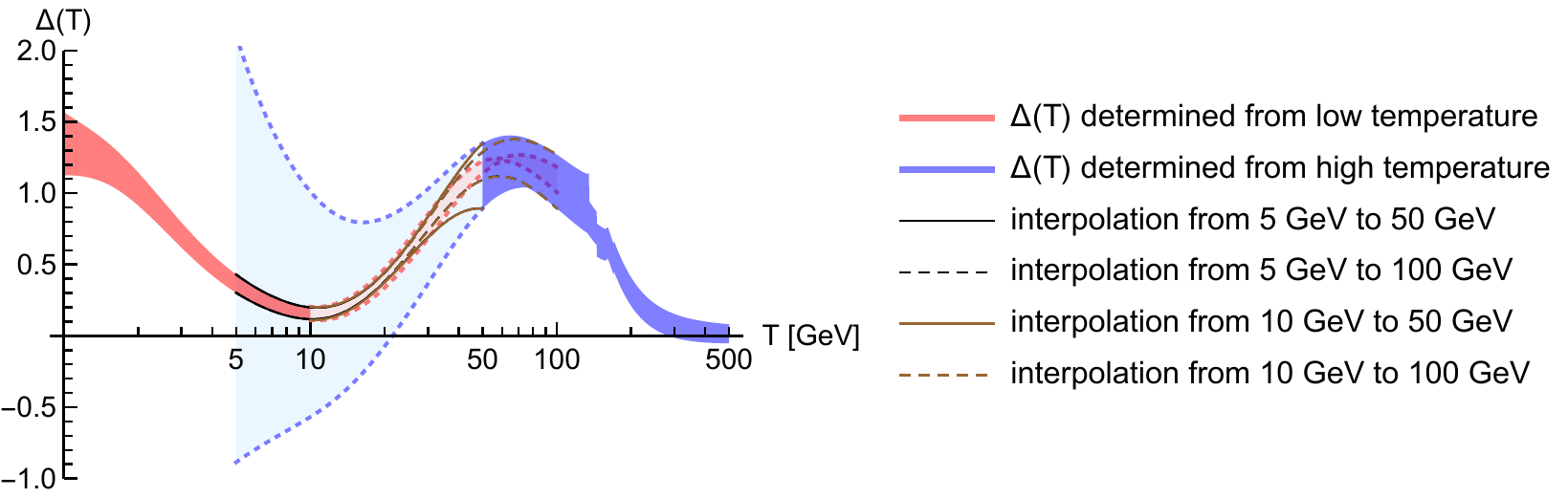}}
\end{array}$
  \caption{The pressure (top panel) and trace anomaly (bottom panel) of the SM at the intermediate temperature range.
  In the top panel, the red (blue) band represents $p(T)/T$ determined by performing the integration~\eqref{pressure_integration} 
  from $T_* = 10\,\mathrm{MeV}$ ($T_* = 10^{17}\,\mathrm{GeV}$) with a boundary condition given by Eq.~\eqref{p_over_T4_low} [Eq.~\eqref{p_over_T4_high}].
  The corresponding values of $\Delta(T)$ are shown in the bottom panel.
  Thin lines show the interpolation obtained by using Eq.~\eqref{p_intermediate} or Eq.~\eqref{delta_intermediate}.}
  \label{fig:compare_integration}
\end{figure}

In Fig.~\ref{fig:compare_integration}, we show the integrated values of $p(T)/T^4$ as well as the integrand $\Delta(T)$
determined from low and high temperature. As expected, the error in $p(T)/T^4$ increases as we proceed with the integration,
and two results overlap each other at $T\sim \mathcal{O}(10)\,\mathrm{GeV}$. Now, let us interpolate these two results 
at this intermediate temperature range. To this end, we introduce the following interpolating function,
\begin{align}
p_{\rm intermediate}(T; q_{c4},q_{c6}) &= p_{\rm leptons}(T) + c_t(T) \left.p_{\rm QCD}(T;q_{c6})\right|_{N_f\,=\,6} \nonumber\\
&\quad + \left(1-c_t(T)\right) p_{\rm QCD}^{(u+d+s+c+b)}(T;q_{c4}) + p_{\rm electroweak,low}(T), \label{p_intermediate}
\end{align}
where $p_{\rm leptons}(T)$ is given by $p_{\rm leptons}(T) = p_{F,e}^0(T) + p_{F,\mu}^0(T) + p_{F,\tau}^0(T)$,
$c_t(T)$ by Eq.~\eqref{c_topmass}, $p_{\rm QCD}(T)$ by Eq.~\eqref{p_QCD_perturbative}, 
$p_{\rm QCD}^{(u+d+s+c+b)}(T)$ by Eq.~\eqref{p_QCD(u+d+s+c+b)}, and $p_{\rm electroweak,low}(T)$ by Eq.~\eqref{p_EW_low}.
We emphasize that the above equation is just one of possible options for the interpolating function, since
perturbative expansion in QCD becomes less controllable in the temperature range considered here.
With this in mind, we take $q_{c4} \equiv q_c(N_f = 4)$ and $q_{c6} \equiv q_c(N_f = 6)$ in Eq.~\eqref{p_intermediate}
as free parameters, which are adjusted such that two integrated results for $p(T)/T^4$ are interpolated at the intermediate temperature range.
In the interpolation procedure, we simply fix $\mu$ to $2\pi T$ for $\left.p_{\rm QCD}(T;q_{c6})\right|_{N_f\,=\, 6}$ and $p_{\rm QCD}^{(u+d+s+c+b)}(T;q_{c4})$
and to $m_Z$ for $p_{\rm electroweak,low}(T)$ in Eq.~\eqref{p_intermediate}.
By varying the values of $q_{c4}$ and $q_{c6}$, $p(T)/T^4$ determined from low and high temperature are glued 
across a temperature range $T_{\rm low} \le T \le T_{\rm high}$ as shown in Fig.~\ref{fig:compare_integration}.
Here we consider four possibilities for the interpolation range, 
$(T_{\rm low},T_{\rm high}) = (5\,\mathrm{GeV},50\,\mathrm{GeV})$, $(5\,\mathrm{GeV},100\,\mathrm{GeV})$, $(10\,\mathrm{GeV},50\,\mathrm{GeV})$,
and $(10\,\mathrm{GeV},100\,\mathrm{GeV})$.
After this interpolation procedure, we identify the uncertainty of $p(T)/T^4$ by using the largest value and the smallest value 
in $p_{\rm intermediate}(T; q_{c4},q_{c6})/T^4$'s as well as $p(T)/T^4$ determined from low temperature for $5\,\mathrm{GeV}< T \le 10\,\mathrm{GeV}$,
in $p_{\rm intermediate}(T; q_{c4},q_{c6})/T^4$'s for $10\,\mathrm{GeV}< T \le 50\,\mathrm{GeV}$, and
in $p_{\rm intermediate}(T; q_{c4},q_{c6})/T^4$'s as well as $p(T)/T^4$ determined from high temperature for $50\,\mathrm{GeV}< T \le 100\,\mathrm{GeV}$.
For $T \le 5\,\mathrm{GeV}$ and $T > 100\,\mathrm{GeV}$, we simply adopt $p(T)/T^4$ determined from low temperature and that determined from high temperature, respectively.

We apply the same procedure to determine $\Delta(T)$ and its uncertainty at the intermediate temperature range.
Here we use the following interpolating function,
\begin{align}
\Delta_{\rm intermediate}(T; q_{c4},q_{c6}) &= \Delta_{\rm leptons}(T) + c_t(T) \left.\Delta_{\rm QCD}(T;q_{c6})\right|_{N_f\,=\,6} \nonumber\\
&\quad + \left(1-c_t(T)\right) \Delta_{\rm QCD}^{(u+d+s+c+b)}(T;q_{c4}) + \Delta_{\rm electroweak,low}(T), \label{delta_intermediate}
\end{align}
where $\Delta_{\rm leptons}(T)$ is given by Eq.~\eqref{delta_leptons}, $\Delta_{\rm QCD}(T)$ by Eq.~\eqref{delta_QCD_perturbative}, 
$\Delta_{\rm QCD}^{(u+d+s+c+b)}(T)$ by Eq.~\eqref{delta_QCD(u+d+s+c+b)}, and $\Delta_{\rm electroweak,low}(T)$ by Eq.~\eqref{delta_EW_low}.
The results of the interpolation are shown in Fig.~\ref{fig:compare_integration}.
Note that the values of $q_{c4}$ and $q_{c6}$ used to interpolate $\Delta(T)$ are different from those used to interpolate $p(T)/T^4$.
This fact implies that $\Delta(T)$ does not exactly correspond to $T\frac{d}{dT}\{p(T)T^{-4}\}$ at the intermediate temperature range.

\begin{figure}[htbp]
\centering
$\begin{array}{cc}
\subfigure{
\includegraphics[width=80mm]{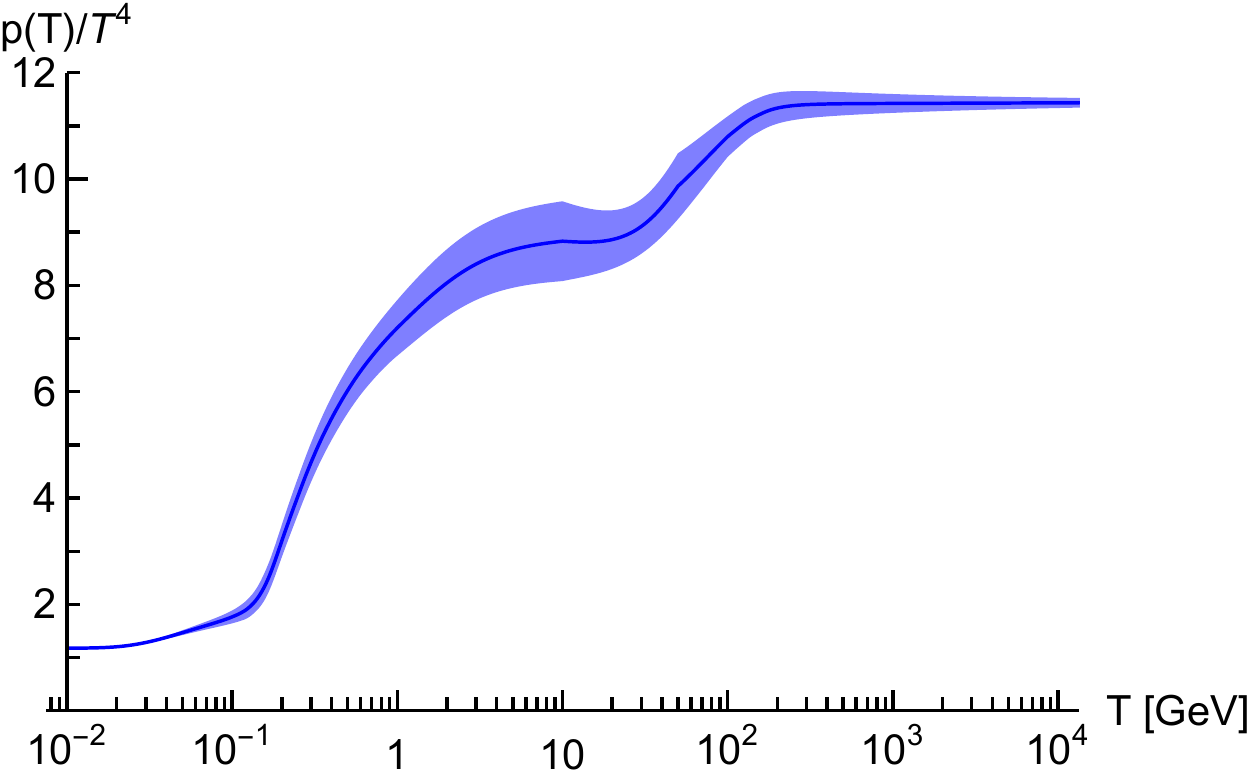}}
\hspace{5mm}
\subfigure{
\includegraphics[width=80mm]{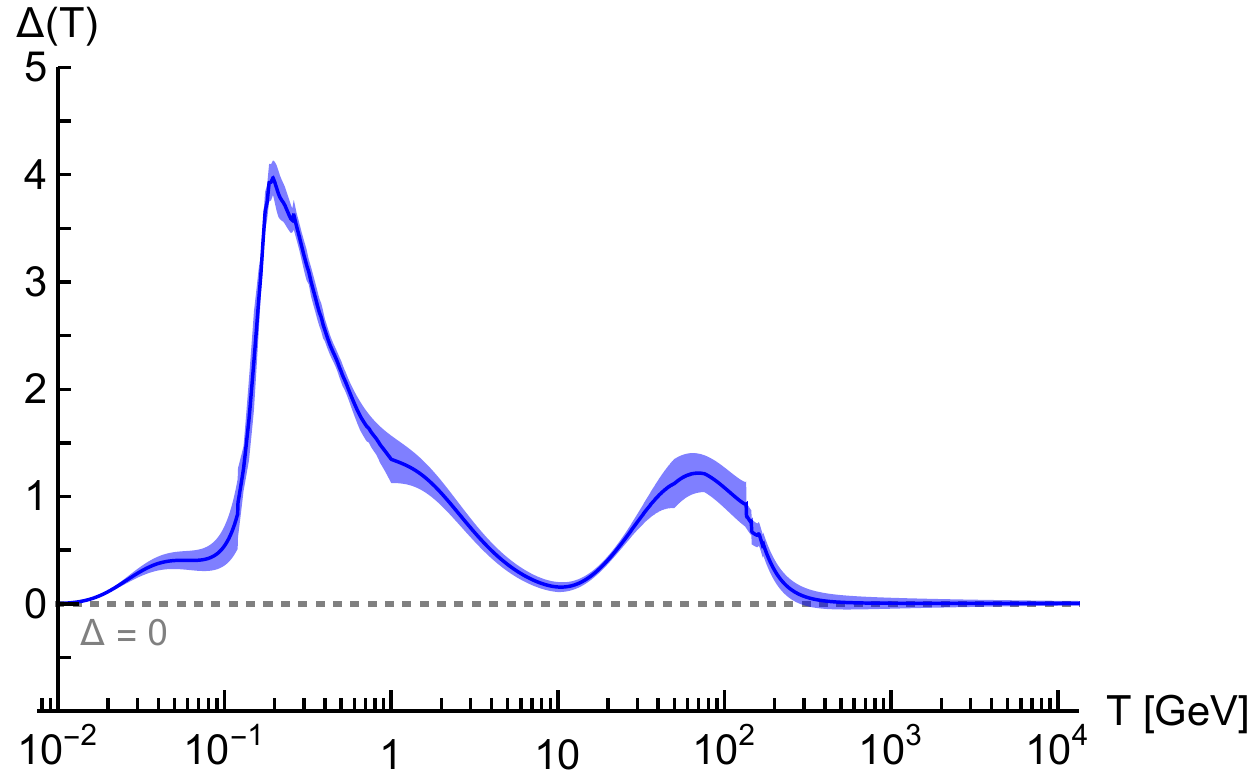}}
\vspace{10mm}
\\
\subfigure{
\includegraphics[width=80mm]{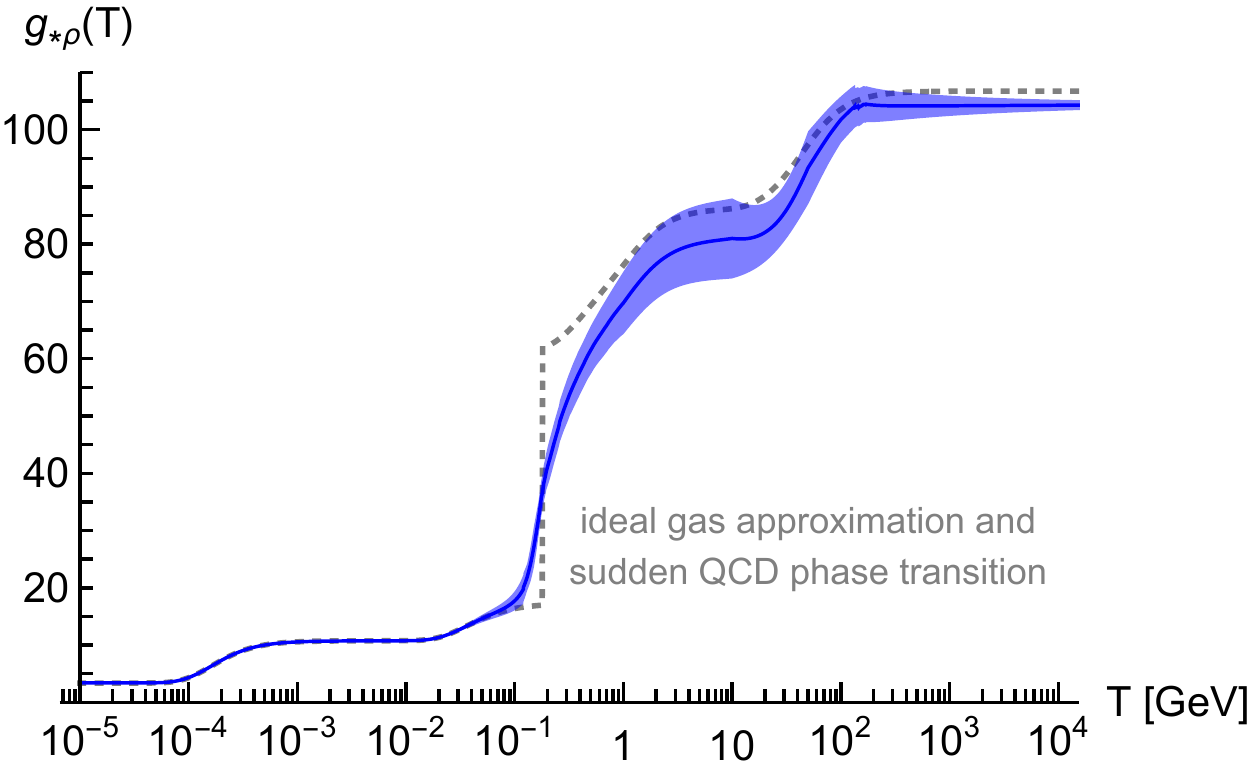}}
\hspace{5mm}
\subfigure{
\includegraphics[width=80mm]{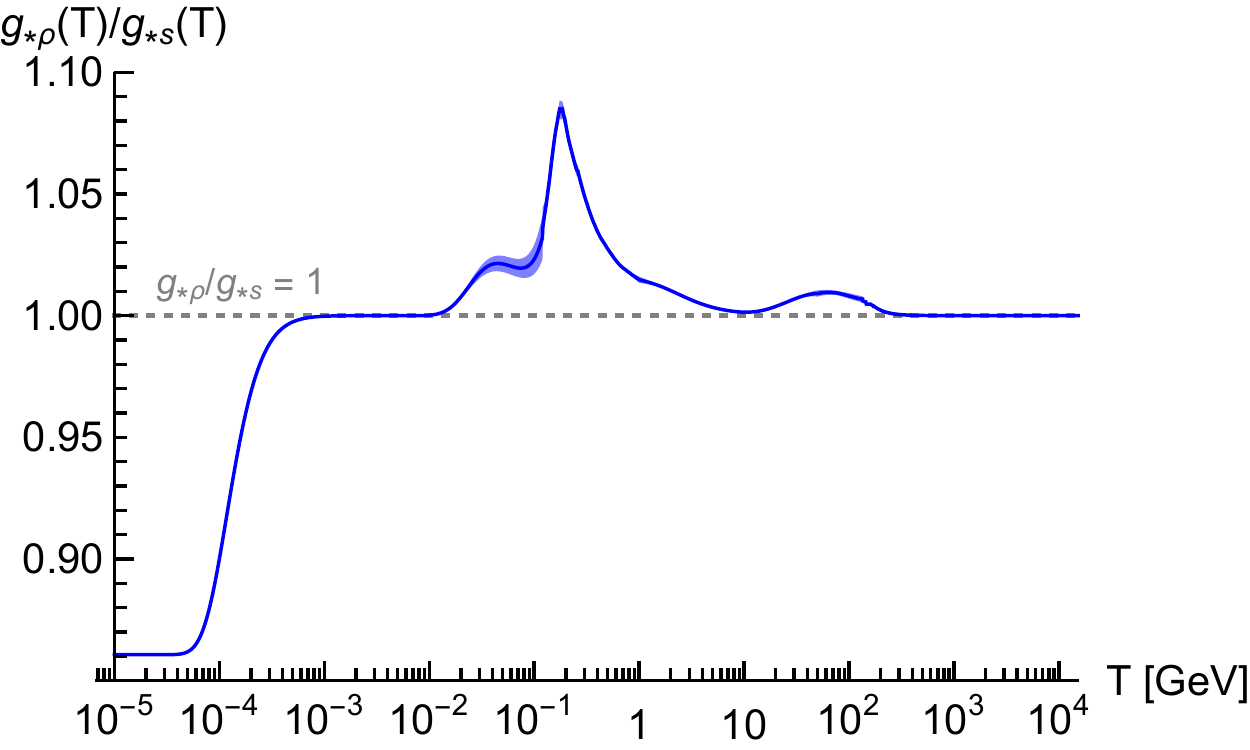}}
\vspace{10mm}
\end{array}$
\subfigure{
\includegraphics[width=80mm]{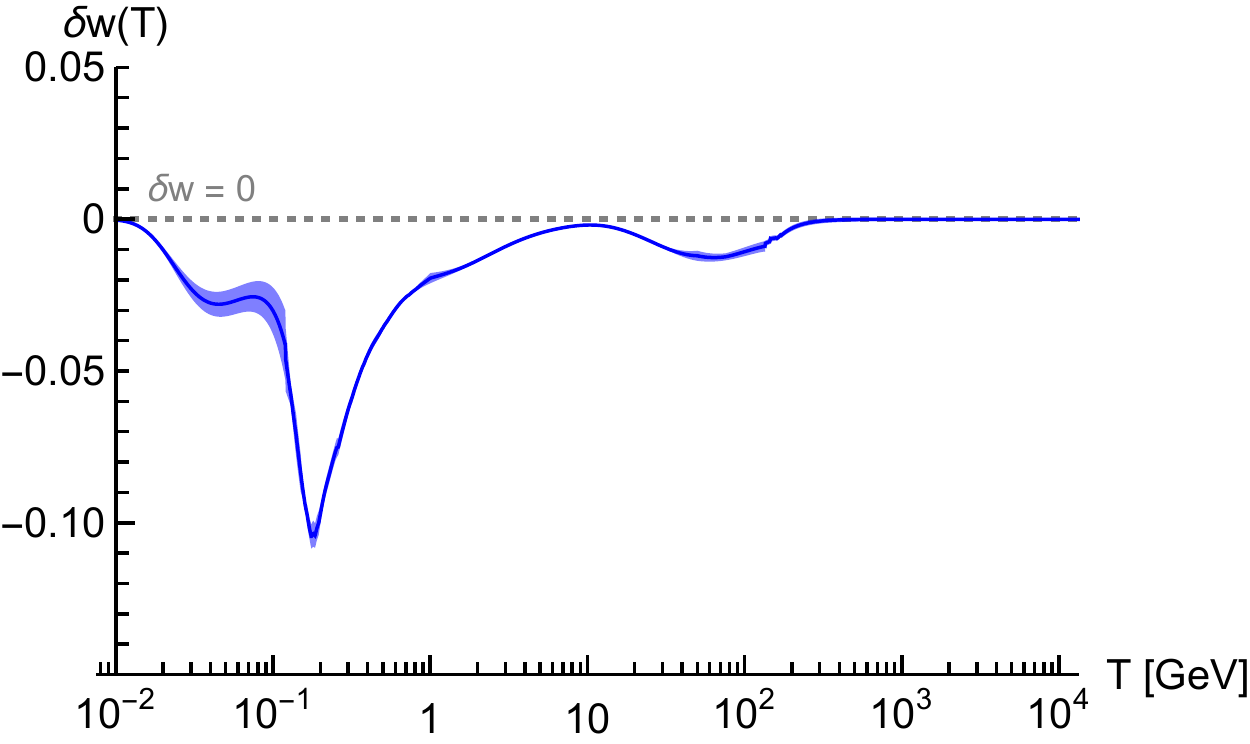}}
  \caption{Temperature dependence of the (renormalized) pressure (top left), trace anomaly (top right),
  effective degrees of freedom for the energy density (middle left), ratio between those for the energy density to those for the entropy density (middle right),
  and equation of state parameter (bottom) in the SM.
  Light blue bands show the uncertainty of the results (see text) and blue solid lines represent medians of them.
  The gray dotted line in the middle left panel shows the estimate of $g_{*\rho}(T)$ 
  based on the ideal gas approximation and assumption of the sudden QCD phase transition,
  similar to what used in Ref.~\cite{Watanabe:2006qe}.}
  \label{fig:eos_results}
\end{figure}

\begin{figure}[htbp]
\centering
$\begin{array}{cc}
\subfigure{
\includegraphics[width=80mm]{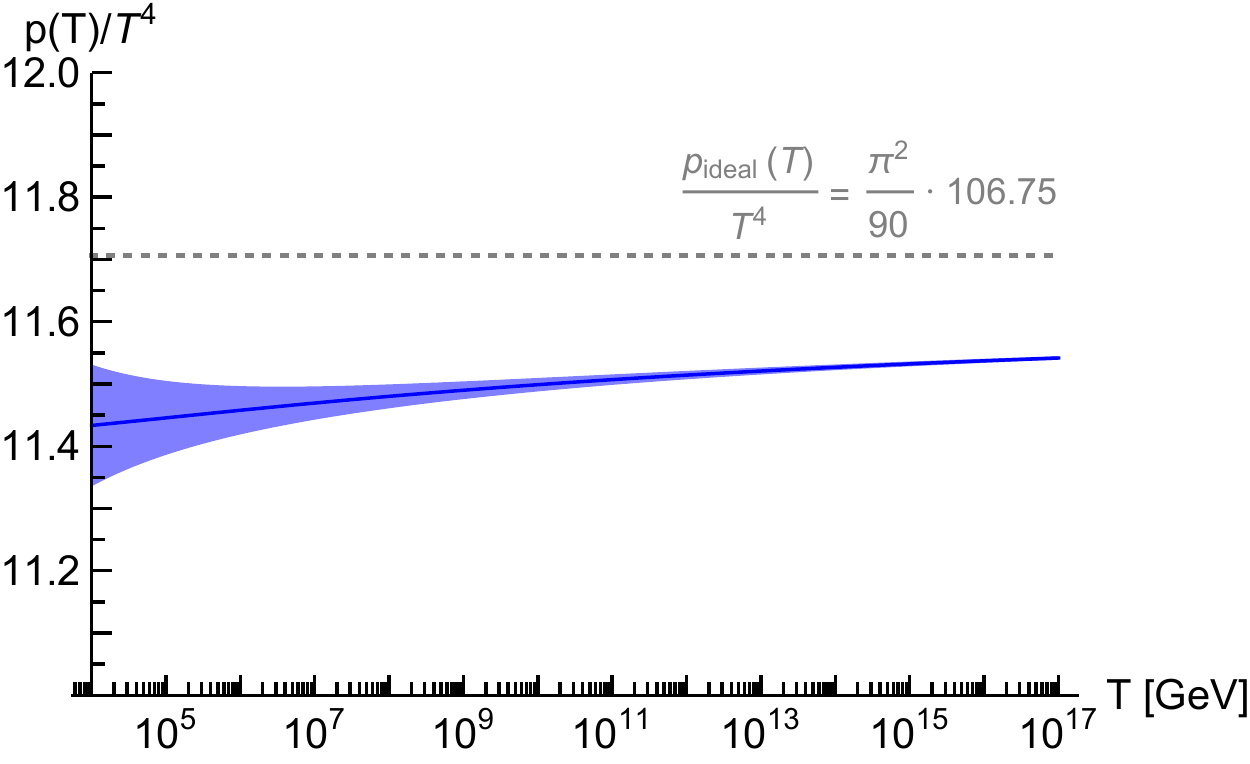}}
\hspace{5mm}
\subfigure{
\includegraphics[width=80mm]{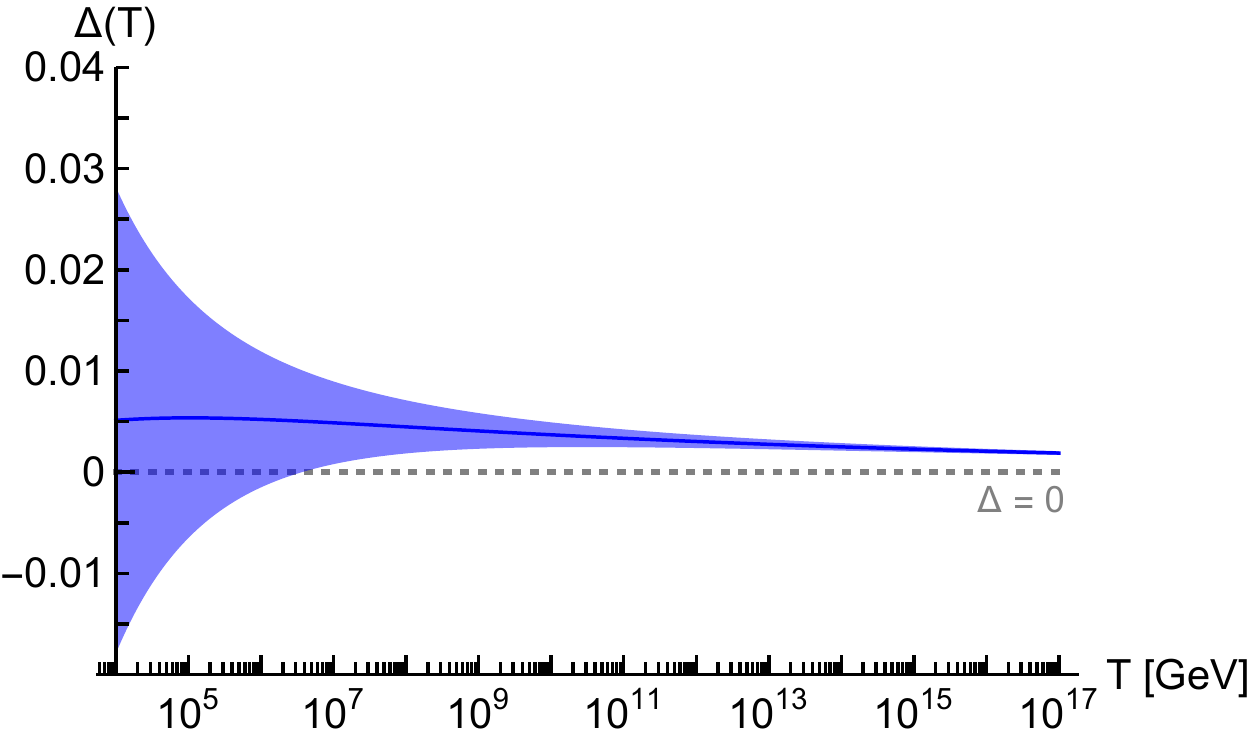}}
\vspace{10mm}
\\
\subfigure{
\includegraphics[width=80mm]{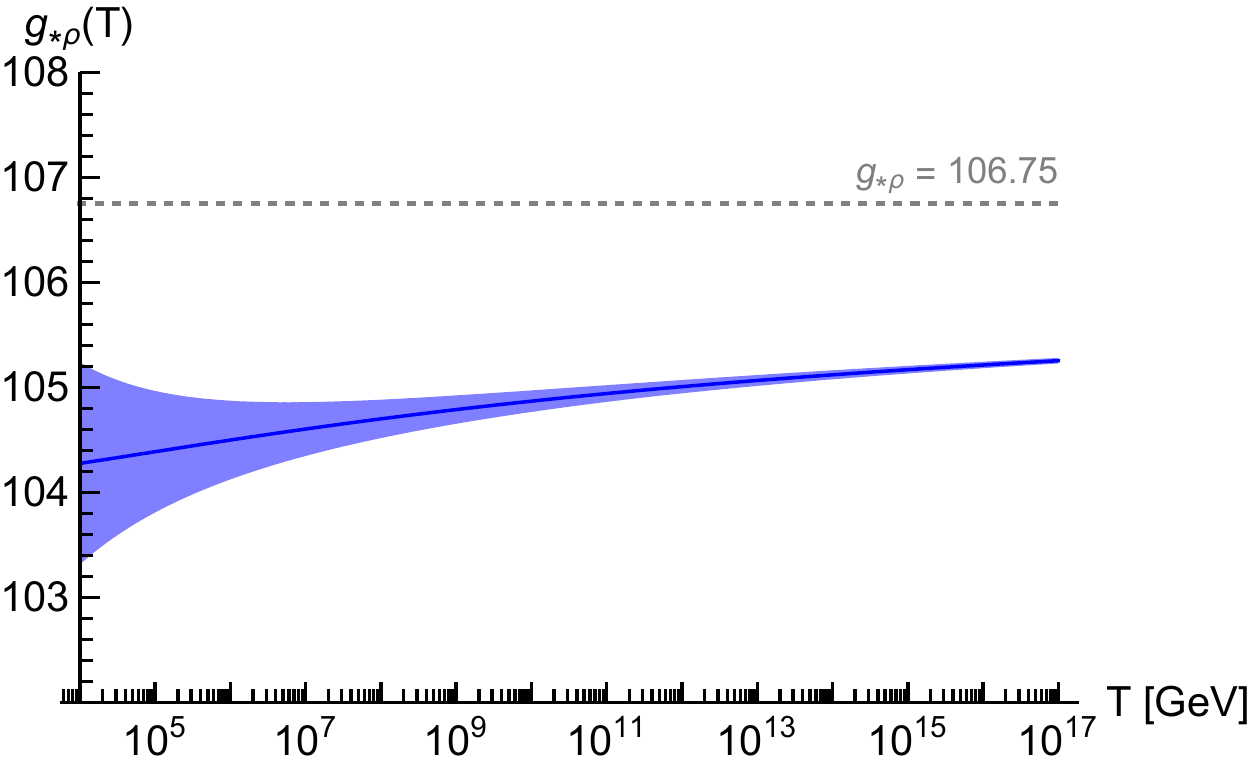}}
\hspace{5mm}
\subfigure{
\includegraphics[width=80mm]{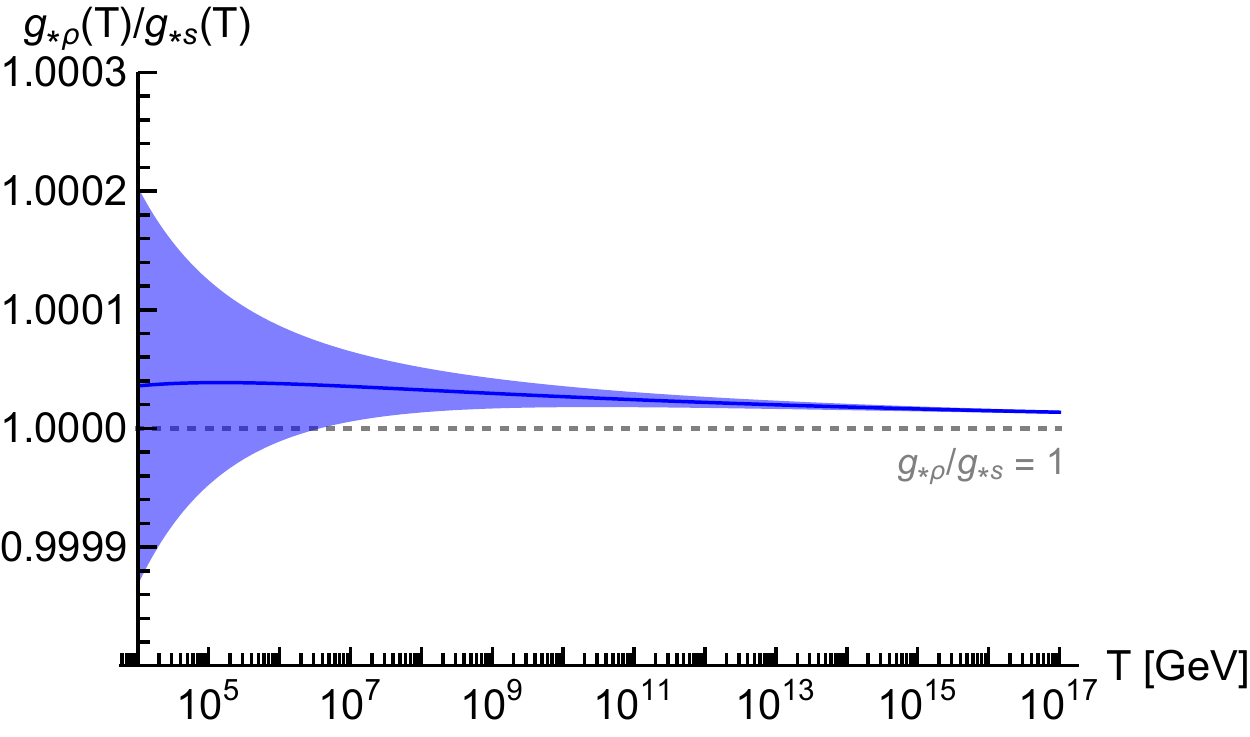}}
\vspace{10mm}
\end{array}$
\subfigure{
\includegraphics[width=80mm]{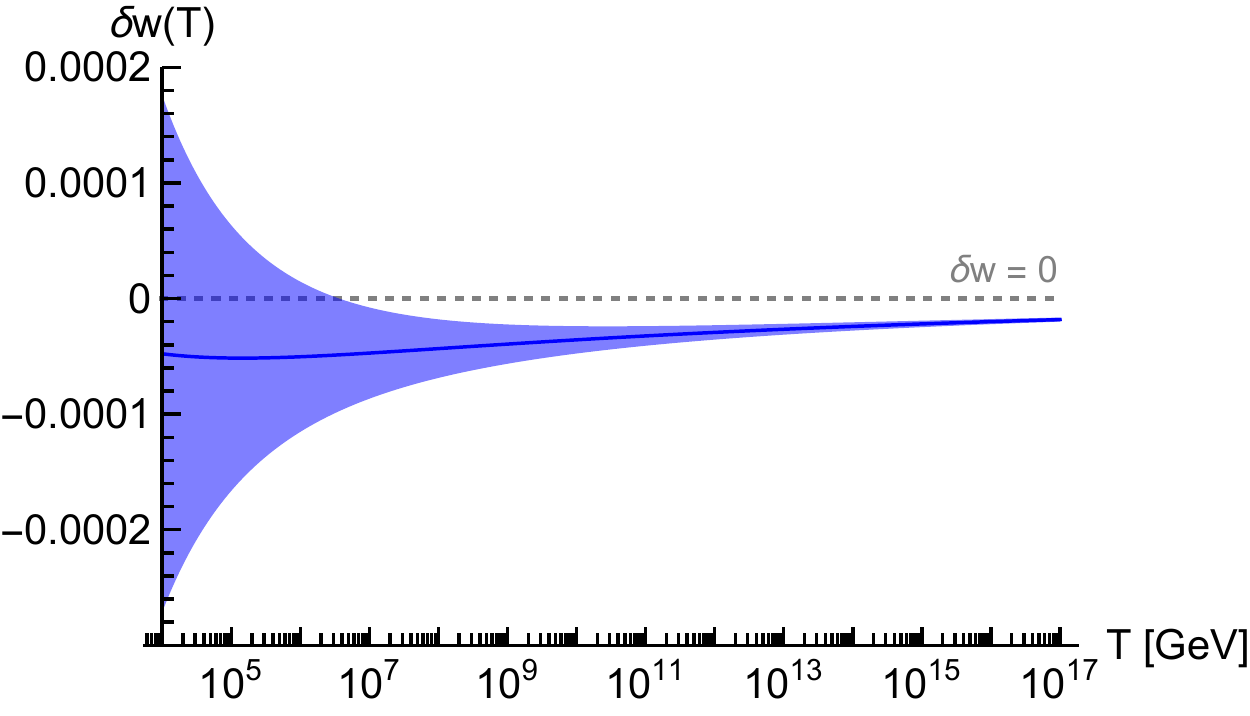}}
  \caption{The same figures as Fig.~\ref{fig:eos_results} but results are plotted 
  in a higher temperature range.}
  \label{fig:eos_results_high}
\end{figure}

Once we obtain $p(T)/T^4$ and $\Delta(T)$ at arbitrary temperatures, it is straightforward to estimate the effective degrees of freedom $g_{*\rho}(T)$
and $g_{*s}(T)$ by using Eqs.~\eqref{gstar_definition} and~\eqref{rho_and_s_from_p_and_delta}.
The results are shown in Figs.~\ref{fig:eos_results} and~\ref{fig:eos_results_high}.
We also show the deviation of the equation of state parameter $w$ from the value for pure radiation $1/3$,
\begin{equation}
\delta w(T) \equiv w(T) - \frac{1}{3} = -\frac{T^4}{3\rho(T)}\Delta(T) = -\frac{\Delta(T)}{\left[\frac{\pi^2g_{*\rho}(T)}{10}\right]}. \label{delta_w_definition}
\end{equation}

For the sake of comparison, in Fig.~\ref{fig:eos_results} we also plot the estimate of $g_{*\rho}(T)$ 
based on the ideal gas approximation in a similar way to Ref.~\cite{Watanabe:2006qe}.
In Ref.~\cite{Watanabe:2006qe}, the effective degrees of freedom were estimated by using the tree-level formulae [Eqs.~\eqref{p_tree_level} and~\eqref{delta_tree_level}],
and the contributions of hadrons whose masses are heavier than pions were simply neglected.
Furthermore, it was assumed that the values of $g_{*\rho}$ and $g_{*s}$ change suddenly at the critical temperature of the QCD phase transition, 
$T = 180\,\mathrm{MeV}$.\footnote{Although we adopt $T_c = 180\,\mathrm{MeV}$ for the transition temperature to reproduce the result of
Ref.~\cite{Watanabe:2006qe}, we note that lattice QCD simulations yielded $T_c \sim 150\,\mathrm{MeV}$ afterwards~\cite{Aoki:2009sc,Bazavov:2011nk}.}
From Fig.~\ref{fig:eos_results}, we clearly see that the new result for $g_{*\rho}(T)$ deviates significantly from the ideal gas result
even if we take account of large uncertainty in the pressure of QCD.

\begin{figure}[htbp]
\begin{center}
\includegraphics[width=140mm]{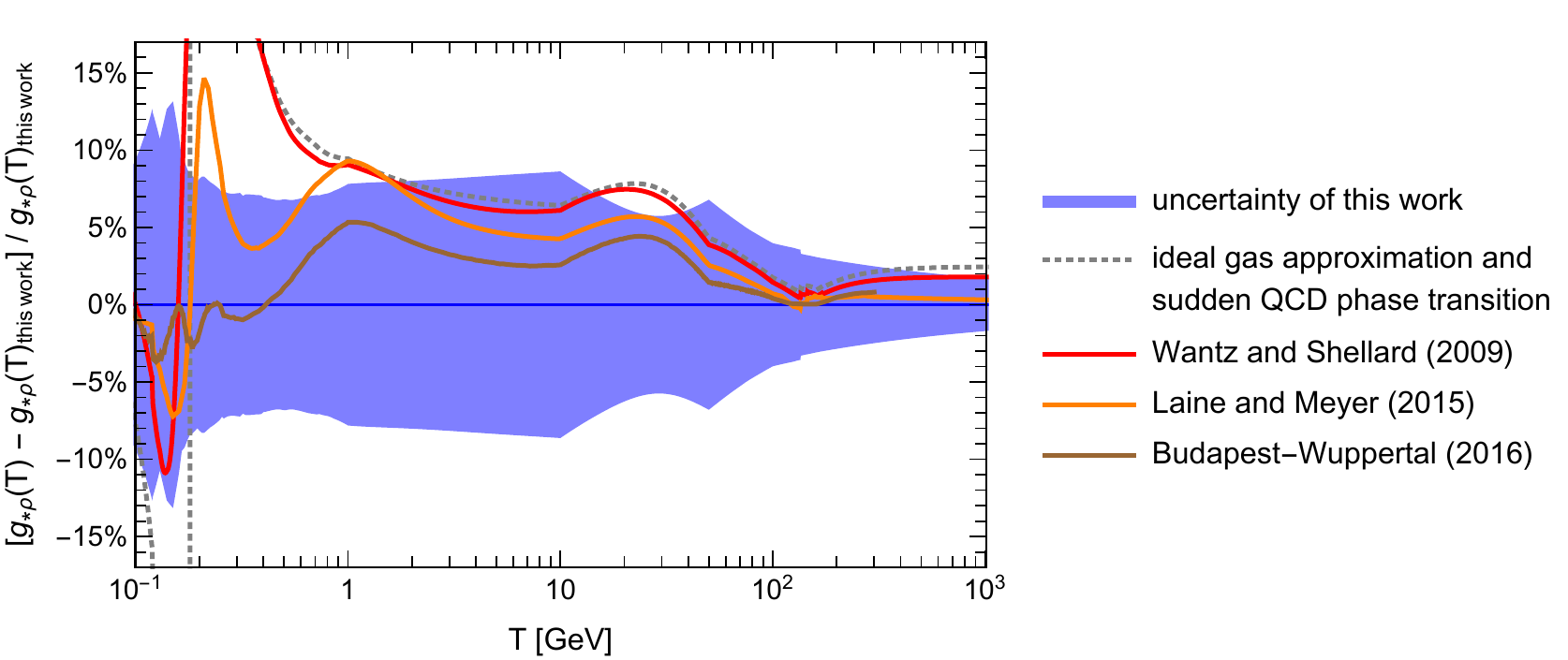}
\end{center}
\caption{The relative difference of $g_{*\rho}(T)$ between our results and other estimates.
The median of the value of $g_{*\rho}(T)$ in this work (blue solid line) is compared to the result based on the ideal gas approximation and
the sudden QCD phase transition (gray dotted line), the fitting function in Ref.~\cite{Wantz:2009it} (red solid line), 
the data set in Ref.~\cite{Laine:2015kra} (orange solid line), and that in Ref.~\cite{Borsanyi:2016ksw} (brown solid line).
The Light blue band shows the uncertainty of the results obtained in this work.}
\label{fig:eos_compare}
\end{figure}

In Fig.~\ref{fig:eos_compare}, we further compare our results to those obtained in previous studies including the fitting function
in Ref.~\cite{Wantz:2009it} and the data sets in Refs.~\cite{Laine:2015kra,Borsanyi:2016ksw}.
First, the fitting function in Ref.~\cite{Wantz:2009it} was obtained based on the ideal gas approximation except for the smoothing at around $T_c = 180\,\mathrm{GeV}$,
and hence it significantly deviates from our result at several temperature intervals. The difference reaches up to $40\,\%$ at $T \sim 200\,\mathrm{MeV}$. 
Second, our result is consistent with the data set in Ref.~\cite{Laine:2015kra} for $T \gtrsim 100\,\mathrm{GeV}$,
since we have used the same $\Delta(T)$ as Ref.~\cite{Laine:2015kra} to calculate the effective degrees of freedom at temperatures above the electroweak crossover.
On the other hand, the difference from the result of Ref.~\cite{Laine:2015kra} becomes significant at $T = \mathcal{O}(0.1\textendash 1)\,\mathrm{GeV}$,
since in Ref.~\cite{Laine:2015kra} the effective degrees of freedom at corresponding temperatures were estimated based on interpolation without using lattice QCD input~\cite{Laine:2006cp}.
Finally, our result is basically consistent with that of Ref.~\cite{Borsanyi:2016ksw}, since our analysis is based on the lattice data obtained in that paper.
However, we have adopted a more conservative estimate of theoretical uncertainty rather than the $1\,\%$ error claimed in Ref.~\cite{Borsanyi:2016ksw}.
It should be noted that the lattice data is only available for temperature up to $1\,\mathrm{GeV}$, and 
that the weak coupling expansion remains insufficient at $T=\mathcal{O}(1\textendash 10)\,\mathrm{GeV}$.
Accordingly, the values of the effective degrees of freedom at these temperatures are sensitive to the interpolation procedure,
and we have found that the corresponding uncertainty can be $6\textendash 9\,\%$.
We also note that the values of the effective degrees of freedom at temperatures much higher than the critical temperature of the electroweak crossover
can be estimated very precisely via the perturbative calculation as long as the description based on the SM remains valid at the corresponding energy scales.

Intriguingly, the values of the effective degrees of freedom do \emph{not} reach the commonly used value $g_{*\rho} = g_{*s} = 106.75$
even at temperature much higher than the critical temperature of the electroweak crossover, as shown in Fig.~\ref{fig:eos_results_high}.
This result can be straightforwardly understood in terms of perturbative corrections in the pressure.
Let us write the leading order expression for the pressure as $p(T,g)/T^4 = \hat{p}_0 + \hat{p}_2g^2 + \dots$,
where dots represent higher order terms in the weak coupling expansion.
As we see below, the non-vanishing contribution to $\Delta(T)$ appears at $\mathcal{O}(g^4)$.
Neglecting such higher order contributions, from Eqs.~\eqref{gstar_definition} and~\eqref{rho_and_s_from_p_and_delta}
we can estimate the leading order correction to the effective degrees of freedom in terms of that to the pressure,
\begin{equation}
\delta g_{*i} = g_{*i} - g_{*i,\mathrm{tree}} = \frac{90}{\pi^2}\hat{p}_2 g^2 + \dots \quad \text{for} \quad i = \rho, s, \label{delta_gstar_leading}
\end{equation}
where $g_{*i,\mathrm{tree}}$ represents the tree-level result (i.e. $g_{*i,\mathrm{tree}} = 106.75$ in the SM).
In QCD, the leading order correction is given by $\hat{p}_2 = (2/45)p_2 = -7/12$ for $N_f = 6$ [see Eq.~\eqref{p_QCD_order_g2}].
Substituting this value together with $g_s^2 \approx 0.29$ at $\mu/\pi = 10^{15}\,\mathrm{GeV}$, we obtain $\delta g_{*i} \approx -1.5$.
Therefore, it is reasonable to expect an $\mathcal{O}(1)$ deviation from the tree-level result in the effective degrees of freedom.
More precise values can be obtained by adding other contributions arising from electroweak and Yukawa interactions as well as higher order corrections,
and the final results are presented in Fig.~\ref{fig:eos_results_high}.

The contribution to the trace anomaly can be understood as follows.
If we assume that there is no dimensionful parameter in theory,\footnote{Note that this assumption is not valid in the SM as there are 
the mass parameter $\nu$ and Higgs expectation value $\langle\phi^{\dagger}\phi\rangle$. Here we ignore these contributions for simplicity.}
we expect that the quantity $p(T)/T^4$ depends on $T$ only through the combination $\mu/T$ because of dimensional reasons.
Since $p(T)/T^4$ should be independent of the renormalization scale $\mu$ at a given order in the weak coupling expansion,
the contribution to the trace anomaly can be extracted from the renormalization group running of the dimensionless coupling parameter $g$~\cite{Drummond:1999he},
\begin{equation}
\Delta(T,g) = \beta(g)\frac{\partial}{\partial g}\left\{\frac{p(T,g)}{T^4}\right\}, \label{delta_from_beta_function}
\end{equation}
where $\beta(g) = \frac{dg}{d\ln\mu}$ denotes the beta function.
If the leading order expressions for the pressure and beta function are given by
$p(T,g)/T^4 = \hat{p}_0 + \hat{p}_2g^2 + \dots$ and $\beta(g) = bg^3 + \dots$, 
from Eq.~\eqref{delta_from_beta_function} we obtain
\begin{equation}
\Delta = 2b\hat{p}_2g^4 + \dots.
\end{equation}
Substituting the value $b = -7/16\pi^2$ in QCD (with $N_f = 6$) together with $\hat{p}_2 = -7/12$ and $g_s^2 \approx 0.29$, we have $\Delta \approx 0.004$.
In practice, there are other contributions arising from electroweak and Yukawa interactions [see Eq.~\eqref{delta_1}] as well as the terms of higher order in $g_s$.
Adding these contributions altogether, we find a slightly smaller value as shown in Fig.~\ref{fig:eos_results_high}.

It is also possible to give an order of magnitude estimate of other quantities.
The ratio between $g_{*\rho}$ to $g_{*s}$ can be written as
\begin{equation}
\frac{g_{*\rho}}{g_{*s}} - 1 = \frac{1-3w}{3(1+w)} \approx -\frac{3}{4}\delta w + \mathcal{O}(\delta w^2), \label{gsr_to_gss_ratio}
\end{equation}
for small $\delta w$. From Eq.~\eqref{delta_w_definition}, we see that $\delta w$ is proportional to $\Delta$, which is a quantity of $\mathcal{O}(g^4)$.
Therefore, we expect that the value of $g_{*\rho}$ can be different from that of $g_{*s}$ at $\mathcal{O}(g^4)$, 
even though they are the same at the leading order as shown in Eq.~\eqref{delta_gstar_leading}.
In the SM, Eqs.~\eqref{delta_w_definition} and~\eqref{gsr_to_gss_ratio} imply that $g_{*\rho}/g_{*s}-1$ and $\delta w$ are positive and negative, respectively,
and that their magnitude is $\mathcal{O}(10^{-5})$ at high temperature.

Although our main focus in this work is the particle interactions in the SM, we can straightforwardly extend above arguments
to arbitrary gauge interactions, which could be relevant to physics beyond the SM.
For example, in SU($N_c$) gauge theory with $N_f$ flavors, we have $\hat{p}_2 = -(N_c^2-1)(N_c+5N_f/4)/144$ 
and $b = -(11N_c/12 - N_f/6)/4\pi^2$~\cite{Kajantie:2002wa}, which lead to
\begin{equation}
\Delta = \frac{g^4}{72(4\pi)^2}\left(N_c^2-1\right)\left(N_c + \frac{5}{4}N_f\right)\left(\frac{11}{3}N_c - \frac{2}{3}N_f\right) + \mathcal{O}(g^5).
\end{equation}
Furthermore, since the tree-level contribution to the effective degrees of freedom is given by 
$g_{*\rho} \approx 2(N_c^2-1)+7N_cN_f/2$, from Eq.~\eqref{delta_w_definition} 
we obtain the following formula for the equation of state~\cite{Davoudiasl:2004gf}\footnote{The right-hand side of Eq.~\eqref{delta_w_general_gauge_theory}
is smaller than the expression shown in Eq.~(4) of Ref.~\cite{Davoudiasl:2004gf} by a factor 2. We believe that the equation contains a typo.} 
\begin{equation}
\delta w = -\frac{5}{36\pi^2}\frac{g^4}{(4\pi)^2}\frac{\left(N_c + \frac{5}{4}N_f\right)\left(\frac{11}{3}N_c - \frac{2}{3}N_f\right)}{2 + \frac{7}{2}\frac{N_cN_f}{N_c^2 - 1}} + \mathcal{O}(g^5). \label{delta_w_general_gauge_theory}
\end{equation}
Hence, for such theory we generically expect that there exists a non-trivial deviation of the equation of state parameter from the value for pure radiation.

The effective degrees of freedom for a wide temperature interval obtained in this section are used as tabulated data 
in numerical calculation of GWs performed in the next section.
Because of brute-force interpolation methods used here, our data contain some discontinuous points, 
which may cause numerical artifacts in the actual computation.
In order to avoid them, we also create fitting functions for $g_{*\rho}(T)$ and $g_{*s}(T)$.
Details are described in Appendix~\ref{app:fitting_functions}.

%%%%%%%%%%%%%%%%%%%%%%%%%%%%%%%%%%%%%%%%%%%%%%%%%%
\section{Spectrum of the gravitational waves}
\label{sec:spectrum_of_GW}
\setcounter{equation}{0}
%%%%%%%%%%%%%%%%%%%%%%%%%%%%%%%%%%%%%%%%%%%%%%%%%%

Using the state-of-the-art results of the equation of state in the SM, we now compute the spectrum of the primordial GWs
based on the method used in Refs.~\cite{Weinberg:2003ur,Watanabe:2006qe}.
Here we numerically solve the following equation obtained from
Eqs.~\eqref{GW_waveequation_kspace},~\eqref{h_prim_T},~\eqref{neutrino_anisotropic_stress}, and~\eqref{photon_anisotropic_stress}:
\begin{align}
& \frac{d^2\chi(u)}{du^2} + \frac{2}{a(u)}\frac{da(u)}{du}\frac{d\chi(u)}{du} + \chi(u) \nonumber \\
& = -24 f_{\gamma}(u)\left[\frac{1}{a(u)}\frac{da(u)}{du}\right]^2\int_{u_{\rm ls}}^udU\left[\frac{j_2(u-U)}{(u-U)^2}\right]\frac{d\chi(U)}{dU} \nonumber\\
& \quad -24 f_{\nu}(u)\left[\frac{1}{a(u)}\frac{da(u)}{du}\right]^2\int_{u_{\nu {\rm dec}}}^udU\left[\frac{j_2(u-U)}{(u-U)^2}\right]\frac{d\chi(U)}{dU},
\label{differential_equation_for_chi}
\end{align}
where $u=k\tau$,
\begin{equation}
f_{\gamma}(u) \equiv \frac{\rho_{\gamma}(u)}{\rho_{\rm crit}(u)}, \label{f_gamma_definition}
\end{equation}
and
\begin{equation}
f_{\nu}(u) \equiv \frac{\rho_{\nu}(u)}{\rho_{\rm crit}(u)}. \label{f_nu_definition}
\end{equation}
The values of $u_{\rm ls}$ and $u_{\nu {\rm dec}}$ in the right-hand side of Eq.~\eqref{differential_equation_for_chi} are fixed such that they correspond
to the temperature at the photon last scattering $T_{\rm ls}$ and that at the neutrino decoupling $T_{\nu {\rm dec}}$, respectively.
In the following analysis, we take $T_{\rm ls} = 3000\,\mathrm{K}$ and $T_{\nu \mathrm{dec}} = 2\,\mathrm{MeV}$.
The initial conditions are specified as
\begin{equation}
\chi(0) = 1, \quad \frac{d\chi}{du}(0) = 0.
\end{equation}
The evolution of the scale factor $a(u)$ is also analyzed by solving the Friedmann equation,
\begin{equation}
\frac{1}{a^2(u)}\frac{da(u)}{du} = \frac{H_0}{k}\sqrt{\frac{\rho_{\rm crit}(u)}{\rho_{\rm crit,0}}} \label{Friedmann_eq}
\end{equation}
with initial conditions
\begin{equation}
a(0) = 0, \quad \frac{da}{du}(0) = \frac{H_0a_0^2}{k}\sqrt{\frac{g_{*\rho,\mathrm{ini}}}{2}\left(\frac{g_{*s,\mathrm{fin}}}{g_{*s,\mathrm{ini}}}\right)^{\frac{4}{3}}\Omega_{\gamma}}, \label{initial_conditions_a}
\end{equation}
where we have used the values of $g_{*\rho}(T)$ and $g_{*s}(T)$ at $T = 10^{17}\,\mathrm{GeV}$ to evaluate 
$g_{*\rho,\mathrm{ini}}$ and $g_{*s,\mathrm{ini}}$, respectively, in Eq.~\eqref{initial_conditions_a}.
The right-hand sides of Eqs.~\eqref{f_gamma_definition},~\eqref{f_nu_definition}, and~\eqref{Friedmann_eq} are evaluated by using $g_{*\rho}(T)$ and $g_{*s}(T)$ obtained in the previous section.
The wave equation~\eqref{differential_equation_for_chi} is solved up to some finite time $u = u_{\rm end}$, and after that time
we extrapolate the solution until the present time by using the WKB solution,
\begin{equation}
\chi(u) = \frac{A}{a(u)}\sin(u+\delta),
\end{equation}
where $A$ and $\delta$ are fixed such that $\chi$ and $d\chi/du$ obtained from the numerical analysis are
matched to those of the WKB solution at $u = u_{\rm end}$.

After obtaining the value of $d\chi/du$ at the present time, we substitute it to Eq.~\eqref{Omega_gw_PT_T} to estimate $\Omega_{\rm gw}h^2$.
At this stage we need to specify the primordial tensor power spectrum $\mathcal{P}_T(k)$.
Regarding the fact that the shape of $\mathcal{P}_T(k)$ strongly depends on the underlying inflationary model,
here we artificially take it as a $k$-independent constant parameterized by 
the inflationary energy scale $V_{\rm inf}^{1/4}$ [i.e. $\mathcal{P}_T(k) = 2V_{\rm inf}/3\pi^2 M_{\rm Pl}^4$],
and focus only on the effects caused by the transfer function $d\chi/du$.
Furthermore, it is implicitly assumed that the reheating temperature is sufficiently high such that we can ignore a feature caused by
the reheating process in the frequency range considered in this section.
It should be straightforward to extend our analysis to a particular inflationary model, in which $\mathcal{P}_T(k)$ has a (small) deviation from scale invariance
and the reheating process gives rise to an additional feature in the spectrum of GWs.

When we solve Eq.~\eqref{differential_equation_for_chi} in the presence of the integral in the right-hand side, 
we fix the time interval $\Delta u$ for each time integration step as $10^{-3}$ and the final time $u_{\rm end}$ as $100$. 
For these choices there remains $\lesssim 2\,\%$ error in the numerical integration.
For the modes with frequencies $f \gtrsim 10^{-9}\,\mathrm{Hz}$, the right-hand side of Eq.~\eqref{differential_equation_for_chi} becomes irrelevant
and we can take $\Delta u = 10^{-5}$ and $u_{\rm end} = 1000$, which enable us to estimate $\Omega_{\rm gw}h^2$ with an accuracy of $\lesssim 0.04\,\%$.

\begin{figure}[htbp]
\begin{center}
\includegraphics[width=120mm]{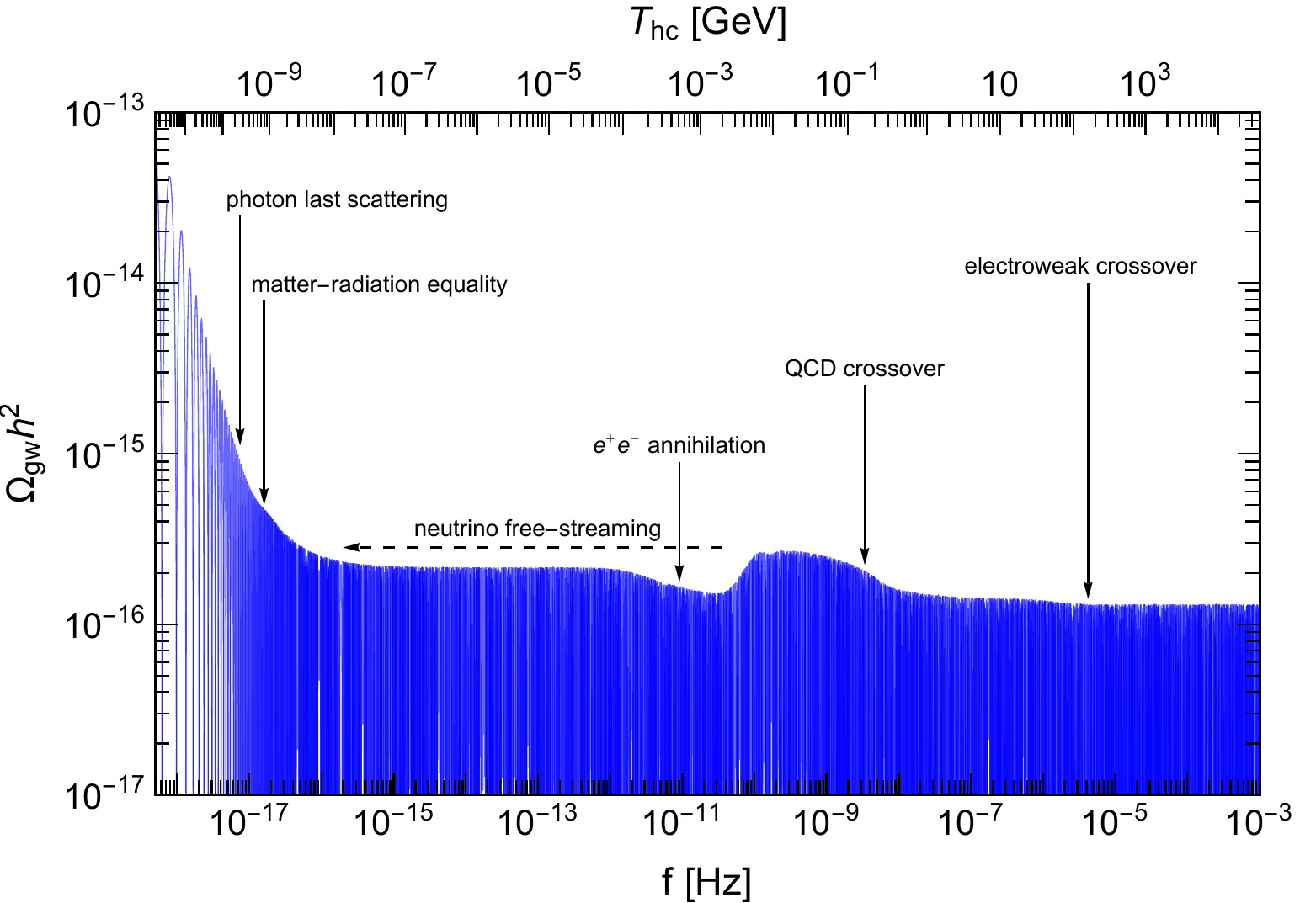}
\end{center}
\caption{The spectrum of inflationary GWs for a broad frequency interval. Here we fix the inflationary scale as $V_{\rm inf}^{1/4} = 1.5\times10^{16}\,\mathrm{GeV}$.
We also show the temperature $T_{\rm hc}$ at which the corresponding mode reenters the horizon.}
\label{fig:gw_spectrum_broad}
\end{figure}

Figure~\ref{fig:gw_spectrum_broad} shows the result of the numerical integration for a broad frequency interval.
We see that various events occurring in the early universe are imprinted on the spectrum of GWs, as mentioned in Sec.~\ref{sec:IGW}.
The amplitude of GWs oscillates with a phase $2k\tau_0$, where $\tau_0$ is the present conformal time.
We note that this oscillation is a genuine feature of inflationary GWs.
All the modes with a fixed wavenumber $k$ enter the horizon at the same time and start to oscillate simultaneously,
and hence they are coherent in the temporal phase.
However, direct detection experiments cannot resolve this oscillation since $k\tau_0 \gg 1$.
For this reason, we replace a rapidly oscillating factor by $1/2$ as we have done in Eq.~\eqref{transfer_function_approximate}.
Hereafter, we use $\Omega_{\rm gw}$ to refer the averaged quantity.

\begin{figure}[htbp]
\centering
$\begin{array}{cc}
\subfigure{
\includegraphics[width=80mm]{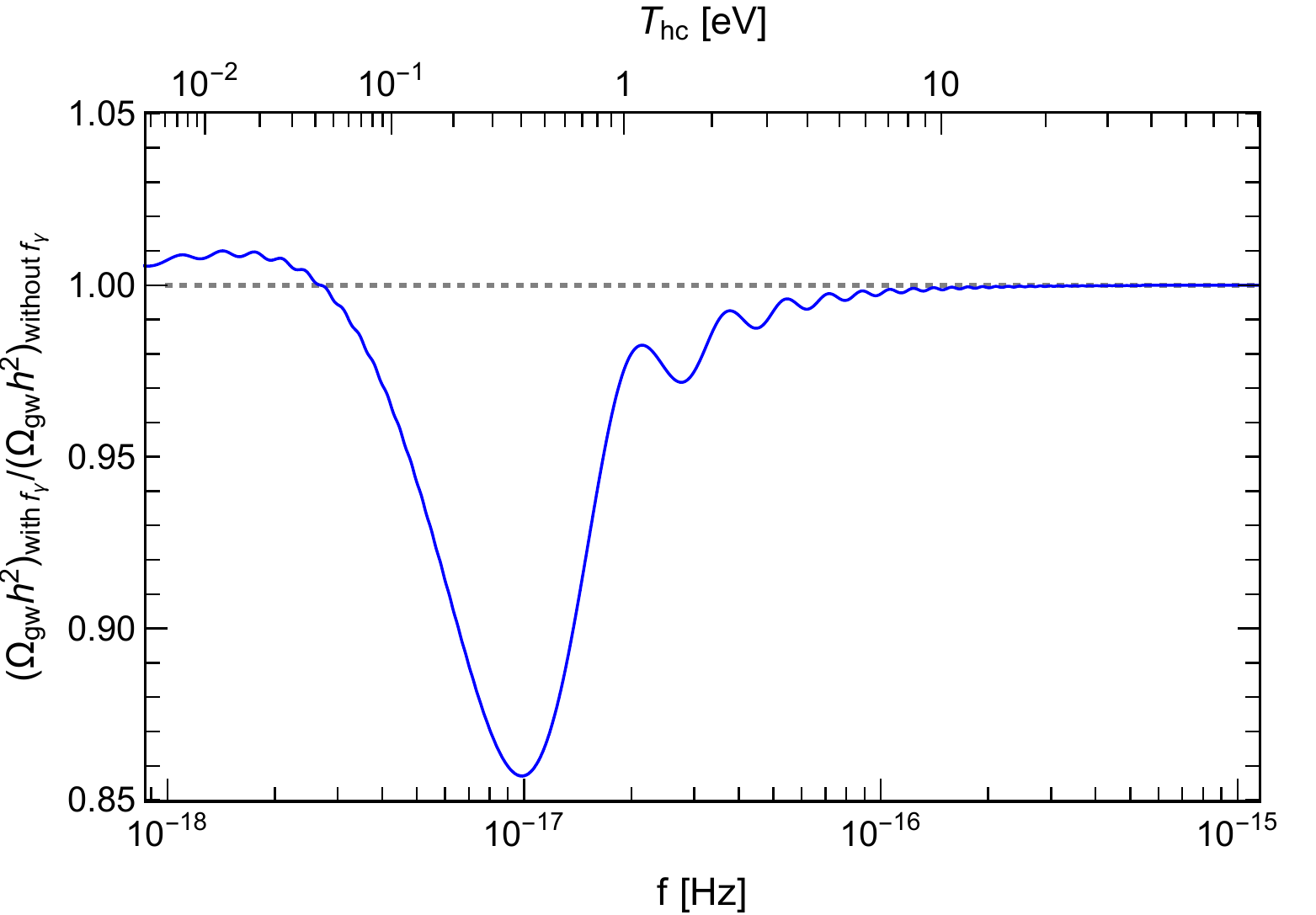}}
\hspace{5mm}
\subfigure{
\includegraphics[width=80mm]{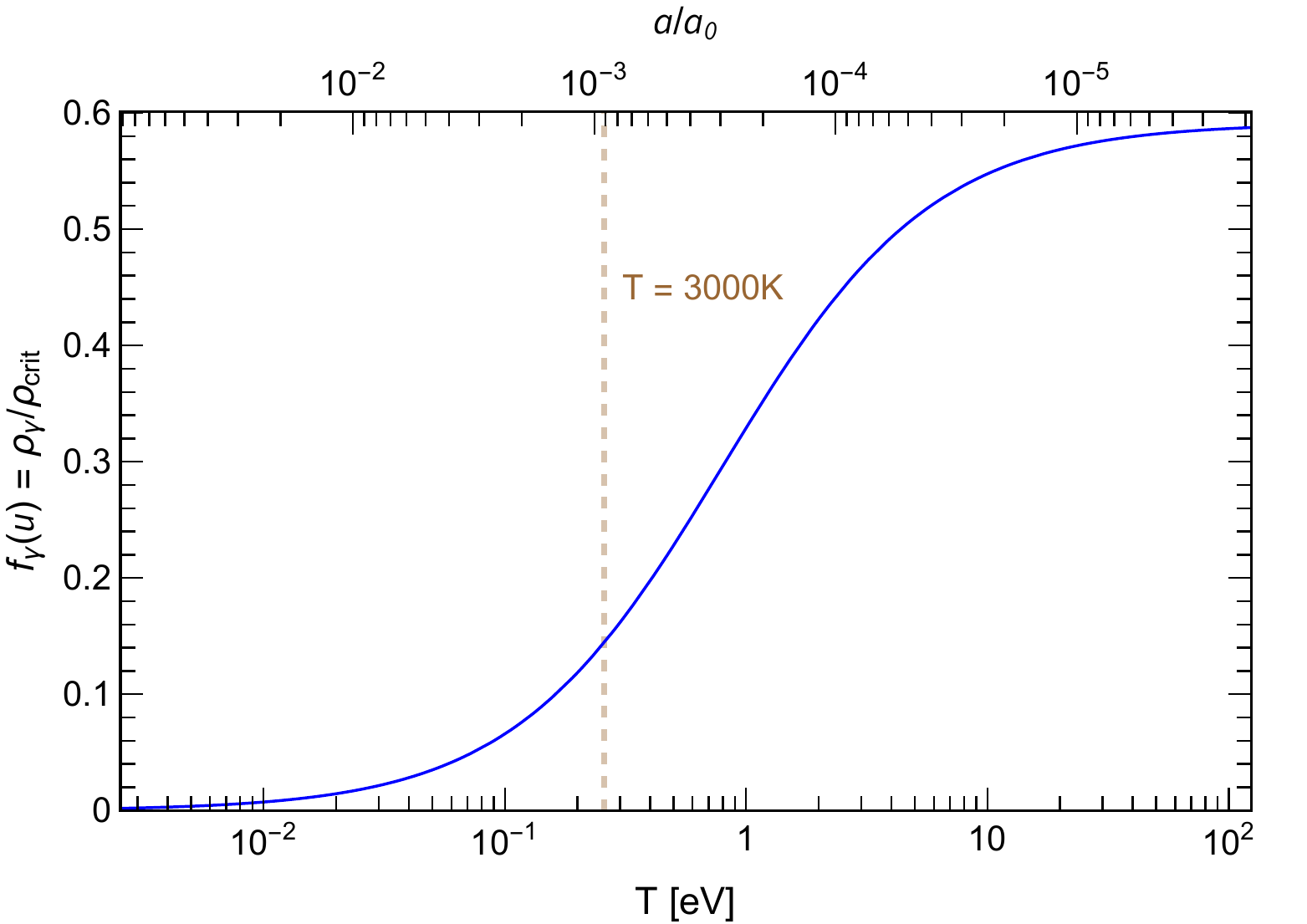}}
\vspace{10mm}
\end{array}$
  \caption{Effect of free-streaming photons on the spectrum of GWs.
  In the left panel, the ratio of $\Omega_{\rm gw}h^2$ obtained by including the anisotropic stress induced by photons~\eqref{photon_anisotropic_stress}
  to that obtained without including Eq.~\eqref{photon_anisotropic_stress} is plotted.
  In the right panel, Eq.~\eqref{f_gamma} is plotted as a function of $T$ or the corresponding value of the scale factor $a$.
  The brown dashed line represents the typical temperature of the photon last scattering, $T = 3000\,\mathrm{K}$.}
  \label{fig:gw_photon_damping}
\end{figure}

As mentioned in Sec.~\ref{sec:IGW_damping}, we have included the contribution of free-streaming photons 
to the anisotropic stress in the second line of Eq.~\eqref{differential_equation_for_chi}.
The efficiency of the damping effect caused by this term is characterized by the coefficient $f_{\gamma}(u)$, which reads
\begin{equation}
f_{\gamma}(u) = \frac{\left(\frac{g_{*s,\mathrm{fin}}}{g_{*s}(T)}\right)^{\frac{4}{3}}\Omega_{\gamma}h^2}{\Omega_Mh^2\left(\frac{a(u)}{a_0}\right)+\frac{g_{*\rho}(T)}{2}\left(\frac{g_{*s,\mathrm{fin}}}{g_{*s}(T)}\right)^{\frac{4}{3}}\Omega_{\gamma}h^2}, \label{f_gamma}
\end{equation}
where $\Omega_Mh^2 \simeq 0.14$~\cite{Ade:2015xua} is the matter density parameter.
In Fig.~\ref{fig:gw_photon_damping}, we show the impact of this term on the spectrum of the primordial GWs.
We see that $f_{\gamma}(u)$ becomes less than 0.145 at $T = 3000\,\mathrm{K}$, 
but this value is not small enough to ignore the contribution to the anisotropic stress completely.
Indeed, we find that the modes reenter the horizon at the epoch of the photon last scattering are subjected to additional damping effects.
The effect is most pronounced at $f \approx 9.88 \times 10^{-18}\,\mathrm{Hz}$ for $T_{\rm ls} = 3000\,\mathrm{K}$,
and the amplitude of GWs is suppressed by about $14\,\%$ at that frequency.

Note that there are some wiggly features at both ends of the dip in the left panel of Fig.~\ref{fig:gw_photon_damping}.
The oscillatory feature appearing at higher frequencies ($f\gtrsim 10^{-17}\,\mathrm{Hz}$) 
is similar to that observed in Refs.~\cite{Watanabe:2006qe,Kuroyanagi:2008ye}
in the context of the damping effect due to free-streaming neutrinos.
As mentioned in Refs.~\cite{Watanabe:2006qe,Kuroyanagi:2008ye}, this feature is an artifact caused by the fact that the anisotropic stress term suddenly appears in
the right-hand side of Eq.~\eqref{differential_equation_for_chi} at $T = T_{\rm ls}$.
We expect that this oscillation is diminished 
when we directly solve the Boltzmann equation for photons together with Eq.~\eqref{GW_waveequation_kspace}
rather than using Eq.~\eqref{differential_equation_for_chi}.
On the other hand, the wiggly feature appearing at lower frequencies ($f\sim 10^{-18}\,\mathrm{Hz}$)
is caused due to the fact that the oscillation phase $2k\tau_0$ of GWs is not rapid enough to ignore the modulation of GWs by taking an average of
the oscillating factor and that the existence of the anisotropic stress due to free-streaming photons shifts the phase of the oscillation.
We also see that the ratio shown in the left panel of Fig.~\ref{fig:gw_photon_damping} becomes larger than 1 at lower frequencies.
This enhancement occurs for a mode that reenters the horizon when the coefficient $f_{\gamma}(u)$ is decreasing.
For such a mode, the second line of Eq.~\eqref{differential_equation_for_chi} decreases when $\chi(u)$ starts to drop from $\chi(0) = 1$, and this mutual reduction
leads to an overshooting of $\chi(u)$. As a result, the amplitude of the corresponding mode becomes slightly larger than that obtained without including
the second line of Eq.~\eqref{differential_equation_for_chi}.

In addition to the contribution of photons to the anisotropic stress, there is also the contribution of neutrinos given by the third line of Eq.~\eqref{differential_equation_for_chi}.
The quantity $f_{\nu}(u)$ defined in Eq.~\eqref{f_nu_definition} represents the significance of the damping effect caused by free-streaming neutrinos.
Regarding the fact that the energy density of photons and neutrinos can be estimated 
by using the first two terms in the right-hand side of Eq.~\eqref{gsr_nu_decoupling_corrected},
we can write this factor as
\begin{equation}
f_{\nu}(u) = \frac{\frac{g_{*\nu}(T)}{2}\left(\frac{g_{*s,\mathrm{fin}}}{g_{*s}(T)}\right)^{\frac{4}{3}}\Omega_{\gamma}h^2}{\Omega_Mh^2\left(\frac{a(u)}{a_0}\right)+\frac{g_{*\rho}(T)}{2}\left(\frac{g_{*s,\mathrm{fin}}}{g_{*s}(T)}\right)^{\frac{4}{3}}\Omega_{\gamma}h^2}, \label{f_nu_new}
\end{equation}
where
\begin{equation}
g_{*\nu}(T) = 2(a_{r1}-1) + \frac{21}{4}\left(\frac{4}{11}\right)^{\frac{4}{3}}a_{r2}\mathcal{S}^{\frac{4}{3}}\left(\frac{m_e}{T}\right).
\end{equation}
We note that the function $f_{\nu}(u)$ used here is different from that used in Refs.~\cite{Weinberg:2003ur,Watanabe:2006qe,Kuroyanagi:2008ye},
\begin{equation}
f_{\nu}(u)_{\rm old} = \frac{0.40523}{1+\frac{a(u)}{a_{\rm eq}}}, \label{f_nu_old}
\end{equation}
where $a_{\rm eq}$ is the scale factor at the time of matter-radiation equality.
The function~\eqref{f_nu_old} underestimates the value of $f_{\nu}(u)$ at $T\gtrsim m_e$ by a factor $\approx 0.83$
since it does not include the contribution of electrons and positrons to the energy density of background radiations
and the variation of the effective neutrino temperature $T_{\nu}$ relative to the photon temperature $T$ at the epoch of the neutrino decoupling.

\begin{figure}[htbp]
\centering
$\begin{array}{cc}
\subfigure{
\includegraphics[width=80mm]{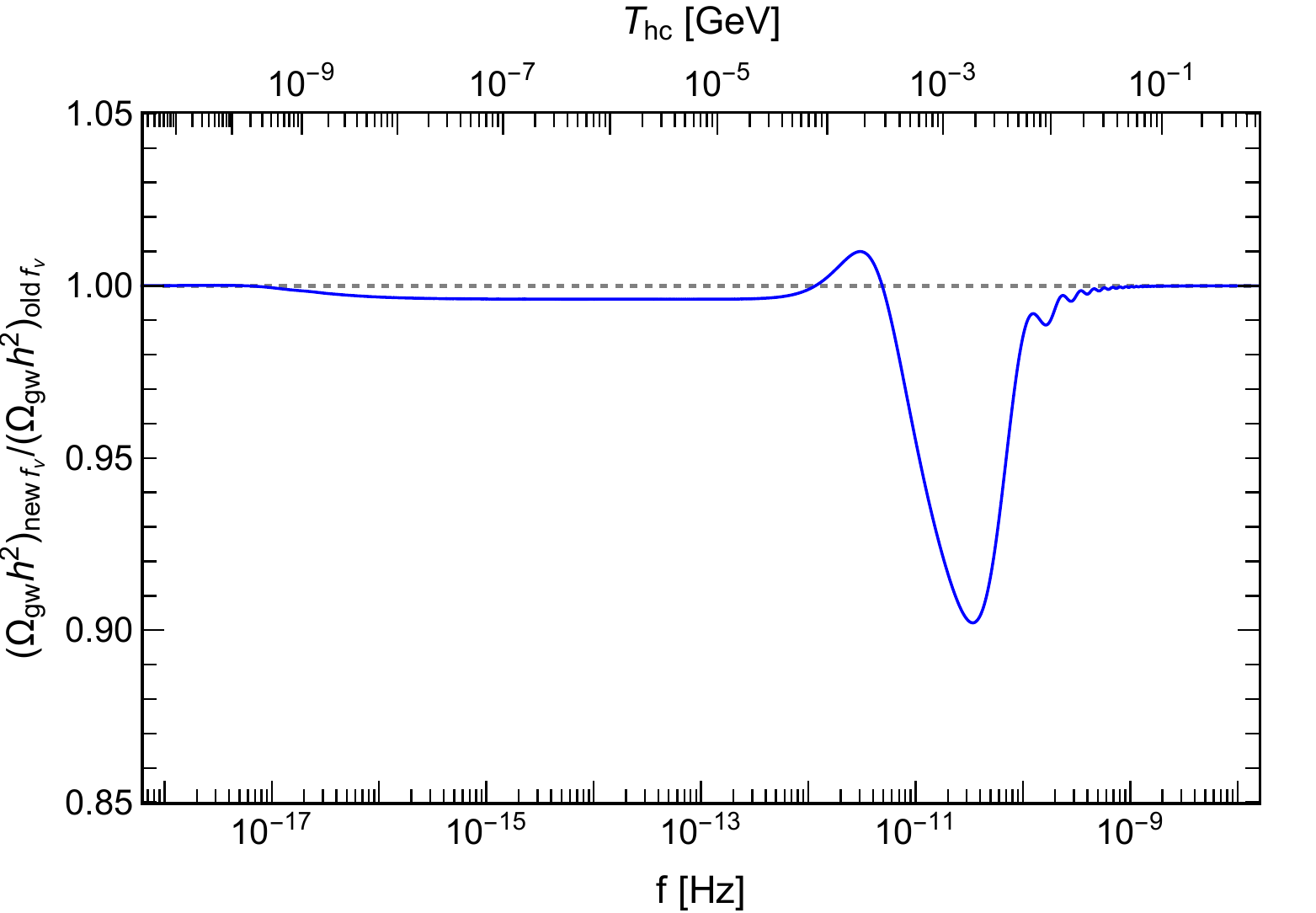}}
\hspace{5mm}
\subfigure{
\includegraphics[width=80mm]{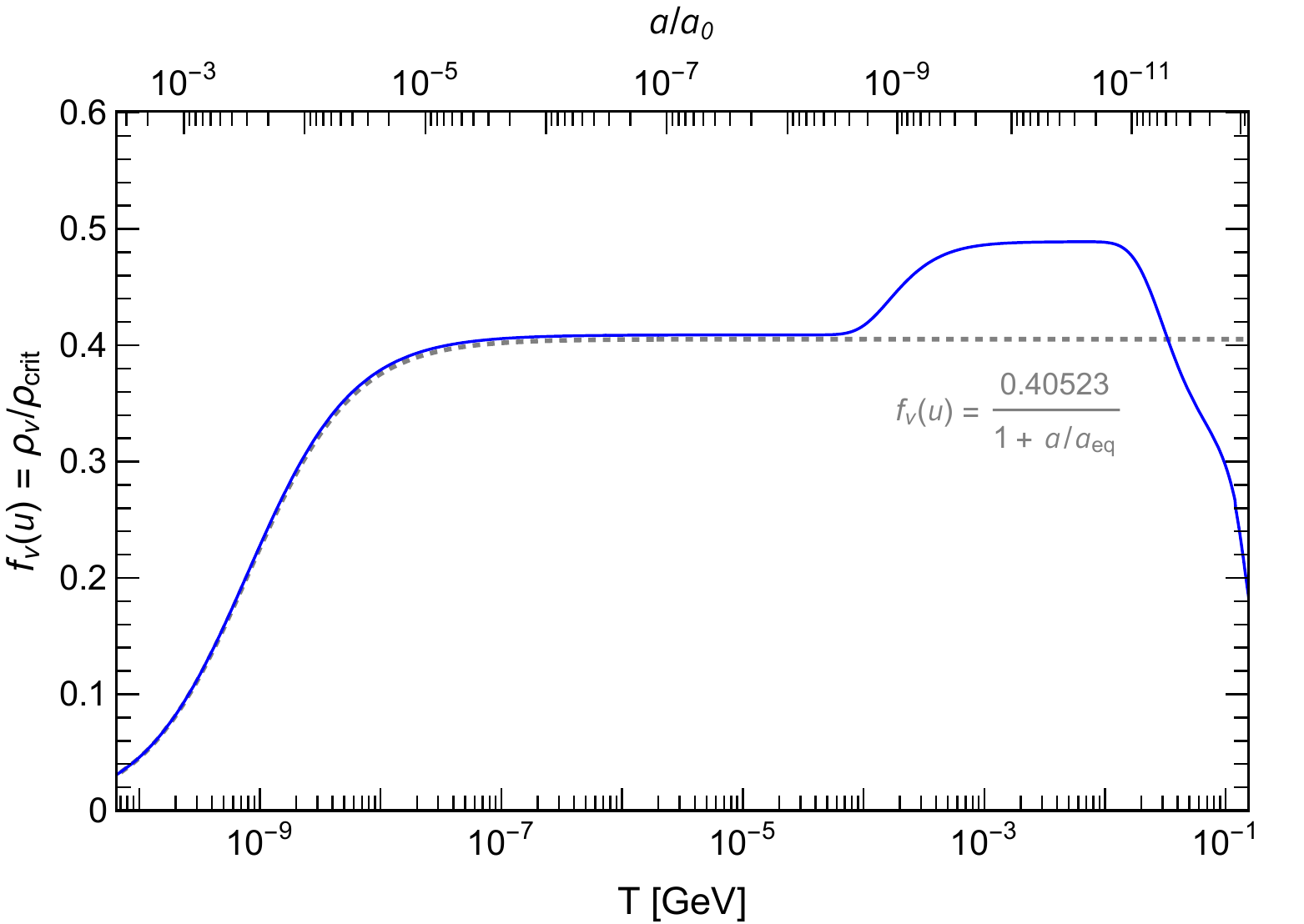}}
\vspace{10mm}
\end{array}$
  \caption{
  Comparison between new and old methods to estimate $f_{\nu}(u)$. 
  In the left panel, the ratio of $\Omega_{\rm gw}h^2$ obtained by using Eq.~\eqref{f_nu_new} to that obtained by using Eq.~\eqref{f_nu_old} is plotted.   
  In the right panel, Eq.~\eqref{f_nu_new} (blue solid line) and Eq.~\eqref{f_nu_old} (gray dotted line) are plotted as functions of $T$ or the corresponding value of the scale factor $a$.}
  \label{fig:f_nu_comparison}
\end{figure}

As shown in Fig.~\ref{fig:f_nu_comparison}, the enhancement of $f_{\nu}(u)$ at $T\gtrsim m_e$ gives rise to an additional $\lesssim 10\,\%$
suppression of the amplitude of GWs at $f\sim 10^{-11}\,\mathrm{Hz}$.
The value of $f_{\nu}(u)$ based on Eq.~\eqref{f_nu_new} is suppressed for $T \gtrsim 20\,\mathrm{MeV}$ due to the contributions of 
muons and hadrons to the energy density of background radiations, but such effect is irrelevant to the damping of GWs
since the corresponding modes reenter the horizon before the neutrino decoupling.
Figure~\ref{fig:f_nu_comparison} also shows that the amplitude of GWs that reenter the horizon after $e^+e^-$ annihilation
becomes slightly smaller than the result based on Eq.~\eqref{f_nu_old}. This is because the factor $0.40523$ in Eq.~\eqref{f_nu_old}
is obtained based on the instantaneous decoupling approximation and does not include the correction due to
the increase in effective degrees of freedom of neutrinos [Eq.~\eqref{N_eff_value}].

In the left panel of Fig.~\ref{fig:f_nu_comparison}, we see an oscillatory feature at $f \sim 10^{-10}\,\mathrm{Hz}$ similar to
what observed in the left panel of Fig.~\ref{fig:gw_photon_damping}.
This oscillation can be regarded as an artifact due to the instantaneous decoupling approximation for neutrinos and 
is expected to be diminished
when we directly solve the Boltzmann equation for neutrinos together with Eq.~\eqref{GW_waveequation_kspace}
rather than using Eq.~\eqref{differential_equation_for_chi}.
We also see that the ratio in the left panel of Fig.~\ref{fig:f_nu_comparison} becomes larger than 1 at $f \gtrsim 10^{-12}\,\mathrm{Hz}$.
Again this feature can be regarded as the enhancement effect caused by the fact that the coefficient $f_{\nu}(u)$ decreases when
the corresponding modes reenter the horizon.

\begin{figure}[htbp]
\centering
$\begin{array}{cc}
\subfigure{
\includegraphics[width=80mm]{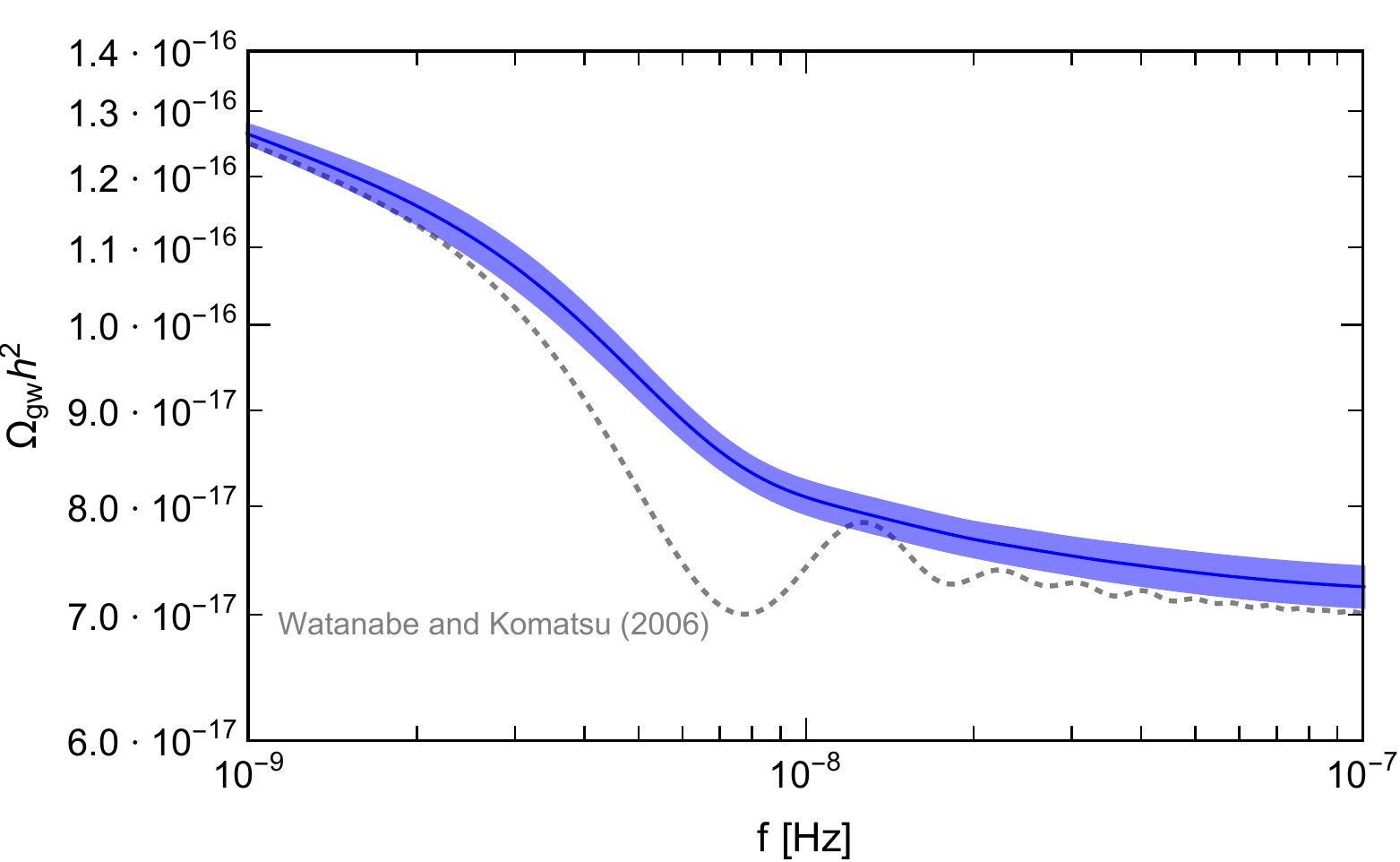}}
\hspace{5mm}
\subfigure{
\includegraphics[width=80mm]{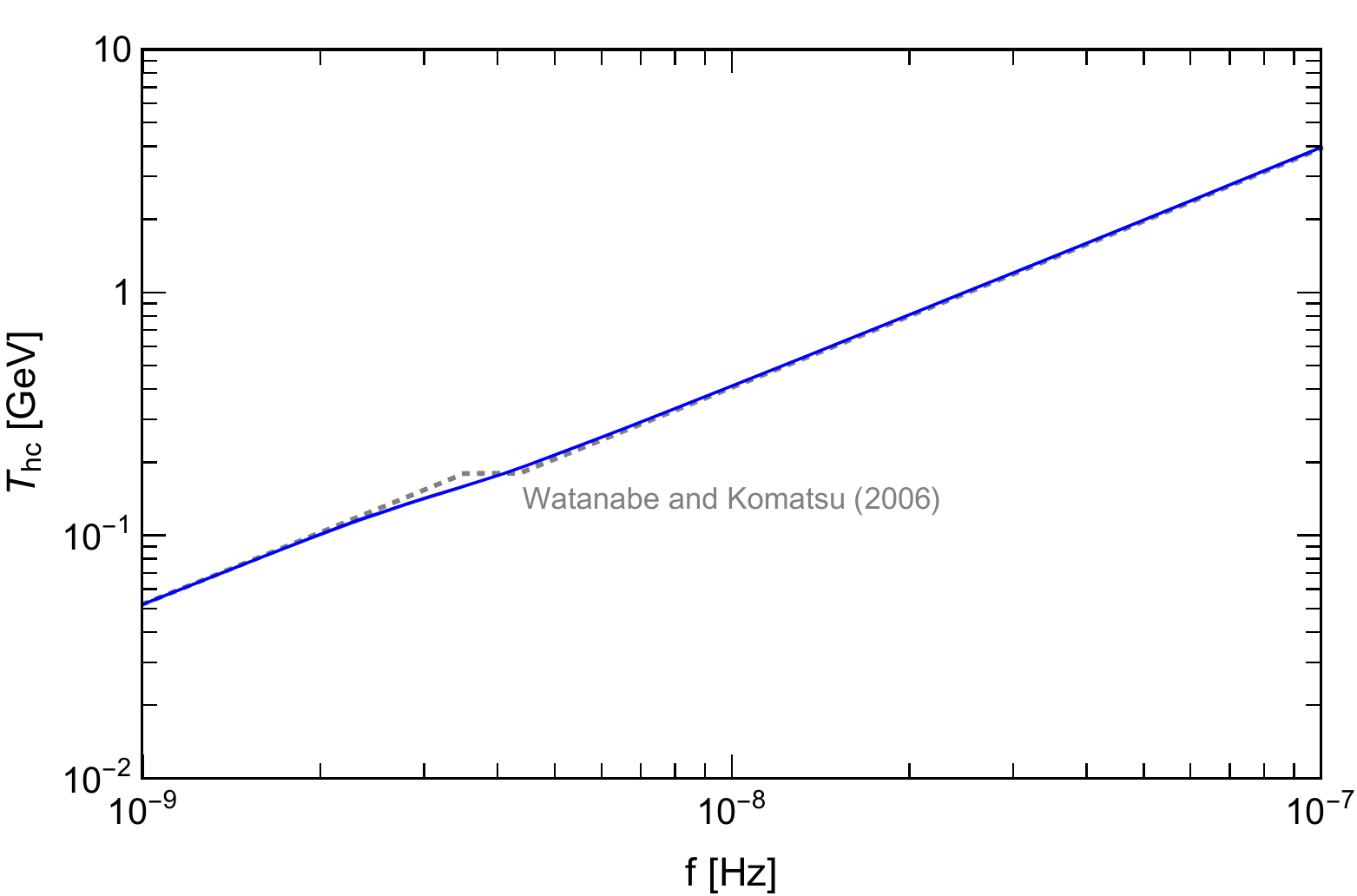}}
\vspace{10mm}
\end{array}$
  \caption{The spectrum of GWs (left panel) and temperature at the horizon crossing (right panel) at a frequency interval $10^{-9}\,\mathrm{Hz}\le f \le 10^{-7}\,\mathrm{Hz}$.
  Light blue bands correspond to the uncertainty of the effective degrees of freedom shown in Fig.~\ref{fig:eos_results} and blue solid lines represent medians.
  Gray dotted lines represent the results obtained by using the effective degrees of freedom based on the ideal gas approximation and the assumption of the sudden
  QCD phase transition (gray dotted line in the middle left panel of Fig.~\ref{fig:eos_results}).
  In these plots the inflationary scale is fixed as $V_{\rm inf}^{1/4} = 1.5\times10^{16}\,\mathrm{GeV}$.}
  \label{fig:gw_QCDPT}
\end{figure}

In Fig.~\ref{fig:gw_QCDPT}, we show the spectrum at frequencies $f = 10^{-9}\textendash 10^{-7}\,\mathrm{Hz}$,
which correspond to the modes reentering the horizon at the epoch of the QCD crossover.
The result shows a smooth spectrum rather than a wiggly feature observed in Ref.~\cite{Watanabe:2006qe}.
This wiggly feature is caused by the fact that the values of $g_{*\rho}(T)$ and $g_{*s}(T)$ suddenly change at $T = T_c$
as shown in the gray dotted line in Fig.~\ref{fig:eos_results}.
Indeed, it can be shown that the wiggle becomes more pronounced if the change of $g_{*\rho}(T)$ and $g_{*s}(T)$ is drastic enough
and that the period of the oscillation is proportional to the inverse of the time interval of the transition $(\Delta \tau)^{-1}$~\cite{Watanabe:2006qe}.
The absence of such a wiggly feature in the updated spectrum is the consequence of the recent lattice QCD analysis
that the QCD phase transition is a smooth crossover rather than a sharp transition assumed in Ref.~\cite{Watanabe:2006qe} [see also Ref.~\cite{Schwarz:1997gv}].
We also note that the sudden changes in $g_{*\rho}(T)$ and $g_{*s}(T)$ make discontinuity in the relation between the frequency and 
temperature at the horizon crossing, as shown in Fig.~\ref{fig:gw_QCDPT}.

\begin{figure}[htbp]
\begin{center}
\includegraphics[width=120mm]{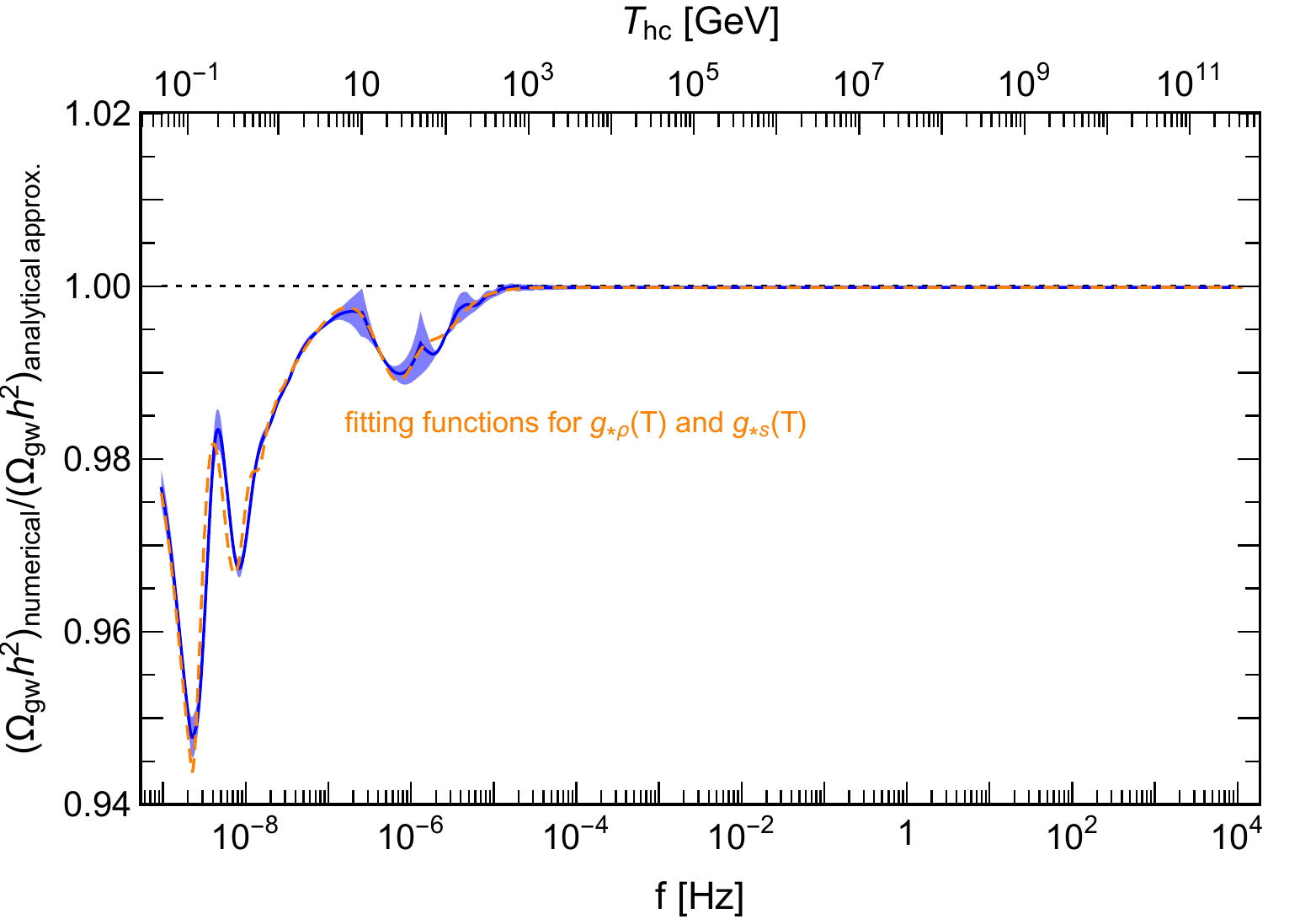}
\end{center}
\caption{The ratio of the amplitude of GWs obtained from the numerical analysis to that based on the analytical approximation [Eq.~\eqref{Omega_gw_gs_PT}].
The light blue band corresponds to the uncertainty of the effective degrees of freedom shown in Fig.~\ref{fig:eos_results} and the blue solid line represents the median.
The orange dashed line represents the result obtained by using the fitting functions for 
the effective degrees of freedom [Eqs.~\eqref{gsr_fitting_function}-\eqref{gss_fitting_function_low}].}
\label{fig:gw_numerical_to_analytical}
\end{figure}

In Fig.~\ref{fig:gw_numerical_to_analytical}, we compare the result based on the numerical analysis with 
that obtained from the analytical approximation in the right-hand side of Eq.~\eqref{Omega_gw_gs_PT}.
The numerical results agree with Eq.~\eqref{Omega_gw_gs_PT} at higher frequencies, while they remain smaller than Eq.~\eqref{Omega_gw_gs_PT}
at lower frequencies. This feature can be understood from the fact that the equation of state parameter $w$ deviates from the value for pure radiation $1/3$ around the time
of the horizon crossing. From Fig.~\ref{fig:eos_results}, we see that the value of $\delta w(T)$ significantly deviates from zero at $T = \mathcal{O}(0.1)\,\mathrm{GeV}$
and $\mathcal{O}(10)\,\mathrm{GeV}$, and the ratio shown in Fig.~\ref{fig:gw_numerical_to_analytical} becomes smaller than $1$ for the modes that reenter
the horizon at the corresponding temperatures.
When $\delta w(T)$ takes some negative (positive) value, the cosmic expansion becomes faster (slower) than the case with $\delta w = 0$,
and we expect that the damping of the amplitude of GWs is more (less) pronounced due to the enhanced (diminished) Hubble friction.
Since the Hubble friction becomes negligible once the mode enters deeply inside the horizon, the damping of the amplitude of GWs
is mostly determined by the behavior of the equation of state parameter at the time of the horizon crossing.

\begin{figure}[htbp]
\begin{center}
\includegraphics[width=160mm]{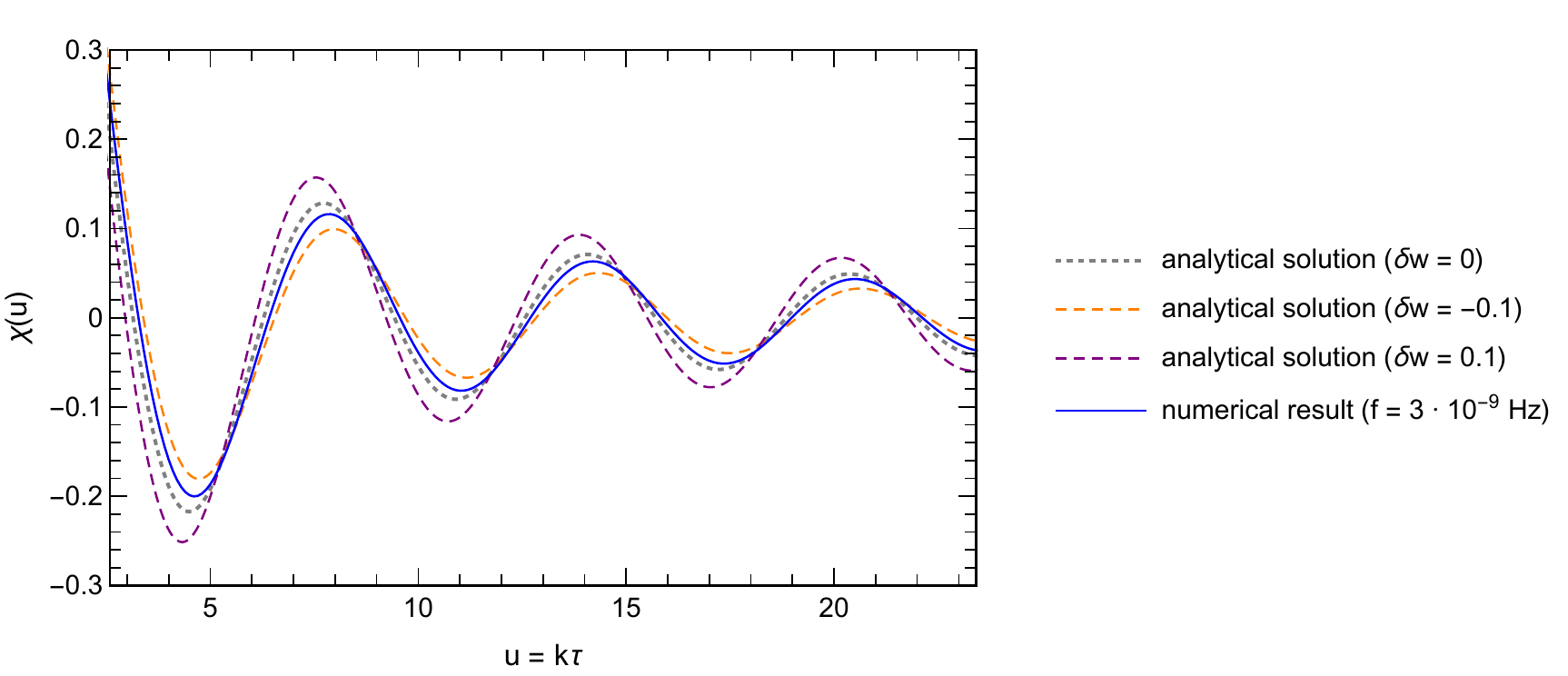}
\end{center}
\caption{Evolution of GWs described by the analytical solutions~\eqref{chi_analytical_solution} 
with $\delta w = 0$ (gray dotted line), $\delta w = -0.1$ (orange dashed line), and $\delta w = 0.1$ (purple dashed line).
Blue line represents the numerical result corresponding to the mode that has a frequency $f = 3\times 10^{-9}\,\mathrm{Hz}$
at the present time.}
\label{fig:chi_evolution}
\end{figure}

We can actually confirm the damping feature mentioned above by using analytical solutions.
For simplicity, let us assume that the equation of state parameter takes a constant value $w = 1/3 + \delta w$.
In the absence of the anisotropic stress, the wave equation~\eqref{differential_equation_for_chi} reads
\begin{equation}
\frac{d^2\chi(u)}{du^2} + \frac{4}{3w+1}\frac{1}{u}\frac{d\chi(u)}{du} + \chi(u) = 0.
\end{equation}
The solution of the above equation satisfying the condition $\left.\chi\right|_{u\to 0} = 1$ is given by
\begin{equation}
\chi(u) = \frac{\Gamma\left(n+\frac{3}{2}\right)}{\Gamma(\frac{3}{2})}\left(\frac{2}{u}\right)^n j_n(u), \label{chi_analytical_solution}
\end{equation}
where $\Gamma(z)$ is the Gamma function, $j_n(z)$ is the spherical Bessel function of the first kind, and
\begin{equation}
n = \frac{1-3w}{1+3w} \simeq -\frac{3}{2}\delta w + \mathcal{O}(\delta w^2).
\end{equation}
If we set $n = 0$ (or $\delta w = 0$), we recover the analytical solution for the pure radiation dominated universe~\cite{Watanabe:2006qe},
\begin{equation}
\left.\chi(u)\right|_{\delta w = 0} = j_0(u) = \frac{\sin u}{u}. \label{chi_solution_pure_radiation}
\end{equation}
In this case, $\chi(u)$ is identical to the WKB solution and Eq.~\eqref{transfer_function_approximate} holds exactly.\footnote{Note that the definition of the horizon crossing
time $k = a_{\rm hc}H_{\rm hc}$ is identical to $u = 1$ for $\delta w = 0$, and hence Eq.~\eqref{chi_solution_pure_radiation} agrees with
the normalization $\chi = (a_{\rm hc}/a)e^{\pm ik\tau}$ used in Eq.~\eqref{transfer_function_approximate}.}
Namely, the amplitude of GWs is exactly given by Eq.~\eqref{Omega_gw_gs_PT} if $\delta w = 0$ at the time of the horizon crossing.
On the other hand, GWs evolve differently from the WKB solution at around the time of the horizon crossing if $n \ne 0$,
as shown in Fig.~\ref{fig:chi_evolution}. From this figure, we see that the amplitude of the oscillation of $\chi(u)$ for $n < 0$ ($n>0$)
becomes smaller (larger) than that for $n=0$ and that a similar tendency appears in the numerical result.
The numerical result shown in Fig.~\ref{fig:chi_evolution} corresponds to the mode that has a frequency $f=3\times 10^{-9}\,\mathrm{Hz}$
at the present time, and it reenters the horizon at $T_{\rm hc} \approx 0.14\,\mathrm{GeV}$.
Since the value of the equation of state parameter becomes $\delta w \approx -0.07$ at that temperature,
the amplitude of the oscillation in the numerical result remains smaller than that of the analytical solution with $\delta w =0$
but larger than that of the analytical solution with $\delta w = -0.1$.
By using the asymptotic behavior of the analytical solution~\eqref{chi_analytical_solution},
\begin{equation}
\chi(u) \xrightarrow{u \gg 1} \frac{\Gamma\left(n+\frac{3}{2}\right)}{\Gamma(\frac{3}{2})}\left(\frac{2}{u}\right)^n \frac{\sin\left(u-\frac{n\pi}{2}\right)}{u},
\end{equation}
we also see that there is a $-n\pi/2$ phase shift in the oscillation of $\chi(u)$ if $n \ne 0$.

In Fig.~\ref{fig:gw_numerical_to_analytical}, we also plotted the result obtained by using fitting functions for the effective degrees of freedom
rather than the tabulated data obtained in Sec.~\ref{sec:full_EoS}.
Although there exist some wiggly features at $f \sim \mathcal{O}(10^{-6})\,\mathrm{Hz}$ in the result based on the tabulated data,
such wiggles are diminished in the result based on the fitting functions.
This fact indicates that the wiggly features at $f \sim \mathcal{O}(10^{-6})\,\mathrm{Hz}$ would be numerical artifacts
caused by some discontinuities in $g_{*\rho}(T)$ and $g_{*s}(T)$ originated from the interpolation at 
$T = \mathcal{O}(10\textendash 100)\,\mathrm{GeV}$ performed in Sec.~\ref{sec:full_EoS}.
On the other hand, we expect that the oscillatory feature at $f \lesssim \mathcal{O}(10^{-8})\,\mathrm{Hz}$ shown in Fig.~\ref{fig:gw_numerical_to_analytical}
corresponds to the genuine effect caused by the fact that the equation of state parameter drastically changes at the epoch of the QCD crossover.
Indeed, we have confirmed that the oscillatory feature becomes more (less) pronounced when we calculate the spectrum of GWs by using some mock data of 
the equation of state whose temperature variation is sufficiently fast (slow).

The numerical results for the transfer function of GWs obtained in this section are made available as tabulated data (see Appendix~\ref{app:supplementary_material}).
Although we focus on the case where the primordial tensor power spectrum $\mathcal{P}_T(k)$ is a $k$-independent constant in this work,
it is possible to use these results to estimate the spectrum of GWs for an arbitrary function of $\mathcal{P}_T(k)$.

\begin{figure}[htbp]
\centering
$\begin{array}{cc}
\subfigure{
\includegraphics[width=80mm]{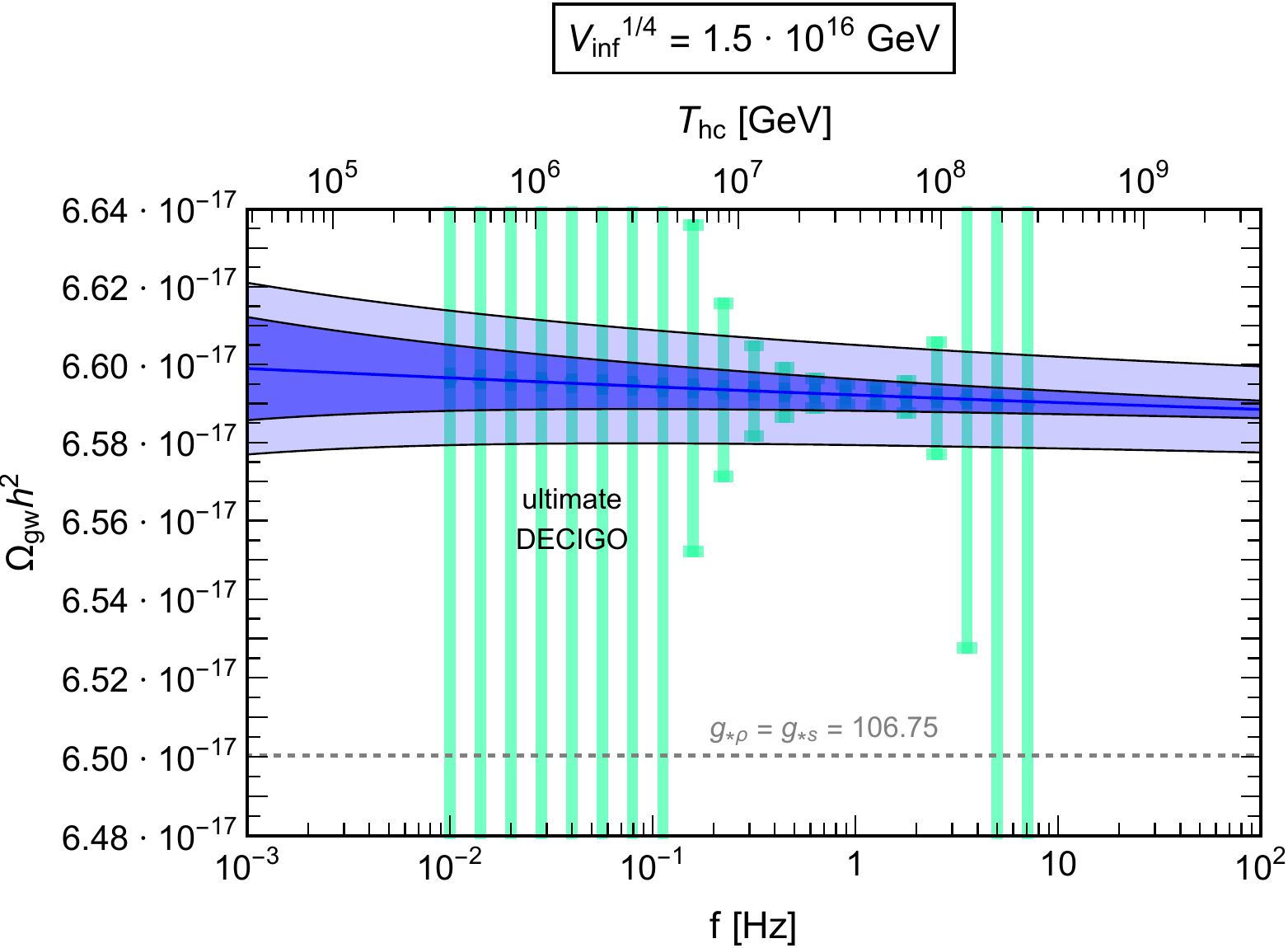}}
\hspace{5mm}
\subfigure{
\includegraphics[width=80mm]{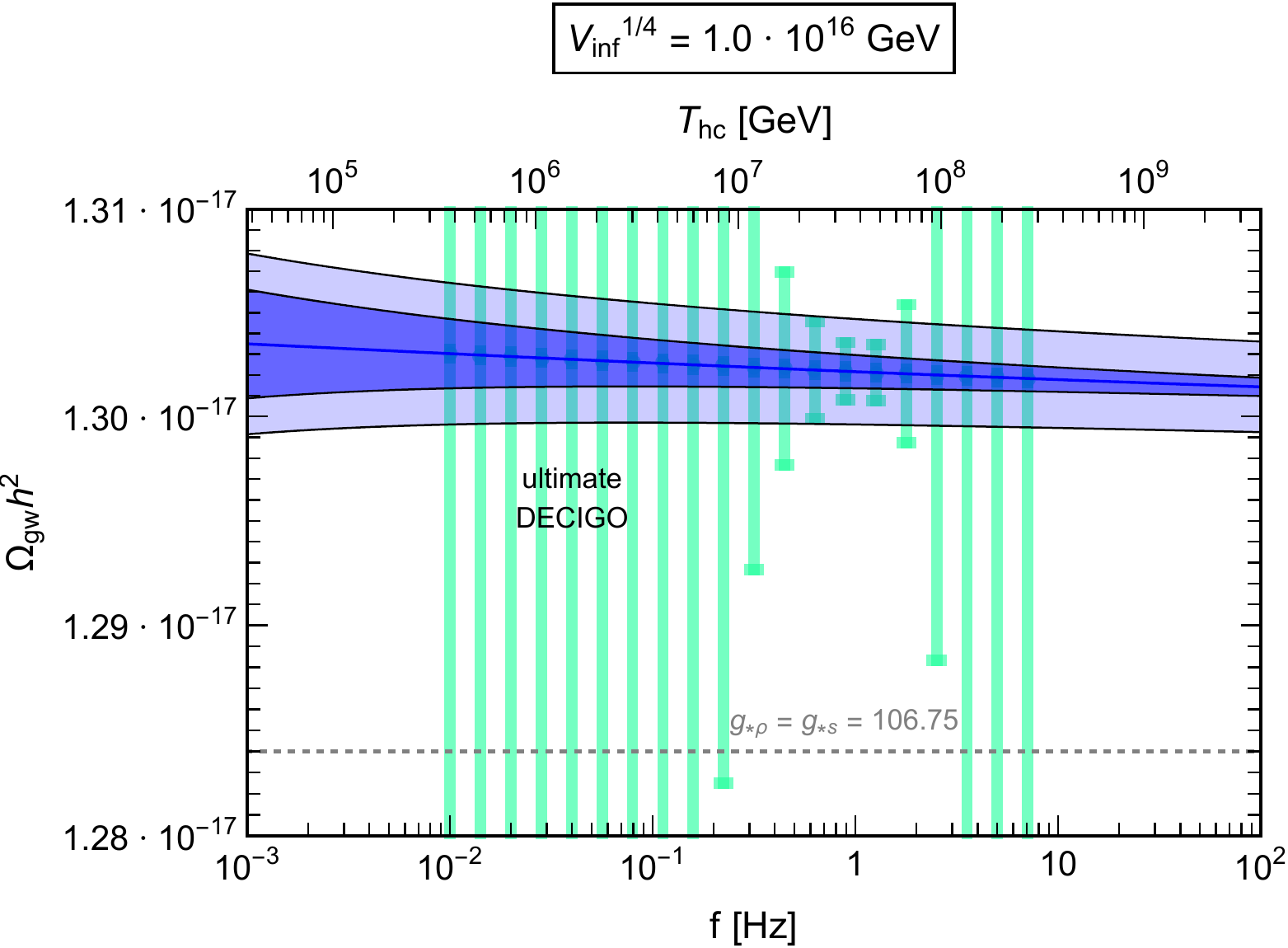}}
\vspace{10mm}
\\
\subfigure{
\includegraphics[width=80mm]{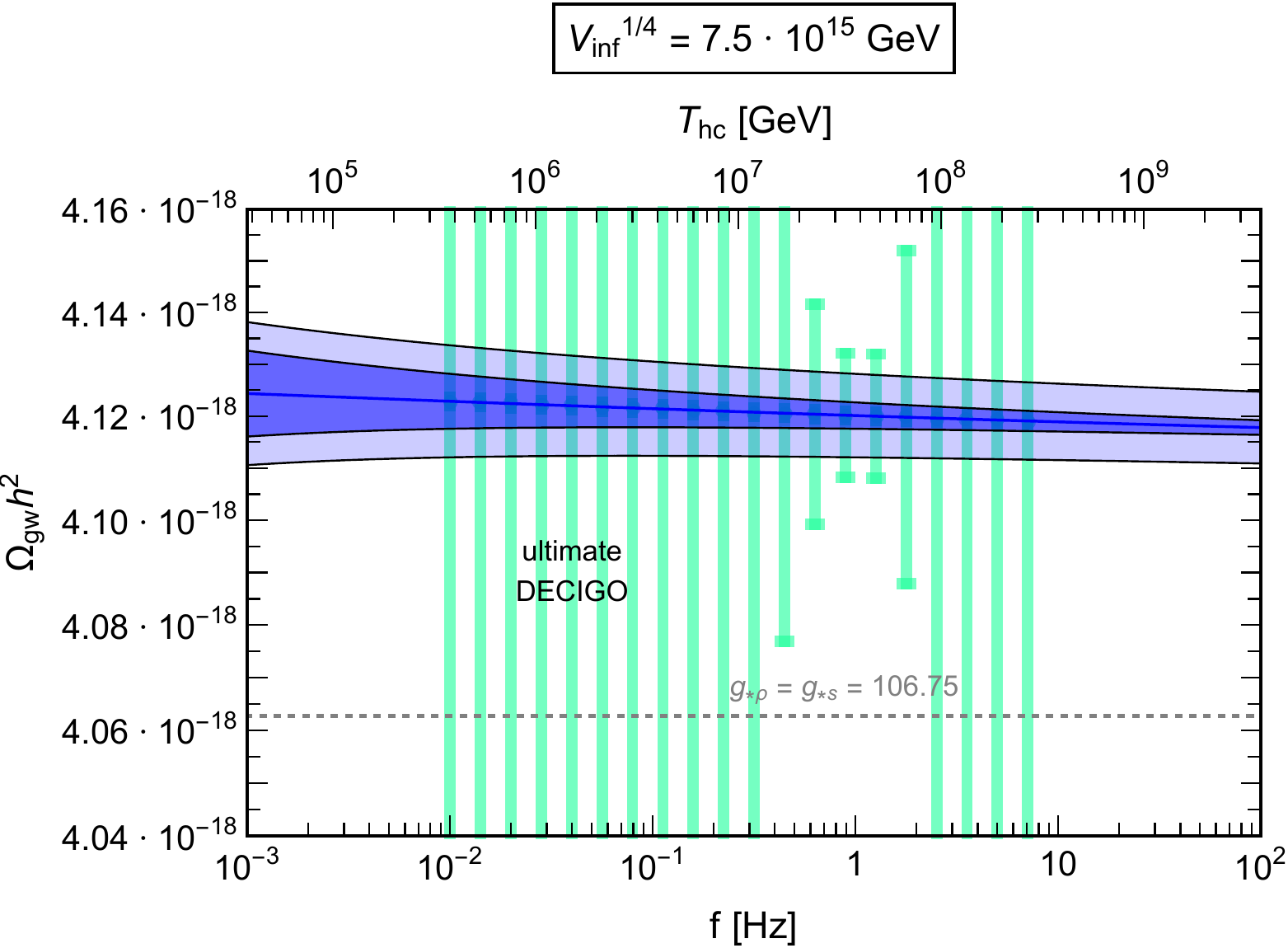}}
\hspace{5mm}
\subfigure{
\includegraphics[width=80mm]{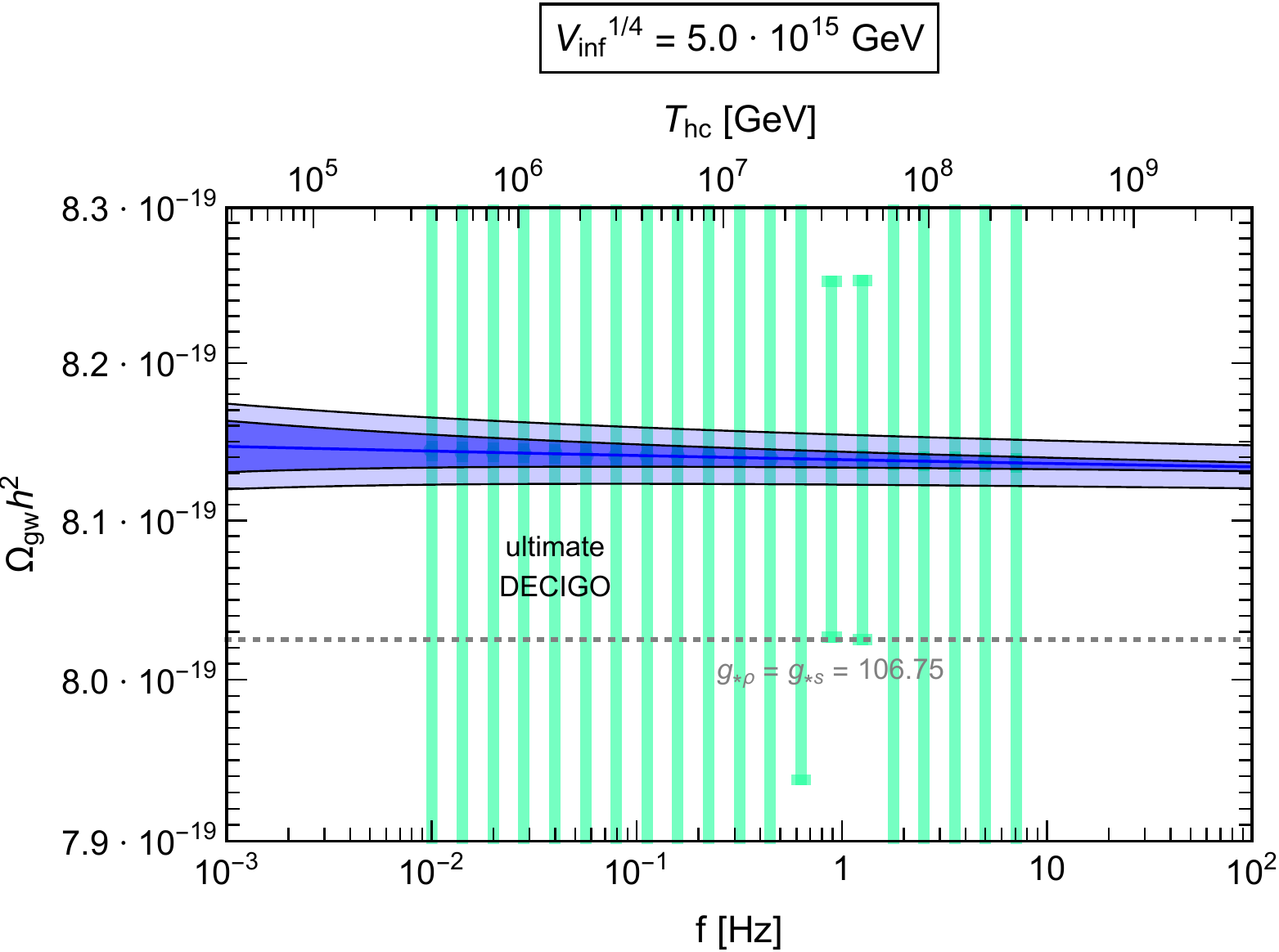}}
\vspace{10mm}\end{array}$
  \caption{The spectrum of inflationary GWs for various choices of the inflationary scale $V_{\rm inf}^{1/4}$.
  Here we adopt $V_{\rm inf}^{1/4} = 1.5\times 10^{16}\,\mathrm{GeV}$ (top left), $1.0\times 10^{16}\,\mathrm{GeV}$ (top right),
  $7.5\times 10^{15}\,\mathrm{GeV}$ (bottom left), and $5.0\times 10^{15}\,\mathrm{GeV}$ (bottom right).
  Blue bands correspond to the uncertainty of the effective degrees of freedom shown in Fig.~\ref{fig:eos_results_high} and the blue solid lines represent medians.
  Lighter blue bands represent additional uncertainties due to the value of $g_{*s,\mathrm{fin}}$ [Eq.~\eqref{gss_fin_value}].
  Gray dotted lines represent the results obtained by using the conventional value $g_{*\rho} = g_{*s} = 106.75$.
  Green bars correspond to the sensitivity of ultimate DECIGO estimated by using Eq.~\eqref{Omega_gw_sensitivity}.}
  \label{fig:gw_spectrum_high_frequency}
\end{figure}

Let us now turn our attention to the spectrum at a higher frequency range, which is relevant to the direct detection experiments of GWs.
In the previous section, we have seen that the value of the effective degrees of freedom do not reach the well-known result $g_{*\rho} = g_{*s} = 106.75$ even at high temperatures.
This fact leads to an important consequence to future GW direct detection experiments, as shown in Fig.~\ref{fig:gw_spectrum_high_frequency}.
We see that the amplitude of GWs at the frequency range shown in Fig.~\ref{fig:gw_spectrum_high_frequency} 
becomes $1.2\textendash 1.8\%$ larger than the conventional estimate based on $g_{*\rho} = g_{*s} = 106.75$.
This correction appears due to the fact that $\Omega_{\rm gw}$ is proportional to $g_{*\rho,\mathrm{hc}}^{-1/3}$ and that the value of $g_{*\rho,\mathrm{hc}}$ remains smaller than
the commonly assumed value $106.75$.
Furthermore, there exists a non-trivial frequency dependence even if the primordial tensor power spectrum is exactly scale-invariant,
since the values of $g_{*\rho}(T)$ and $g_{*s}(T)$ weakly depend on $T$ due to the renormalization group running of SM gauge and Yukawa couplings.
Denoting this scale dependence as a correction $\Delta n_T$ to the tilt of the primordial tensor power spectrum $n_T$ 
(e.g. $\Omega_{\rm gw} \propto f^{n_T+\Delta n_T}$), we find $\Delta n_T\simeq -(1.3\pm 0.4)\times 10^{-4}$ at $f = 1\,\mathrm{Hz}$.

In Fig.~\ref{fig:gw_spectrum_high_frequency}, we also show some uncertainty of the GW spectrum
corresponding to that of the effective degrees of freedom.
In the frequency range shown in Fig.~\ref{fig:gw_spectrum_high_frequency}, 
Eq.~\eqref{Omega_gw_gs_PT} with $g_{*\rho} = g_{*s}$ holds with good accuracy, and the amplitude of GWs depends on 
the effective degrees of freedom as $\Omega_{\rm gw} \propto g_{*s,\mathrm{fin}}^{4/3}g_{*\rho,\mathrm{hc}}^{-1/3}$.
The uncertainty of $g_{*\rho,\mathrm{hc}}$ becomes less important at higher frequencies since the perturbative expansion of the free energy
becomes more accurate at higher temperatures.
Meanwhile, there exists some uncertainty in the value of $g_{*s,\mathrm{fin}}$ determined at very low temperature, as discussed in Sec.~\ref{sec:photons_and_leptons}.
We see that the corresponding uncertainty can be even larger than that of $g_{*\rho,\mathrm{hc}}$, because of the fact that the exponent $4/3$ in Eq.~\eqref{Omega_gw_gs_PT}
enhances the effect of $g_{*s,\mathrm{fin}}$ relative to that of $g_{*\rho,\mathrm{hc}}$.
Note that the resulting amplitude of GWs is significantly larger than that obtained based on $g_{*\rho} = g_{*s} = 106.75$
even if we take account of such uncertainty.

It has been argued that future space-borne GW interferometers can survey the GW background in the frequency range $\lesssim 1\,\mathrm{Hz}$,
and that the ultimate sensitivity of DECIGO may become comparable to $\Omega_{\rm gw} \sim 10^{-20}$~\cite{Seto:2001qf}.
If this is the case, the correction due to the equation of state in the SM has actual impact on the observational results.
In Fig.~\ref{fig:gw_spectrum_high_frequency}, we plot the sensitivity of ultimate DECIGO for the sake of comparison (see Appendix~\ref{app:sensitivity} for details).
The figures clearly show that the difference between the result including the effect of particle interactions in the SM 
and that based on the crude approximation $g_{*\rho} = g_{*s} = 106.75$ is larger than the experimental sensitivity
as long as the inflationary scale for the relevant modes is sufficiently high ($V_{\rm inf}^{1/4} \gtrsim 5\times 10^{15}\,\mathrm{GeV}$).
Therefore, in principle it is possible to observe the non-trivial evolution of the effective degrees of freedom in the SM through high-sensitivity GW experiments.

%%%%%%%%%%%%%%%%%%%%%%%%%%%%%%%%%%%%%%%%%%%%%%%%%%
\section{Conclusions and discussion}
\label{sec:conclusion}
\setcounter{equation}{0}
%%%%%%%%%%%%%%%%%%%%%%%%%%%%%%%%%%%%%%%%%%%%%%%%%%

In this paper, we have investigated an impact of the equation of state in the SM on the spectrum of primordial GWs originated from inflation.
We estimated the effective degrees of freedom $g_{*\rho}(T)$ and $g_{*s}(T)$ as well as the equation of state parameter $w(T)$
for a wide temperature interval by collecting state-of-the-art results of perturbative and non-perturbative analysis in the SM and identified their uncertainty.
It was found that the resulting values of $g_{*\rho}(T)$ and $g_{*s}(T)$ including the effects of particle interactions in the SM
deviate from those obtained based on the ideal gas approximation, as shown in Figs.~\ref{fig:eos_results} and~\ref{fig:eos_results_high}.
After obtaining the values of $g_{*\rho}(T)$ and $g_{*s}(T)$, we applied them to the estimation of the spectrum of inflationary GWs.

We have seen that there exist several corrections on the spectrum of primordial GWs due to the effects that were overlooked in previous studies.
First, we showed that the inclusion of the contribution of free-streaming photons to the anisotropic stress leads to an additional damping of the amplitude
of GWs by $\lesssim 14\,\%$ at $f \sim 10^{-17}\,\mathrm{Hz}$.
Second, we pointed out that the revision of the function $f_{\nu}(u)$ for the evaluation of the collisionless damping effect due to free-streaming neutrinos [Eq.~\eqref{f_nu_new}]
gives rise to an additional $\lesssim 10\,\%$ suppression of the amplitude of GWs at $f \sim 10^{-11}\,\mathrm{Hz}$.
Third, we found that the amplitude of GWs changes smoothly at $f \sim 10^{-8}\,\mathrm{Hz}$ as the recent results of lattice QCD simulations imply that
the nature of the QCD phase transition is a smooth crossover rather than a sharp transition.
This result is in contrast to that of previous analysis~\cite{Watanabe:2006qe,Kuroyanagi:2008ye} that there exists a wiggly 
feature at the corresponding frequency due to the assumption that
the values of $g_{*\rho}(T)$ and $g_{*s}(T)$ drastically change at the critical temperature of the QCD phase transition.
Finally, we showed that the amplitude of GWs at higher frequencies becomes $\gtrsim 1\,\%$ larger than the estimate 
based on the ideal gas approximation for the effective degrees of freedom due to the fact that the values 
of $g_{*\rho}(T)$ and $g_{*s}(T)$ remain smaller than the commonly used value
$g_{*\rho} = g_{*s} = 106.75$ even at temperature much higher than the critical temperature of the electroweak crossover.
In principle, such a correction can be observed by future high-sensitivity GW experiments such as ultimate DECIGO.

The existence of the above-mentioned damping of the amplitude of GWs due to free-streaming photons implies that
there is some energy flow between GWs and photons around the epoch of the photon last scattering,
and hence it should also affect the spectrum of CMB polarization observed today.
We believe that this effect was already included in the well-established formalism to compute 
the CMB polarization power spectrum [see, e.g. Refs.~\cite{Hu:1997hp,Tram:2013ima}]
and implemented in the public code such as {\tt CLASS}~\cite{CLASS}.
Instead of analyzing the effect on the CMB polarization power spectrum,
in this paper we focused on the effect on the spectrum of GWs itself, which was not explicitly analyzed in previous literature.

In Sec.~\ref{sec:EoS_in_the_SM}, we saw that the dominant source of uncertainty in thermodynamic quantities in the SM
is the estimation of the pressure of QCD at $T = \mathcal{O}(1\textendash 10)\,\mathrm{GeV}$.
In particular, the perturbative expansion in terms of the gauge coupling $g_s$ shows a poor convergence at the corresponding temperatures,
and there exists an unknown constant $q_c(N_f)$ that cannot be estimated based on the perturbative approach.
In this paper, we have not improved these situations but aimed to quantify the corresponding uncertainty
by interpolating the pressure and trace anomaly at $T = 5\textendash 100\,\mathrm{GeV}$ (see Sec.~\ref{sec:full_EoS}).
The resolution of these ambiguities should be achieved by direct non-perturbative analysis at high temperatures.

Since our analysis to determine the temperature evolution of $g_{*\rho}(T)$ and $g_{*s}(T)$ is based on the SM of particle physics,
the effects discussed in this paper always exist and are relevant to various sources that lead to cosmological backgrounds of GWs.\footnote{An exceptional case
is the production of GWs via first order phase transitions~\cite{Witten:1984rs,Kamionkowski:1993fg}, 
where we cannot apply the results obtained based on equilibrium thermodynamics.}
In particular, several alternative scenarios can lead to cosmological backgrounds of GWs 
whose amplitude at high frequencies is much larger than the nearly scale invariant spectrum predicted in the standard inflationary scenario,
and such scenarios are explored by planned experiments such as LISA in the near future~\cite{Bartolo:2016ami}.
It is probable that we misinterpret the results of such forthcoming experiments if we do not include the corrections induced by 
the equation of state in the SM in the evaluation of the GW signatures.

In the analysis performed in this paper, we artificially fixed the primordial tensor power spectrum as a $k$-independent constant
and focused on the effects caused by the late time evolution of GWs (or the transfer function) in the SM.
In actual cases, we must include the contribution of the primordial tensor power spectrum as well as the transfer function.
Furthermore, the spectrum of GWs could be further corrected if there exists some physics beyond the SM.
For instance, contributions of new particles and their interaction properties to the effective degrees of freedom at high temperature
can lead to further modification in the spectrum of GWs at high frequency,
and the amplitude of GWs can also be damped if there exist some free-streaming particles other than the SM photons and neutrinos.
It would be interesting to investigate the spectrum of primordial GWs in the context of 
concrete models of inflation and thermal history of the universe based on physics beyond the SM,
such as a model based on the spontaneous breaking of the $B-L$ symmetry~\cite{Buchmuller:2012wn} 
or that of the Peccei-Quinn symmetry~\cite{Ballesteros:2016xej}.

The effective degrees of freedom in the SM constructed in Sec.~\ref{sec:EoS_in_the_SM} 
are made available as tabulated data (see Appendix~\ref{app:supplementary_material})
and also as fitting functions (see Appendix~\ref{app:fitting_functions}).
These results can be used not only in the analysis of primordial GWs but also in that of other cosmological relics.
For instance, in the standard freeze-out production mechanism of WIMP dark matter,
the freeze-out temperature is estimated to be $T_f \sim m_{\chi}/30$~\cite{Kolb:1990vq}, 
which reads $T_f \sim 3\textendash 30\,\mathrm{GeV}$ for a WIMP mass of $m_{\chi} \sim 100\textendash 1000\,\mathrm{GeV}$,
and hence the uncertainty of pressure in QCD at the corresponding temperature has an impact on the WIMP dark matter abundance.
Such kind of uncertainty was already pointed out in Refs.~\cite{Hindmarsh:2005ix,Drees:2015exa}, 
but it is reasonable to review this issue in light of updated results of the equation of state in the SM as it 
might have a relevance to future dark matter searches with improved sensitivity.
The same is true of the estimation of the axion dark matter abundance.
The number of axions produced via the realignment mechanism and/or the decay of topological defects
is fixed at the epoch of the QCD crossover~\cite{Kawasaki:2014sqa}, and the resulting abundance is subjected to the uncertainty in 
the equation of state as well as that in the topological susceptibility~\cite{Borsanyi:2016ksw}.
On top of these issues, there would be a potential application of the equation of state in the SM
to several topics in cosmology such as leptogenesis, sterile neutrinos, dark radiations, and primordial black holes. 
We hope that our approach in this paper makes a small step to improve all the analysis in precision cosmology.

%%%%%%%%%%%%%%%%%%%%%%%%%%%%%%%%%%%%%%%%%%%%%%%%%%
\section*{Acknowledgments}
%%%%%%%%%%%%%%%%%%%%%%%%%%%%%%%%%%%%%%%%%%%%%%%%%%
KS would like to thank Takeshi Chiba, Thomas Konstandin, Alexander Merle, Andreas Ringwald, and Masahide Yamaguchi for discussions and comments.
Numerical computations in this work were partially carried out at the Max Planck Computing and Data Facility (MPCDF).
This work is supported by Grant-in-Aid for Scientific Research from the Ministry of Education, Culture, Sports, Science, and Technology (MEXT), Japan, No. 17H02878 (SS) and by World Premier International Research Center Initiative (WPI), MEXT, Japan (SS).
KS acknowledges partial support by the Deutsche Forschungsgemeinschaft through Grant No.\ EXC 153 (Excellence
Cluster ``Universe'') and Grant No.\ SFB 1258 (Collaborative Research
Center ``Neutrinos, Dark Matter, Messengers'') as well as by the
European Union through Grant No.\ H2020-MSCA-ITN-2015/674896 (Innovative Training Network ``Elusives'').
%%%%%%%%%%%%%%%%%%%%%%%%%%%%%%%%%%%%%%%%%%%%%%%%%%
\appendix
%%%%%%%%%%%%%%%%%%%%%%%%%%%%%%%%%%%%%%%%%%%%%%%%%% 

%%%%%%%%%%%%%%%%%%%%%%%%%%%%%%%%%%%%%%%%%%%%%%%%%% 
\section{Supplementary material}
\setcounter{equation}{0}
\label{app:supplementary_material}
%%%%%%%%%%%%%%%%%%%%%%%%%%%%%%%%%%%%%%%%%%%%%%%%%%

Supplementary material containing tabulated data for the effective degrees of freedom 
obtained in Sec.~\ref{sec:EoS_in_the_SM} is available online as a filename {\tt standardmodel2018.dat}.
It is placed in the directory {\tt anc/} included in the source of the arXiv manuscript.
It can also be downloaded from \url{http://member.ipmu.jp/satoshi.shirai/EOS2018}.

We also provide a data set for the transfer function of primordial GWs:
\begin{equation}
T(k) \equiv \frac{1}{12a_0^2 H_0^2}\overline{\left[\chi'(\tau_0,k)\right]^2}h^2,
\end{equation}
where the overbar represents the average of rapidly oscillating functions in $\left[\chi'(\tau_0,k)\right]^2$.
The function $T(k)$ can be used to produce the spectrum of primordial GWs for a given primordial tensor power spectrum $\mathcal{P}_T(k)$
[see Eq.~\eqref{Omega_gw_PT_T}],
\begin{equation}
\Omega_{\rm gw}h^2(k) = T(k)\mathcal{P}_T(k).
\end{equation}
The corresponding data is made available as a filename {\tt transferfunction.dat}.

%%%%%%%%%%%%%%%%%%%%%%%%%%%%%%%%%%%%%%%%%%%%%%%%%% 
\section{Notes on thermodynamic quantities}
\label{app:thermodynamic}
\setcounter{equation}{0}
%%%%%%%%%%%%%%%%%%%%%%%%%%%%%%%%%%%%%%%%%%%%%%%%%%

In this appendix, we review some basic formulae for thermodynamic quantities used in the main text
and give an explicit proof of the equivalence between two expressions of the damping factor of GWs shown in Eq.~\eqref{Omega_gw_damp_gs}.
We start from two fundamental assumptions: One is the conservation of entropy,
\begin{equation}
s(T)a^3 = \text{const}., \label{S_conservation}
\end{equation}
and the other is the first law of thermodynamics,
\begin{equation}
d\left(s(T)V\right)=\frac{d\left(\rho(T)V\right)+p(T)dV}{T}, \label{first_law}
\end{equation}
where $V$ is the volume of the system.
By comparing the coefficients of $dT$ and $dV$ in Eq.~\eqref{first_law}, we obtain
\begin{align}
s(T) &= \frac{\rho(T)+p(T)}{T}, \label{s_rho_p_relation} \\
\rho(T) &= T\frac{dp}{dT}(T) - p(T). \label{rho_p_relation}
\end{align}

By differentiating Eq.~\eqref{rho_p_relation}, we have
\begin{equation}
\frac{d\ln\rho}{d\ln a} = \frac{d\ln T}{d\ln a}\frac{T^2}{\rho}\frac{d^2p}{dT^2}. \label{dlnrho_dlna}
\end{equation}
On the other hand, the derivative of Eq.~\eqref{S_conservation} reads
\begin{equation}
0 = \frac{d}{dT}\left(s a^3\right) = 3a^2\frac{da}{dT}s + a^3 \frac{ds}{dT}. \nonumber
\end{equation}
Substituting Eqs.~\eqref{s_rho_p_relation} and~\eqref{rho_p_relation} into the above equation, we have
\begin{equation}
\frac{d\ln a}{d\ln T} = -\frac{T^2}{3(\rho + p)}\frac{d^2p}{dT^2}. \label{dlna_dlnT}
\end{equation}
From Eqs.~\eqref{dlnrho_dlna} and~\eqref{dlna_dlnT}, we obtain the energy conservation law,
\begin{equation}
d\ln\rho = -3(1+w)d\ln a, \label{energy_conservation}
\end{equation} 
where 
\begin{equation}
w = \frac{p}{\rho}.
\end{equation}

The evolution of the energy density $\rho$ can also be described in terms of $g_{*\rho}$ and $g_{*s}$ defined in Eq.~\eqref{gstar_definition},
\begin{equation}
\rho \propto g_{*\rho} T^4 \propto g_{*\rho} g_{*s}^{-\frac{4}{3}} a^{-4}, \label{rho_gstar_dependence}
\end{equation}
where the right-hand side follows from the conservation of entropy~\eqref{S_conservation}.
By using the equivalence between Eqs.~\eqref{energy_conservation} and~\eqref{rho_gstar_dependence},
we actually see that the damping factor shown in the first line of Eq.~\eqref{Omega_gw_damp_gs} corresponds to that shown in the second line,
\begin{align}
\exp\left[\int^{a_{\rm hc,1}}_{a_{\rm hc,2}}(3w-1)d\ln a\right] &= \exp\left[-\int^{a_{\rm hc,1}}_{a_{\rm hc,2}}d\ln \rho\right]\exp\left[-4\int^{a_{\rm hc,1}}_{a_{\rm hc,2}}d\ln a\right] \nonumber\\
&= \left(\frac{\rho(\tau_{\rm hc,2})}{\rho(\tau_{\rm hc,1})}\right)\left(\frac{a_{\rm hc,2}}{a_{\rm hc,1}}\right)^4 \nonumber\\
&= \left(\frac{g_{*\rho,\mathrm{hc,2}}}{g_{*\rho,\mathrm{hc,1}}}\right) \left(\frac{g_{*s,\mathrm{hc,2}}}{g_{*s,\mathrm{hc,1}}}\right)^{-\frac{4}{3}}. \label{equivalence_damping_factor}
\end{align}
One may naively think that the quantities $g_{*\rho}$, $g_{*s}$, and $a$ evolve independently over time, but they are tied to each other via thermodynamics equations.
Actually, $g_{*\rho}$ and $g_{*s}$ can be determined from a single quantity $p$ as shown in Eqs.~\eqref{s_rho_p_relation} and ~\eqref{rho_p_relation},
and the scale factor $a$ is related to $g_{*s}$ via Eq.~\eqref{S_conservation}.
These relations guarantee the equivalence between two expressions for the damping factor of GWs.

%%%%%%%%%%%%%%%%%%%%%%%%%%%%%%%%%%%%%%%%%%%%%%%%%% 
\section{Fitting functions for effective degrees of freedom}
\label{app:fitting_functions}
\setcounter{equation}{0}
%%%%%%%%%%%%%%%%%%%%%%%%%%%%%%%%%%%%%%%%%%%%%%%%%%

Here we give fitting functions to reproduce our estimates of $g_{*\rho}$ and $g_{*s}$ for $ T\leq 10^{16}\,\mathrm{GeV}$.
These functions can be fitted to $g_{*\rho}$ and $g_{*s}$ obtained in Sec.~\ref{sec:EoS_in_the_SM} 
within the present uncertainty.

The fitting functions for $120\,\mathrm{MeV}\leq T\leq 10^{16}\,\mathrm{GeV}$ are approximately given as:
\begin{align}
g_{*\rho}(T) &\simeq  \frac{\sum_{i=0}^{11} a_i t^i}{\sum_{i=0}^{11} b_i t^i},\label{gsr_fitting_function}\\ 
\frac{g_{*\rho}(T)}{g_{*s}(T)} &\simeq 1 + \frac{\sum_{i=0}^{11} c_i t^i}{\sum_{i=0}^{11} d_i t^i},\label{gsr_to_gss_fitting_function}
\end{align}
where $t = \ln(T\mathrm{[GeV]})$. The coefficients $a_i, b_i, c_i$, and $d_i$ are given in Table~\ref{tab:fit}.

\begin{table}[h]
\caption{Coefficients for the fitting functions for  $120\,\mathrm{MeV}\leq T \leq 10^{16}\,\mathrm{GeV}$. \label{tab:fit}}
\centering \begin{tabular}{|c|rrrr|}
\hline
$i$&
\multicolumn{1}{c}{$a_i$}&
\multicolumn{1}{c}{$b_i$}&
\multicolumn{1}{c}{$c_i$}&
\multicolumn{1}{c|}{$d_i$}\\
\hline
\hline
0&1&1.43382E$-$02&1&7.07388E+01\\
1&1.11724E$+$00&1.37559E$-$02&6.07869E$-$01&9.18011E+01\\
2&3.12672E$-$01&2.92108E$-$03&$-$1.54485E$-$01&3.31892E+01\\
3&$-$4.68049E$-$02&$-$5.38533E$-$04&$-$2.24034E$-$01&$-$1.39779E+00\\
4&$-$2.65004E$-$02&$-$1.62496E$-$04&$-$2.82147E$-$02&$-$1.52558E+00\\
5&$-$1.19760E$-$03&$-$2.87906E$-$05&2.90620E$-$02&$-$1.97857E$-$02\\
6&1.82812E$-$04&$-$3.84278E$-$06&6.86778E$-$03&$-$1.60146E$-$01\\
7&1.36436E$-$04&2.78776E$-$06&$-$1.00005E$-$03&8.22615E$-$05\\
8&8.55051E$-$05&7.40342E$-$07&$-$1.69104E$-$04&2.02651E$-$02\\
9&1.22840E$-$05&1.17210E$-$07&1.06301E$-$05&$-$1.82134E$-$05\\
10&3.82259E$-$07&3.72499E$-$09&1.69528E$-$06&7.83943E$-$05\\
11&$-$6.87035E$-$09&$-$6.74107E$-$11&$-$9.33311E$-$08&7.13518E$-$05\\
\hline
\end{tabular}
\end{table}

For $T < 120\,\mathrm{MeV}$, the fitting functions are given as:
\begin{align}
g_{*\rho}(T) &\simeq 2.030 +1.353 \mathcal{S}_{\rm fit}^{\frac{4}{3}}\left(\frac{m_e}{T}\right) + 3.495 
 f_\rho\left(\frac{m_e}{T}\right)+ 3.446 f_\rho\left(\frac{m_{\mu}}{T}\right) + 1.05b_\rho\left(\frac{m_{\pi^0}}{T}\right)\nonumber \\
 &\quad 
 +2.08 b_\rho\left(\frac{m_{\pi^{\pm}}}{T}\right)
  +4.165 b_\rho\left(\frac{m_{1}}{T}\right)
  +30.55 b_\rho\left(\frac{m_{2}}{T}\right)
  +89.4 b_\rho\left(\frac{m_{3}}{T}\right)
+ 8209 b_\rho\left(\frac{m_{4}}{T}\right)  
  ,\label{gsr_fitting_function_low}\\  
g_{*s}(T) &\simeq 2.008 +1.923 \mathcal{S}_{\rm fit}\left(\frac{m_e}{T}\right) + 3.442
 f_s\left(\frac{m_e}{T}\right)+ 3.468 f_s\left(\frac{m_{\mu}}{T}\right) + 1.034b_s\left(\frac{m_{\pi^0}}{T}\right) \nonumber \\
 & \quad
 +2.068  b_s\left(\frac{m_{\pi^{\pm}}}{T}\right)
  +4.16 b_s\left(\frac{m_{1}}{T}\right)
  +30.55 b_s\left(\frac{m_{2}}{T}\right)
  +90 b_s\left(\frac{m_{3}}{T}\right)
  +6209 b_s\left(\frac{m_{4}}{T}\right),\label{gss_fitting_function_low}
  \end{align}
where $m_e = 511\times10^{-6}$ GeV, $m_\mu = 0.1056$ GeV, 
$m_{\pi^0} = 0.135$ GeV, $m_{\pi^{\pm}} = 0.140$ GeV,
$m_{1} = 0.5$ GeV, $m_{2} = 0.77$ GeV, $m_3 = 1.2$ GeV, and $m_4 = 2$ GeV.
Functions used in the right-hand sides of Eqs.~\eqref{gsr_fitting_function_low} and~\eqref{gss_fitting_function_low} are given as:
\begin{align}
f_{\rho}(x) &= \exp(- 1.04855 x)(1  + 1.03757 x + 0.508630 x^2 + 0.0893988x^3 ),\\
b_{\rho}(x) &=  \exp(-1.03149 x)(1  +1.03317 x + 0.398264 x^2 + 0.0648056x^3 ),\\
f_{s}(x) &=\exp(-1.04190 x)(1  +1.03400x + 0.456426 x^2 + 0.0595248x^3 ),\\
b_{s}(x) &=\exp(-1.03365 x)(1  +1.03397 x + 0.342548 x^2 + 0.0506182x^3 ),\\
\mathcal{S}_{\rm fit}(x) &=  1 +   \frac{7}{4}  \exp(- 1.0419  x)(1  + 1.034 x + 0.456426x^2 + 0.0595249x^3 ).
\end{align}

%%%%%%%%%%%%%%%%%%%%%%%%%%%%%%%%%%%%%%%%%%%%%%%%%% 
\section{Estimation of the sensitivity of gravitational wave detectors}
\label{app:sensitivity}
\setcounter{equation}{0}
%%%%%%%%%%%%%%%%%%%%%%%%%%%%%%%%%%%%%%%%%%%%%%%%%%

In this appendix, we discuss the methodology to estimate the projected sensitivity of GW experiments.
Here we follow the correlation analysis method developed in Refs.~\cite{Allen:1997ad,Maggiore:1999vm}.

In order to measure the stochastic GW background, we correlate the output of two (or more) GW detectors.
Let us denote the output of two detectors (labeled by $a = 1,2$) by
\begin{equation}
\hat{S}_a = \hat{s}_a + \hat{n}_a,
\end{equation}
where $\hat{n}_a$ represent intrinsic noises of the detectors and $\hat{s}_a = D_a^{ij}h_{ij}$ represent the strains of GWs
specified by detector tensors $D_a^{ij}$.
Here we write the transverse-traceless components of the metric perturbation in terms of the plane wave expansion,
\begin{equation}
h_{ij}(x) = \sum_{\lambda}\int^{\infty}_{-\infty}df \int d\Omega \tilde{h}_{\lambda}(f,{\bf \hat{n}})\epsilon_{ij}^{\lambda}({\bf \hat{n}})e^{-2\pi if(t-{\bf \hat{n}\cdot x})}, \label{h_ij_plane_wave_expansion}
\end{equation}
where $\epsilon_{ij}^{\lambda}({\bf \hat{n}})$ are the polarization tensors for GWs propagating along a direction ${\bf \hat{n}}$,
and $d\Omega$ is the measure of the integration over the direction of ${\bf \hat{n}}$.
Note that the coefficient should satisfy $\tilde{h}^*_{\lambda}(f,{\bf \hat{n}}) = \tilde{h}_{\lambda}(-f,{\bf \hat{n}})$.
The power of the signal is given by the spectral density $S_h(f)$, 
\begin{equation}
\langle \tilde{h}^*_{\lambda}(f,{\bf \hat{n}})\tilde{h}_{\lambda'}(f',{\bf \hat{n}'})\rangle 
= \frac{1}{4\pi}\delta(f-f')\delta^{(2)}({\bf \hat{n}},{\bf \hat{n}'})\delta_{\lambda\lambda'}\frac{1}{2}S_h(f),
\end{equation}
where $\langle\dots\rangle$ represents the ensemble average.
Here, the factor $1/2$ is included because of the fact that we formally consider the negative frequency range 
$-\infty < f < 0$ in Eq.~\eqref{h_ij_plane_wave_expansion}.
The spectral density can be related to the energy density of GWs:
\begin{align}
\Omega_{\rm gw}(f) &= \frac{4\pi^2}{3H_0^2}f^3 S_h(f). \label{Omega_gw_S_h_relation}
\end{align}
It is assumed that there is no correlation between intrinsic noises of two detectors, and the noise power
is characterized by the spectral noise density $S_{n,a}(f)$,
\begin{equation}
\langle\hat{\tilde{n}}^*_a(f)\hat{\tilde{n}}_{a'}(f')\rangle = \delta_{aa'}\delta(f-f')\frac{1}{2}S_{n,a}(f),
\end{equation}
where $\hat{\tilde{n}}_a(f)$ denotes the Fourier transform of $\hat{n}_a(t)$.
We also assume that signals and noises are uncorrelated: $\langle \hat{s}_a \hat{n}_{a'}\rangle = 0$.

The cross-correlation of the output of two detectors is given by
\begin{equation}
S_{12} = \int^{T_{\rm obs}/2}_{-T_{\rm obs}/2}dt \int^{T_{\rm obs}/2}_{-T_{\rm obs}/2}dt' \hat{S}_1(t)\hat{S}_2(t')Q(t-t'),
\end{equation}
where $Q(t-t')$ denotes a filter function, which can be optimized according to the details of detector configurations and the spectrum of the GW background.
Assuming that the observation time $T_{\rm obs}$ is much longer than the time scale in which $Q(t-t')$ falls off, 
we can rewrite the above equation as
\begin{equation}
S_{12} = \int^{\infty}_{-\infty}df \int^{\infty}_{-\infty}df' \delta_{T_{\rm obs}}(f-f')\hat{\tilde{S}}_1^*(f)\hat{\tilde{S}}_2(f')\tilde{Q}(f'),
\end{equation}
where $\delta_{T_{\rm obs}}(f-f') = \int^{T_{\rm obs}/2}_{-T_{\rm obs}/2}dt \exp[-2\pi(f-f')t] = \sin[\pi(f-f')T_{\rm obs}]/\pi(f-f')$, and the tilde denotes the Fourier transform,
\begin{equation}
\hat{S}_a(t) = \int^{\infty}_{-\infty}df e^{2\pi ift} \hat{\tilde{S}}_a(f), \quad Q(t) =  \int^{\infty}_{-\infty}df e^{2\pi ift} \tilde{Q}(f). \nonumber
\end{equation}
Then, the mean signal reads
\begin{equation}
\mu \equiv \langle S_{12} \rangle = T_{\rm obs}F_{12}\int^{\infty}_0 df S_h(f)\gamma(f)\tilde{Q}(f), \label{signal_integrated}
\end{equation}
where $\gamma(f) = \Gamma(f)/F_{12}$ denotes the overlap reduction function,
$F_{aa'} = \sum_{\lambda}\int\frac{d\Omega}{4\pi}F_{\lambda a}F_{\lambda a'}$ the angular efficiency factor, 
$F_{\lambda a}({\bf \hat{n}}) = D_a^{ij}\epsilon_{ij}^{\lambda}({\bf \hat{n}})$ the detector pattern function,
and $\Gamma(f) = \sum_{\lambda}\int\frac{d\Omega}{4\pi}F_{\lambda1}F_{\lambda2}e^{2\pi if{\bf \hat{n}}\cdot \Delta {\bf x}}$
with $\Delta {\bf x}$ being a distance between two detectors.
We can also estimate the variance as
\begin{equation}
\sigma^2 \equiv \langle S_{12}^2 \rangle - \langle S_{12} \rangle^2 = \frac{T_{\rm obs}}{2}\int^{\infty}_0 df |\tilde{Q}(f)|^2 R(f), \label{noise_integrated}
\end{equation}
where
\begin{equation}
R(f) = [F_{11}F_{22} + F_{12}^2\gamma^2(f)] S_h^2(f) + [F_{22}S_{n1}(f) + F_{11}S_{n2}(f)]S_h(f) + S_{n1}(f)S_{n2}(f).
\end{equation}

Equations~\eqref{signal_integrated} and~\eqref{noise_integrated} can be used to estimate the signal-to-noise ratio.
The quantities $F_{aa'}$, $\gamma(f)$, $\tilde{Q}(f)$, and $S_{n,a}(f)$ should be specified once the detailed configuration of detectors is fixed.
Instead of completely specifying them, for the moment we simply adopt the following approximations,
\begin{equation}
\tilde{Q}(f) = 1, \quad \gamma(f) = 1, \quad F_{11} = F_{22} = F_{12} \equiv F, \quad S_{n1}(f) = S_{n2}(f) \equiv S_n(f). \label{detector_simplification}
\end{equation}
These approximations correspond to the assumptions that two detectors possess exactly identical properties with their relative distance neglected
($\Delta {\bf x} \simeq 0$) and that the filter function $Q(t-t')$ is equal to $\delta(t-t')$.
There must be some corrections to the above equations, but they are still reasonable as they provide at least order of magnitude estimates.
By using Eq.~\eqref{detector_simplification}, Eqs.~\eqref{signal_integrated} and~\eqref{noise_integrated} are simplified as
\begin{align}
\mu & = T_{\rm obs}F\int^{\infty}_0 dfS_h(f), \label{signal_integrated_approximate}\\
\sigma^2 &= \frac{T_{\rm obs}}{2}\int^{\infty}_0 df \left[2F^2 S_h^2(f) + 2FS_h(f)S_n(f) + S_n^2(f)\right]. \label{noise_integrated_approximate}
\end{align}

When we consider the frequency dependence of the GW background, we may use a discrete set of frequency bins $\{f_i\}$
labeled by an integer $i$ and define the mean signal and variance at each frequency bin~\cite{Saito:2012bb}
rather than performing the integration over a whole frequency range in Eqs.~\eqref{signal_integrated_approximate} and~\eqref{noise_integrated_approximate}.
Namely, we define the mean signal and variance at $i$-th bin as
\begin{align}
\mu_i &\equiv T_{\rm obs}F\Delta f \bar{S}_h(f_i), \\
\sigma^2_i &\equiv \frac{T_{\rm obs}}{2}\Delta f \bar{R}(f_i),
\end{align}
where
\begin{align}
\bar{S}_h(f_i) &= \frac{1}{\Delta f}\int_{F_i}dfS_h(f), \\
\bar{R}(f_i) &= \frac{1}{\Delta f}\int_{F_i}df \left[2F^2 S_h^2(f) + 2FS_h(f)S_n(f) + S_n^2(f)\right],
\end{align}
and $F_i =[f_i-\Delta f/2, f_i+\Delta f/2]$ denotes a frequency domain for $i$-th bin.
Using the relation~\eqref{Omega_gw_S_h_relation}, we can estimate the corresponding uncertainty of $\Omega_{\rm gw}$:
\begin{equation}
\Delta \Omega_{\rm gw} (f_i) = \frac{4\pi^2}{3H_0^2}\frac{f_i^3}{FT_{\rm obs}\Delta f}\sigma_i = \frac{4\pi^2}{3H_0^2} \frac{f_i^3}{F\sqrt{2T_{\rm obs}\Delta f}}\sqrt{\bar{R}(f_i)} \label{Omega_gw_sensitivity}
\end{equation}
with
\begin{equation}
\bar{R}(f_i) = \frac{1}{\Delta f}\int_{F_i}df \left[2F^2\left(\frac{3H_0^2}{4\pi^2}\right)^2\frac{\Omega_{\rm gw}^2(f)}{f^6} 
+ 2F\left(\frac{3H_0^2}{4\pi^2}\right)\frac{\Omega_{\rm gw}(f)}{f^3}S_n(f) + S_n^2(f)\right]. \label{bar_R_f_i_Ogw}
\end{equation}

Let us apply the above formalism to the case of ultimate DECIGO.
Regarding the possibility that the sensitivity is diminished at $f\lesssim \mathcal{O}(0.1)\,\mathrm{Hz}$ by 
GWs from white-dwarf binaries, we use the following spectral noise density,
\begin{align}
S_n(f) &= S_{\rm WD}(f) + S_{\rm inst}(f), \label{S_n_f}
\end{align}
where we adopt the fitting formula for the white-dwarf confusion noise obtained in Refs.~\cite{Barack:2003fp,Nishizawa:2011eq} for $S_{\rm WD}(f)$
and use the parameters specified in Ref.~\cite{Kuroyanagi:2014qza} to estimate the instrumental noise $S_{\rm inst}(f)$.
Using Eqs.~\eqref{Omega_gw_sensitivity},~\eqref{bar_R_f_i_Ogw}, and~\eqref{S_n_f}, we estimate the sensitivity of ultimate DECIGO, 
which is shown in Fig.~\ref{fig:gw_spectrum_high_frequency}.
Here we take $F = 2/5$, which is the value for the detector with perpendicular arms~\cite{Maggiore:1999vm},
and $\Delta f = f/10$. The observation time is chosen as $T_{\rm obs} = 5\,\mathrm{year}$.

\begin{figure}[htbp]
\begin{center}
\includegraphics[width=120mm]{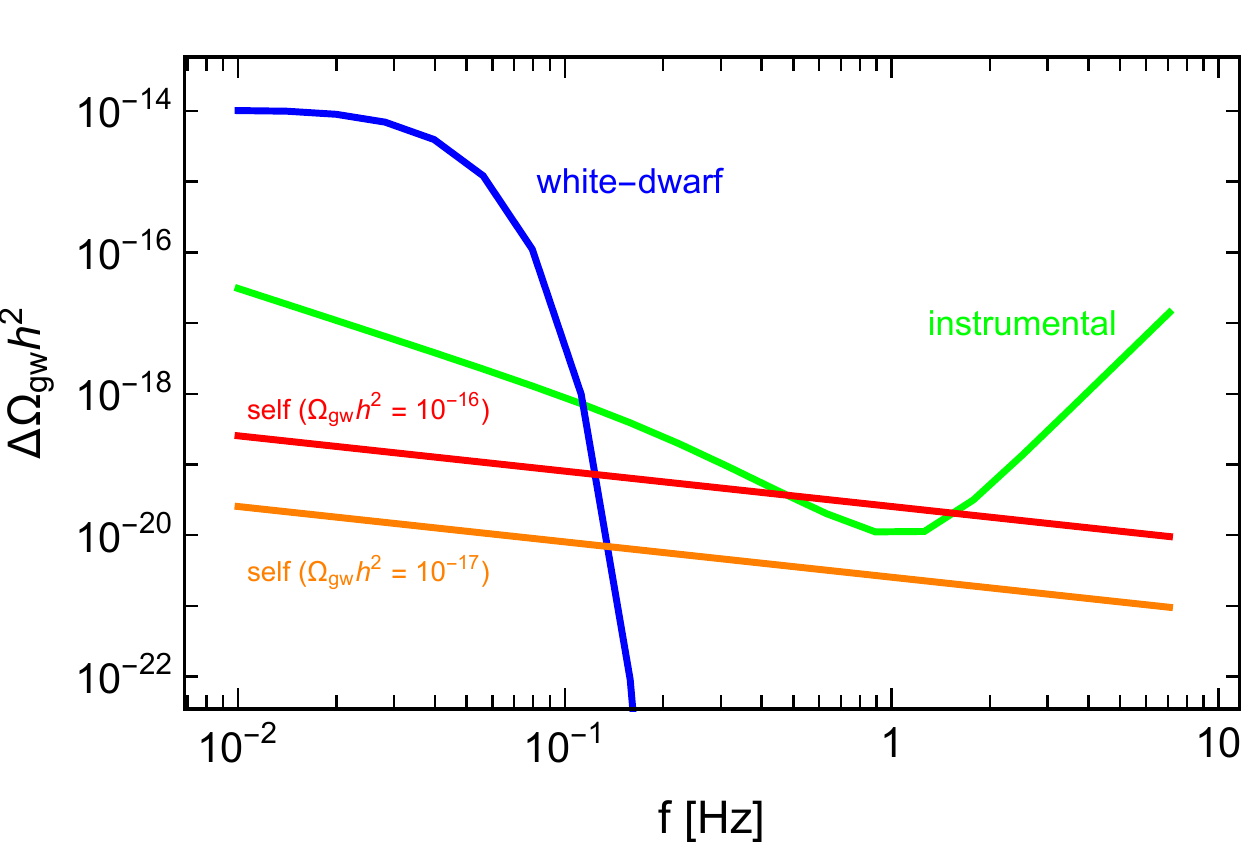}
\end{center}
\caption{Sensitivity of ultimate DECIGO estimated based on Eq.~\eqref{Omega_gw_sensitivity}.
Contributions of the instrumental noises of ultimate DECIGO (green) and the white-dwarf confusion noise (blue)
are estimated by including only the third term in Eq.~\eqref{bar_R_f_i_Ogw} and specifying $S_n(f) = S_{\rm inst}(f)$ and $S_n(f) = S_{\rm WD}(f)$, respectively.
Contribution of the self noise is estimated by including only the first term in Eq.~\eqref{bar_R_f_i_Ogw}.
Here we simply assume that $\Omega_{\rm gw}$ does not depend on $f$, taking a constant value $\Omega_{\rm gw}h^2 = 10^{-16}$ (red) or $10^{-17}$ (orange).}
\label{fig:sensitivity}
\end{figure}

We note that there exists an additional contribution proportional to $\Omega_{\rm gw}$ due to the first and second terms in the bracket 
in Eq.~\eqref{bar_R_f_i_Ogw} (hereinafter, referred to as the ``self" noise).
Usually, these terms are omitted based on the assumption that they are much smaller than the third term $S_n^2(f)$~\cite{Maggiore:1999vm}.
However, this is not the case for ultimate DECIGO, since the instrumental noise is so small that there exists a frequency interval in which
the self noise dominates in Eq.~\eqref{bar_R_f_i_Ogw}.
We show the contributions of self, white-dwarf, and instrumental noises in Fig.~\ref{fig:sensitivity}.
If the amplitude of the GW background $\Omega_{\rm gw}$ is sufficiently large, the sensitivity at $f \sim 1\,\mathrm{Hz}$ is determined by 
the self noise rather than $S_{\rm WD}(f)$ and $S_{\rm inst}(f)$.
From Eqs.~\eqref{Omega_gw_sensitivity} and~\eqref{bar_R_f_i_Ogw}, we see that this self noise is smaller than $\Omega_{\rm gw}$ itself by a factor $1/\sqrt{T_{\rm obs}\Delta f}$.

%%%%%%%%%%%%%%%%%%%%%%%%%%%%%%%%%%%%%%%%%%%%%%%%%%%%
%%%%%%%%%%%%%%%%%%%%%%%%%%%%%%%%%%%%%%%%%%%%%%%%%%%%

%%%%%%%%%%%%%%%%%%%%%%%%%%%%%%%%%%%%%%%%%%%%%%%%%%%%%%%%%%%%%%%%%%%%%%%%%%%%%%%%%%%%%%%%%%%%%%%%%%%%%%%%%%%%%%%%%%%%%%%%%%%%%%%%%
%%%%%%%%%%%%%%%%%%%%%%%%%%%%%%%%%%%%%%%%%%%%%%%%%%%%%%%%%%%%%%%%%%%%%%%%%%%%%%%%%%%%%%%%%%%%%%%%%%%%%%%%%%%%%%%%%%%%%%%%%%%%%%%%%
%%%%%%%%%%%%%%%%%%%%%%%%%%%%%%%%%%%%%%%%%%%%%%%%%%%%%%%%%%%%%%%%%%%%%%%%%%%%%%%%%%%%%%%%%%%%%%%%%%%%%%%%%%%%%%%%%%%%%%%%%%%%%%%%%

\end{document}